\documentclass{LMCS}
\usepackage{epsfig,graphicx,color}
\usepackage{enumerate,amsmath,amssymb,latexsym,smallmath}
\newcommand{\hide}[1]{}

\newcommand{\funtype}[3]{{#1}:{#2}\rightarrow{#3}}
\newcommand{\set}[1]{\left\{#1\right\}}

\newcommand{\tuple}[1]{\left(#1\right)}
\newcommand{\setcomp}[2]{\left\{#1|\; #2\right\}}
\newcommand{\denotationof}[1]{[\![#1]\!]}
\newcommand{\bigdenotationof}[1]{\left[\!\!\!\left[#1\right]\!\!\!\right]}

\newcommand{\sizeof}[1]{|{#1}|}
\newcommand{\maxsize}[1]{{\it maxsize}({#1})}
\newcommand{\fract}{{\it frac}}
\newcommand{\arcs}{{\it A}}

\newcommand{\nat}{{\mathbb N}}
\newcommand{\integers}{{\mathbb Z}}
\newcommand{\preals}{{\mathbb R}^{> 0}}
\newcommand{\nnreals}{{\mathbb R}^{\geq 0}}

\newcommand{\lang}{L}
\newcommand{\langof}[1]{L(#1)}

\newcommand{\states}{Q}
\newcommand{\state}{q}
\newcommand{\instr}{{\it instr}}
\newcommand{\counters}{C}
\newcommand{\counter}{c}
\newcommand{\cmtrans}{\leadsto}
\newcommand{\cmtransclosure}{\stackrel {*} {\cmtrans}}
\newcommand{\ctransitions}{\delta}
\newcommand{\cvalue}{{\it Val}}
\newcommand{\lcm}{L}
\newcommand{\lcmtuple}{\tuple{\states,\state_0,\counters,\ctransitions}}
\newcommand{\conf}{\gamma}
\newcommand{\inc}[1]{{#1}{++}}
\newcommand{\dec}[1]{{#1}{--}}

\newcommand{\val}{\it Val}

\newcommand{\comp}{\pi}

\newcommand{\Intervals}{{\it Intrv}}
\newcommand{\interval}{{\mathcal I}}

\newcommand{\jinterval}{{\mathcal J}}

\newcommand{\oointrvl}[2]{({#1}:{#2})}
\newcommand{\ointrvl}[2]{({#1}:{#2})}
\newcommand{\mset}{{\it b}}
\newcommand{\aset}{{\it A}}
\newcommand{\lst}[1]{[{#1}]}
\newcommand{\mleq}{\leq}
\newcommand{\emptystring}{\epsilon}

\newcommand{\tpn}{N}
\newcommand{\tpntuple}{\tuple{\places,\transitions,\inputs,\outputs}}
\newcommand{\places}{P}
\newcommand{\place}{p}
\newcommand{\qplace}{q}
\newcommand{\transitions}{T}
\newcommand{\transition}{t}
\newcommand{\ltransition}{\ell}
\newcommand{\wtrans}{\longrightarrow}
\newcommand{\inputs}{\it In}

\newcommand{\outputs}{\it Out}

\newcommand{\marking}{M}
\newcommand{\markings}{{\sf M}}
\newcommand{\markingtuple}{\lst{\tuple{\place_1,x_1},\ldots,\tuple{\place_n,x_n}}}

\newcommand{\trans}[1]{\stackrel{#1}{\longrightarrow}}
\newcommand{\timed}{\delta}
\newcommand{\timedtrans}{\longrightarrow_\timed}
\newcommand{\disc}{D}
\newcommand{\disctrans}{\longrightarrow_\disc}
\newcommand{\ttrans}{\longrightarrow_{\transition}}
\newcommand{\transclosure}{\stackrel {*} {\longrightarrow}}

\newcommand{\post}{\textit{Post}}

\newcommand{\pre}{\textit{Pre}}

\newcommand{\region}{R}

\newcommand{\regiontuple}{\tuple{\zerobag, b_1\ldots b_m, \maxbag}}
\newcommand{\zerobag}{b_0}
\newcommand{\maxbag}{b_{m+1}}

\newcommand{\word}{w}

\newcommand{\maxval}{{\it max}}

\newcommand{\w}{{\it w}}
\newcommand{\z}{{\it z}}

\newcommand{\msets}[1]{{#1}^\odot}

\newcommand{\reachable}{{\it Reach}}

\newcommand{\morder}{\preceq}

\newcommand{\st}{{\it ST}}
\newcommand{\symbols}{{\it Sym}}
\newcommand{\sym}{{\it sym}}
\newcommand{\age}{{\it x}}
\newcommand{\subInP}{{\mathcal P}_{\it in}}
\newcommand{\subOutP}{{\mathcal P}_{\it out}}
\newcommand{\xtimedtrans}[1]{\longrightarrow_{{#1}}}

\newcommand{\wstar}[1]{\left\{{#1}\right\}^*}
\newcommand{\wprod}{\;\bullet\;}

\newcommand{\regs}{{\it Reg}}
\newcommand{\rleq}{\preceq}
\newcommand{\ucdenotationof}[1]{[\![#1]\!]^{\uparrow}}

\newcommand{\eqdenotationof}[1]{[\![#1]\!]}

\newcommand{\mprod}{+}

\newcommand{\wleq}{\leq^\word}

\newcommand{\vect}{{\it v}}
\newcommand{\vecdef}{\vec}

\newcommand{\automaton}{\mathcal A}
\newcommand{\vmin}{{\mathcal V}}
\newcommand{\pred}{{\it pred}}
\newcommand{\vset}{{\it W}}
\newcommand{\covergraph}{{\mathcal C}}
\newcommand{\qmarking}{{\mathcal M}}
\newcommand{\effect}{{\it Effect}}

\newcommand{\parikh}{\varphi}

\newcommand{\presburger}{\rho}
\newcommand{\xrultrans}[1]{\longrightarrow_{{#1}}}

\usepackage{hyperref}

\def\doi{3 (1:1) 2007}
\lmcsheading%
{\doi}
{1--61}
{}
{}
{Jan.~12, 2006}
{Jan.~23, 2007}
{}   

\begin{document}
\title[Dense-Timed Petri Nets]{Dense-Timed Petri Nets: Checking Zenoness, 
Token liveness and Boundedness}
\author[P.~A.~Abdulla]{Parosh Aziz Abdulla\rsuper{a}}
\address{{\lsuper{a}}Uppsala University, Department of Information Technology, 
Box 337, SE-751 05 Uppsala, Sweden}
\email{parosh@it.uu.se}
\author[P.~Mahata]{Pritha Mahata\rsuper{b}}
\address{{\lsuper{b}}University of Newcastle, School of Electrical Engineering and
  Computer Science, University Drive, Callaghan NSW 2308, Australia}
\email{pritha.mahata@newcastle.edu.au}
\author[R.~Mayr]{Richard Mayr\rsuper{c}}
\address{{\lsuper{c}}North Carolina State University,
Department of Computer Science,
Campus Box 8206,
Raleigh, NC 27695, USA}
\urladdr{http://www4.ncsu.edu/\~{}rmayr}

\keywords{Real-time systems, Timed Petri nets, Verification, Zenoness}
\subjclass{F1.1, F3.1, F4.1, F4.3}
\titlecomment{An extended abstract (without proofs) of some parts of this paper 
(sections 3, 7 and 8) has appeared in FST\&TCS 2004 \cite{AMM:FSTTCS2004}.}

\begin{abstract}
\noindent 
We consider {\it Dense-Timed Petri Nets (TPN)},
an extension of Petri nets in which each
token is equipped with a real-valued clock and where the semantics is lazy
(i.e., enabled transitions need not fire; time can pass and disable transitions).
We consider the following verification problems
for TPNs.

(i) {\it Zenoness:}
whether there exists a zeno-computation from a given
marking, i.e., an infinite computation which takes only 
a finite amount of time.
We show decidability of zenoness for TPNs,
thus solving an open problem from \cite{Escrig:etal:TPN}.
Furthermore, the related question if there exist arbitrarily fast
computations from a given marking is also decidable. 

On the other hand, {\em universal zenoness}, i.e., the question if {\em all}
infinite computations from a given marking are zeno, is undecidable.

(ii)
{\it Token liveness:}
whether a token is {\it alive} in a marking, i.e.,
whether there is a computation from the marking which eventually consumes
the token.
We show decidability of the problem by reducing it to the 
{\it coverability problem},
which is decidable for TPNs.

(iii)
{\it Boundedness:}
whether the size of the reachable markings is bounded.
We consider two versions of the problem; namely
{\it semantic boundedness} 
where only live tokens are taken into consideration in the
markings,
and 
{\it syntactic boundedness} 
where also dead tokens are considered.
We show undecidability of semantic boundedness,
while we prove that syntactic boundedness is decidable
through an extension of the Karp-Miller algorithm.
\end{abstract}

\maketitle
\vfill\eject

\section{Introduction}

\noindent
Petri nets \cite{Petri,Peterson:survey,Murata:IEEE89}
are one of the most widely used models for
analysis and verification of concurrent systems.
Many different formalisms have been proposed which extend 
Petri nets with clocks and real-time constraints, leading to 
various definitions of {\it Timed Petri nets (TPNs)}.
A complete discussion of all these formalisms is beyond the scope of this 
paper and the interested reader is referred to the survey by Bowden
\cite{Bowden:TPN:Survey} and a more recent overview in \cite{BCHLR-atva2005}.

In this paper we consider the TPN model used in 
\cite{Parosh:Aletta:bqoTPN} where each token has an age which is represented
by a real-valued clock, and the firing-semantics is lazy (like 
in standard untimed Petri nets).
This dense time TPN model of
\cite{Parosh:Aletta:bqoTPN} is an adaption of the
discrete time model of Escrig et
al. \cite{Escrig:etal:TPN:nondecidability,Escrig:etal:TPN}.

The main difference between dense time TPN 
and discrete time TPN is the following.
In discrete time nets, time is interpreted as
being incremented in discrete steps and thus the ages of tokens are in a countable domain,
commonly the natural numbers. Such discrete time nets have been studied 
in, e.g., \cite{Escrig:etal:TPN:nondecidability,Escrig:etal:TPN}.
In dense time nets, time is interpreted as continuous, and the ages of
tokens are real numbers. Some problems for dense time nets have been
studied in \cite{Parosh:Aletta:bqoTPN,Parosh:Aletta:infinity,ADMN:dlg}.

In this paper we mainly consider the dense time case. However, 
we also solve some open questions for discrete time nets, since
they follow as corollaries from our more general results on the dense time case.

The main characteristics of our TPN model (i.e., the model of
\cite{Parosh:Aletta:bqoTPN}) are the following.
\begin{enumerate}[$\bullet$]
\item
Our TPNs are {\em not} bounded. The number of tokens
present in the net may grow beyond any finite bound.
\item
Each token has an age which is represented by a real-valued clock, i.e., time
is continuous.
\item
A transition is enabled iff there are enough tokens of the right ages on its
input places. The right ages are specified by labeling the input arcs of
transitions with time intervals.
\item
The semantics is {\em lazy}, just like in standard untimed Petri nets.
This means that an enabled transition need not fire immediately. 
It is possible that more time will pass and disable the transition again.
(This is in contrast to many other classes of Petri nets with time, which have
an eager semantics where transitions must fire when they are enabled; see
\cite{BCHLR-atva2005} for an overview.)
\item
When a transition fires, the clocks of the consumed tokens are not
preserved. Tokens which are newly created by a transition have their own
new clocks.
\end{enumerate}
The formal definition of this TPN model is given in
Section~\ref{defs:section}.

TPN can, among other things, model
parameterized timed systems (systems consisting of
an unbounded number of timed processes) 
\cite{Parosh:Aletta:bqoTPN}.

Our TPN model is computationally more powerful than timed 
automata \cite{Alur:Dill:Region:Graph,AD:timedautomata}, since it 
operates on a potentially unbounded number of clocks.
In particular, TPN subsume normal untimed Petri 
nets w.r.t.\ the semantics of fired transition sequences, while finite timed automata 
do not subsume Petri nets. 
Furthermore, both the reachability problem
\cite{Escrig:etal:TPN:nondecidability}
and several liveness problems \cite{Escrig:etal:TPN,Parosh:Aletta:infinity}
are undecidable for TPNs (even in the discrete time case).

Most verification problems for TPNs are extensions of both
classical problems previously studied for standard (untimed) Petri nets, 
and problems for finite-state timed models like timed automata.
We consider several verification problems for TPNs.

\medskip

\noindent{\bf Zenoness.}
A fundamental progress property for timed systems is that it should be possible
for time to {\it diverge} \cite{tripakis}.
This requirement is justified by the fact that timed processes
cannot be infinitely fast.
Computations violating this property are called {\it zeno}.
Given a TPN and a marking $M$, we check whether
$M$ is a {\it zeno-marking}, i.e., whether there is an infinite computation
from $M$ with a finite duration.
The zenoness problem is solved 
in \cite{Alur:Dill:Region:Graph,Alur:thesis} for timed automata  using the region graph
construction.
Since region graphs only deal with
a finite number of clocks, the algorithm of \cite{Alur:Dill:Region:Graph,Alur:thesis} cannot
be extended to check zenoness for TPNs.
In Section~\ref{sec:zeno}, we solve the zenoness problem for TPNs.
To do this, we consider a subclass of transfer nets
\cite{FS98} which we call {\it simultaneous-disjoint 
transfer net (SD-TN)}.
This class is an extension of standard Petri nets, in which 
we also have {\it transfer} transitions 
which may move all tokens in one place to another with 
the restriction that (a) all such transfers  take 
place simultaneously and (b) the sources and targets of 
 all transfers are disjoint.

Given a TPN $N$, we perform the following three
steps:
\begin{enumerate}[-]
\item Derive a corresponding SD-TN $N'$.
\item Characterize the set of markings 
in $N'$ from which there are infinite computations\footnote{
In contrast to SD-TN, such a characterization
is not computable for general transfer nets \cite{Mayr:LCM:TCS}.}.
\item Re-interpret the set computed above as a characterization of the set
of zeno-markings in $N$.
\end{enumerate}
In fact,
the above procedure solves a more general problem than that
of checking  whether a given marking is zeno; 
namely it gives a symbolic characterization of 
the set of zeno-markings.

The zenoness problem was left open 
in  \cite{Escrig:etal:TPN}
both for dense TPNs
(the model we consider in this paper) and for discrete TPNs
(where behavior is interpreted over the discrete time domain).
The construction given in this paper considers the more general dense time case.
The construction can easily be modified (in fact simplified)
to deal with the discrete time case. (In the discrete time case, unlike for
dense time, every zeno computation must have an infinite suffix that takes
zero time.)

\medskip

\noindent{\bf Arbitrarily Fast Computations.}
In Section~\ref{sec:all_zeno} we consider a 
question related to zenoness: `Given a marking $M$,
is it the case that for every $\epsilon > 0$ there is an $M$-computation
which takes at most $\epsilon$ time?' 
This is a stronger requirement than zenoness, and we call markings which
satisfy it {\em allzeno}-markings. Like for zeno-markings, one can compute a 
symbolic characterization of the set of allzeno-markings, and thus the problem
is decidable. 

Markings from which there are computations which take no time at all are
called {\em zerotime}-markings. For discrete time nets, allzeno-markings and
zerotime-markings coincide, but for general dense time nets zerotime-markings
are (in general) a strict subset. Again one can compute a 
symbolic characterization of the set of zerotime-markings.

\medskip

\noindent{\bf Universal Zenoness.}
In the zenoness problem, the question was whether there existed at least one
zeno run, i.e., an infinite computation which takes finite time.
The universal zenoness problem is the question whether {\em all} infinite runs are zeno.
The negation of this question is the following:
Given
some marking $M$, does there exist some non-zeno $M$-computation, i.e., an
infinite computation from $M$ which takes an infinite amount of time?
In Section~\ref{sec:non_zeno} we show that this question (and thus 
universal zenoness) is undecidable, by a 
reduction from lossy counter machines \cite{Mayr:LCM:TCS}.

\medskip

\noindent{\bf Token Liveness.}
Markings in TPNs may contain tokens which cannot 
be used by any future computations
of the TPN. 
Such tokens do not affect the behavior of the TPN and
are therefore called {\em dead tokens}. 
We give an algorithm to check, given a token and a marking,
whether the token is dead (or alive).
We do this by reducing the problem to the problem of {\it coverability}
in TPNs.
An algorithm to solve the coverability problem is given in
\cite{Parosh:Aletta:bqoTPN}.

Token liveness for dense TPNs was left open
in  \cite{Escrig:etal:TPN}.

\medskip

\noindent{\bf Boundedness.}
We consider the {\it boundedness problem} for TPNs:
given a TPN and an initial marking, check whether
the size of  reachable markings is bounded.
The decidability of this problem depends on whether we 
take dead tokens into consideration.
In {\em syntactic boundedness} one considers 
dead tokens as part of the (size of the) marking, while in
{\em semantic boundedness} 
we disregard dead tokens; that is we check whether
we can reach markings with unboundedly many live tokens.
Using techniques similar to \cite{Escrig:etal:TPN:nondecidability}
it can be shown that semantic boundedness is undecidable.
On the other hand we show decidability of syntactic boundedness.
This is achieved through an extension of the Karp-Miller algorithm
where each node represents a region (rather than a single marking).
The underlying ordering on the nodes (regions) inside
the Karp-Miller tree is  
a {\it well quasi-ordering} \cite{Higman:divisibility}.
This guarantees termination of the procedure.

Decidability of syntactic boundedness was shown for
the simpler discrete time case in 
\cite{Escrig:etal:TPN}, while the problem was left open for the dense case.

\section{Timed Petri Nets and Regions}
\label{defs:section}

\subsubsection*{Timed Petri Nets}

We consider {\em Timed Petri Nets} ({\it TPN}s) where each token is
equipped with a real-valued clock representing the age of the token.
The firing conditions of a transition include the usual ones for Petri nets.
Additionally, each arc between a place and a transition is labeled with a 
time-interval whose bounds are natural numbers (or possibly $\infty$ as upper
bound). 
These intervals can be open, closed or half
open. When firing a transition, tokens which 
are removed (added)  from (to) places must have ages lying in the 
intervals of the corresponding transition arcs.

We use $\nat,\nnreals, \preals$ to denote the sets
of natural numbers (including 0), nonnegative reals, 
and strictly positive reals, respectively. 
For a natural number $k$, we use $\nat^k$ and $\nat^k_\omega$ 
to denote the set of vectors of size $k$ over $\nat$ and  $\nat
\cup\set{\omega}$, respectively
($\omega$ represents the first limit ordinal).

We use a set {\it Intrv} of intervals. An open interval is 
written as $\oointrvl  w z$ 
 where $\w\in\nat$ and $\z\in\nat\cup\set{\infty}$. Intervals can also 
 be closed in one or both directions, 
 e.g. $[w:z]$ is closed in both directions  and
  $[w:z)$ is closed to the left and open to the right.

\begin{defi}\label{def:multiset}
For a set $A$, we use $A^*$ and $\msets\aset$
to denote the set of finite words and finite multisets over $A$, respectively.
We view a multiset $\mset$ over $A$ as
a mapping $\mset: A \mapsto \nat$.
Sometimes, we write finite multisets as lists with multiple occurrences, so 
$\lst{2.4^3\;,\;5.1^2}$ represents a multiset $\mset$ over $\nnreals$
where $\mset(2.4)=3$, $\mset(5.1)=2$ and $\mset(x)=0$ for $x\neq
2.4,5.1$.
%
%
For multisets $\mset_1$ and $\mset_2$ over $\nat$, we say that
$\mset_1\mleq\mset_2$ if $\mset_1(a)\leq\mset_2(a)$ for each $a\in A$.
The multiset union $\mset = \mset_1 \cup \mset_2$ is defined by
$\mset(a)=\max{(\mset_1(a), \mset_2(a))}$ for each $a \in A$ 
and the multiset intersection $\mset = \mset_1 \cap \mset_2$ is defined by
$\mset(a)=\min{(\mset_1(a), \mset_2(a))}$ for each $a \in A$. 

We define $\mset_1+\mset_2$ to be the multiset $\mset$ where
$\mset(a)=\mset_1(a)+\mset_2(a)$, and (assuming $\mset_1\mleq\mset_2$) we
define $\mset_2 - \mset_1$ to be the multiset $\mset$ where
$\mset(a)=\mset_2(a)-\mset_1(a)$, for each $a\in A$.

For a multiset $\mset: A \mapsto \nat$, we 
write $\sizeof\mset := \sum_{a \in A} \mset(a)$ for the number of elements in $\mset$. 
We use $\emptyset$ to denote the empty multiset and $\emptystring$ 
to denote the empty word.

Given a set $A$ with partial order $\leq$, we define a partial order
$\wleq$ on $A^*$ as follows. We have $a_1\dots a_n \wleq b_1\dots b_m$ iff
there is a subsequence $b_{j_1}\dots b_{j_n}$ of 
$b_1\dots b_m$ s.t. $\forall k \in \{1,\dots,n\}.\, a_k \le b_{j_k}$.

Given a set $A$ with an ordering $\preceq$ and 
a subset $B\subseteq A$, $B$ is  said to be {\it upward closed}
in $A$
if $a_1\in B, a_2 \in A$ and $a_1\morder a_2$ implies
$a_2\in B$.
Given a set $B\subseteq A$,
we define the {\it upward closure} $B\uparrow$ to be the set
$\setcomp{a\in A}{\exists a'\in B:\; a'\preceq a}$.
A {\it downward closed} set $B$  and the {\it downward closure} 
$B\downarrow$ are defined in a similar manner.
We also use $a\uparrow$, $a\downarrow$, $a$ instead of 
$\set{a}\uparrow$, $\set{a}\downarrow$, $\set a$, respectively.
\end{defi} 

\begin{defi} \cite{Parosh:Aletta:bqoTPN}
 A {\em Timed Petri Net (TPN)} is a tuple $\tpn=\tpntuple$ where 
$\places$ is a finite set of places, $\transitions$ is a 
finite set of transitions and  $\inputs,\outputs$ are partial functions
from $\transitions\times\places$ to $\Intervals$.
\end{defi}

If $\inputs(\transition,\place)$ 
(respectively $\outputs(\transition,\place)$) is defined, we say that $\place$ is an 
{\em input (respectively output) place} of $\transition$.

We let $\maxval$ denote the maximum integer appearing on the arcs of a given TPN.

A {\it marking} $\marking$ of $\tpn$ is a finite multiset over 
$\places\times\nnreals$. 
%
%
The marking $\marking$ defines the numbers  and ages
of tokens in each place in the net.
We identify a token in a marking
$\marking$ by the pair $(\place,x)$ representing its place and age
in $\marking$.
Then, $\marking((\place,x))$ 
defines the number of tokens with age $x$ in place $\place$.
%
Abusing notation again, we define, for each place $\place$, a multiset $\marking(\place)$ 
over $\nnreals$,
where $\marking(\place)(x)=\marking((p,x))$.

For a marking $\marking$ of the form
$\lst{(\place_1,x_1)\;,\;\ldots\;,\;(\place_n,x_n)}$
 and $x\in\preals$, we use $\marking^{+x}$
to denote the marking
$\lst{(\place_1,x_1+x)\;,\;\ldots\;,\;(\place_n,x_n+x)}$.

\medskip
\noindent{\bf Transitions:} We define two transition relations on the 
set of markings: timed transition
and discrete transition.
A {\em timed transition} increases the age of each token
by the same real number.
Formally, for $x\in\preals$, $\marking_1\longrightarrow_x\marking_2$ 
if $\marking_2=\marking_1^{+x}$.
We use $\marking_1\timedtrans\marking_2$ to denote that
$\marking_1\longrightarrow_x\marking_2$ for some $x\in\preals$.

We define the set of {\em discrete transitions $\disctrans$} as
$\bigcup_{\transition\in\transitions}\ttrans$, where
$\ttrans$ represents the effect of {\it firing} the discrete 
transition $\transition$.
More precisely, 
$\marking_1\ttrans\marking_2$ if
 the set of input arcs 
 $\setcomp{(\place,\interval)}{\inputs(\transition,\place)=\interval}$
 is of the form 
 $\set{(\place_1,\interval_1),\ldots,(\place_k,\interval_k)}$, the set of output 
arcs $\setcomp{(\place,\interval)}{\outputs(\transition,\place)=\interval}$
is of the form 
$\set{(\qplace_1,\jinterval_1),\ldots,(\qplace_{\ell},\jinterval_{\ell})}$, 
and there are multisets 
$\mset_1=\lst{(\place_1,x_1)\;,\ldots\;,(\place_k,x_k)}$ and
$\mset_2=\lst{(\qplace_1,y_1)\;,\ldots\;,(\qplace_{\ell},y_{\ell})}$ 
over $\places\times\nnreals$
such that the following holds:
\newline - 
$\mset_1\mleq\marking_1$
\newline -
$x_i\in\interval_i$, for $i:1\leq i\leq k$.
\newline -
$y_i\in\jinterval_i$, for $i:1\leq i\leq \ell$.
\newline - 
$\marking_2= (\marking_1 - \mset_1) + \mset_2$.
\newline 
We say that $\transition$ is  {\it enabled} in $\marking$ if there
is a $b_1$ such that the first two conditions 
are satisfied.
A transition $\transition$ may be fired only if for each
incoming arc, there is a token with the right age
in the corresponding input place.
These tokens will be removed when the transition is fired.
The newly produced tokens have ages which are chosen nondeterministically from the
relevant intervals on the transitions' output arcs.

We write $\longrightarrow \, =\, \timedtrans\,\cup\,\disctrans$ to denote all 
transitions, $\trans{*}$ to denote the reflexive-transitive closure 
of $\longrightarrow$ and $\disctrans^+$ to denote the transitive closure 
of $\disctrans$. It is easy to extend $\trans{*}$ for sets of markings.
We define $\reachable(M) := \{M'\,|\, M \trans* M'\}$ as the set of markings 
reachable from $M$. 

\medskip
\noindent{\bf Computations:} 
Generally, a computation from a given marking is just a (finite or
infinite) sequence of enabled transitions.
For technical reasons, we need to distinguish two types of computation:
disc-computations where the first transition is a discrete transition
and time-computations where the first transition is a timed transition.

A {\em $\marking_0$-disc-computation} $\pi$ from a marking $\marking_0$ is a 
computation that starts with a discrete transition. It is a  
(finite or infinite) sequence
\[
\marking_0 \disctrans^+ \marking_0'
\longrightarrow_{x_0} \marking_1
\disctrans^+ \marking_1'
\longrightarrow_{x_1} \marking_2
\disctrans^+ \marking_2'
\longrightarrow_{x_2} \marking_3
\disctrans^+ \ldots
\]
of markings and transitions where $x_i \in \preals$. (If the sequence is
infinite but contains only finitely many timed transitions then the infinite
suffix has the form $\disctrans^\omega$.)
It follows that
\begin{enumerate}[$\bullet$]
\item
The first transition is a discrete transition. Thus
$\marking_0 \disctrans^+ \marking_0'$.
\item
Every timed transition has a non-zero delay, i.e., $x_i \in \preals$.
\item
Without restriction, timed transitions cannot directly follow each other. 
We can require this, since $\longrightarrow_{x_1}\longrightarrow_{x_2}$ has
the same effect as $\longrightarrow_{(x_1+x_2)}$.
Therefore, timed transitions
must be separated by at least one discrete transition. Thus we require
$\marking_i \disctrans^+ \marking_i'$ for $i \ge 0$.
\item
This implies that every infinite computation $\pi$ must contain infinitely many
discrete transitions $\disctrans$. An infinite computation may contain either
finitely many or infinitely many timed transitions.
\end{enumerate}
The {\it delay} of the disc-computation $\pi$ is defined as 
\[
\Delta(\pi) := \sum_{i=0}^\infty x_i
\]
A {\em $\marking_0$-time-computation} $\pi$ from a marking $\marking$
has the form
\[
M \longrightarrow_{x} M_0 \dto{\pi'} \ldots
\]
where $x \in \preals$ and $\pi'$ is a $\marking_0$-disc-computation.
In this case the delay $\Delta(\pi) := x + \Delta(\pi')$.

Intuitively, the delay
is the total amount of time passed in all timed transitions in the 
sequence. For infinite computations $\pi$, the delay $\Delta(\pi)$ can be
either infinite or finite. In the latter case the computation $\pi$ is called
a {\em zeno computation} (see Section~\ref{sec:zeno}). 
By $\marking \dto{\pi}$ we denote the fact that $\pi$ is an $\marking$-computation.

\begin{figure}[htbp]
\begin{center}
\scalebox{0.9}{
\input{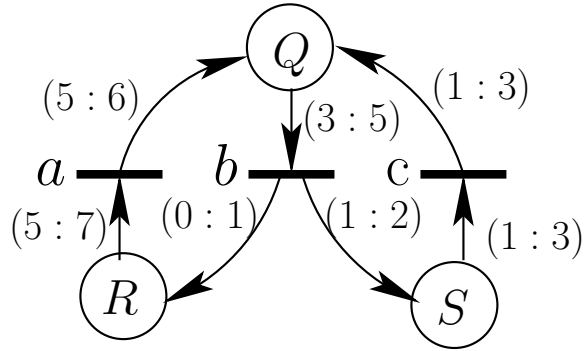}
}
\caption{A small timed Petri net.}
\label{fig:small:example}
\end{center}
\end{figure}

Figure~\ref{fig:small:example} shows an example of a TPN where
$\places=\set{Q,R,S}$ and $\transitions=\set{a,b,c}$. For instance,
$\inputs(b,Q)=\ointrvl 3 5$ and
$\outputs(b, R)=\ointrvl 0 1$ and $\outputs(b, S) =  \ointrvl 1 2$.
A marking of the given net is
$\marking_0=\lst{({Q},{2.0}),\;(R,{4.3}),\;(R, {3.5})}$. 
A timed transition from $\marking_0$ is given by 
$\marking_0 \longrightarrow_{1.5} \marking_1$ where 
$\marking_1 = \lst{(Q, {3.5}), \;(R,{5.8}),\; (R,{5.0})}$.  
An example of a discrete transition is given by 
$\marking_1 \longrightarrow_b \marking_2$ where 
$\marking_2 = \lst{(R, {0.2}),\;(S,{1.6}), \;(R, {5.8}), \;(R, {5.0})}$.

Our model subsumes untimed Petri nets in the following sense.
If all intervals are of the form $[0:\infty)$ then the age of the tokens does
not matter for the transitions, and thus the possible behavior (i.e.,
sequences of fired transitions) is the same as that of an untimed Petri net with the
same structure. However, there cannot be any bijection between the sets of
markings of a timed- and the corresponding untimed net, since the former is
(in general) uncountable.

\smallskip
Next, we  recall a constraint system 
called {\it regions}  defined for 
{\it Timed automata} \cite{Alur:Dill:Region:Graph}.  

\medskip
\noindent{\bf Regions:}
A {\it region} defines the integral  
parts of clock values up to $\maxval$
(the exact age of a token is irrelevant if it is greater
than $\maxval$),  and also
the ordering of the fractional parts.
For TPNs, we need to use a variant which also
defines the place in which each token (clock) resides. 
Following Godskesen \cite{Godskesen:thesis}, we represent 
a region in the following manner.

\begin{defi}
\label{region:definition}
A {\it region} is  a triple
$\tuple{\mset_0,\word,\mset_{\maxval}}$ where 
\begin{enumerate}[$\bullet$]
\item $\mset_0\in\msets{(\places\times\set{0,\ldots,\maxval})}$.
$\mset_0$ is a multiset of pairs. A pair of the form $\tuple{\place,n}$ 
represents a token with age exactly $n$ in place $\place$.
\item $\word\in\left(\msets{\left(\places\times\set{0,\ldots,\maxval
          -1}\right)} - \{\emptyset\}\right)^*$.
This means that $\word$ is a word over the set 
$\msets{(\places\times\set{0,\ldots,\maxval-1})}- \{\emptyset\}$, i.e.,
$\word$ is a word where each element in the word is a non-empty multiset
over 
$\places\times\set{0,\ldots,\maxval-1}$.
The pair $\tuple{\place,n}$ represents a token in place $\place$ 
with age $x$ such that $x\in\oointrvl n {n+1}$. 
Pairs in the
same multiset represent tokens whose ages have equal fractional
parts. 
The order of the multisets in $\word$ corresponds to the order
of the fractional parts (i.e., smaller fractional parts come first in the 
word $\word$).
\item
$\mset_{\maxval}\in\msets{\places}$.
$\mset_{\maxval}$ is a multiset over $\places$ representing tokens with ages
strictly greater than $\maxval$. 
Since the actual ages of these tokens
are irrelevant, the information about their ages is omitted in the 
representation. (This is because the transitions in the net cannot distinguish
between different ages of tokens if these are strictly bigger than $\maxval$.
Note that tokens with age exactly $\maxval$ are represented in $\mset_0$.)
\end{enumerate}
The semantic of a region $\tuple{\mset_0,\word,\mset_{\maxval}}$ 
would not change if we allowed empty multisets to
appear in $\word$. Therefore we forbid this in order to obtain a unique
representation.
However, the multisets $\mset_0$ and $\mset_{\maxval}$ can be empty. 

Formally, each region $\region$ characterizes an infinite set of markings 
$\denotationof\region$ as follows.
Assume a marking $\marking=\markingtuple$ and a region
$\region=\tuple{\mset_0,\mset_1\mset_2\cdots\mset_m,\mset_{m+1}}$.
Let each multiset $\mset_j$ be of the form 
$\left[\tuple {q_{(j,1)},y_{(j,1)}},\ldots,
 \tuple {q_{(j,\ell_j)},y_{(j,\ell_j)}}\right]$ for $j:0\leq j\leq m$ and 
 $\mset_{m+1}$ is of the form $\lst{q_{(m+1,1)},\ldots,q_{(m+1,l_{m+1})}}$.
We say that $\marking$ {\it satisfies} $\region$,
i.e., $\marking\in\denotationof\region$, iff
there is a bijection $h$ from the set
 $\set{1,\ldots,n}$ 
to the set of pairs 
$\setcomp{\tuple{j,k}}{(0\leq j \leq
m+1) \wedge (1\leq k\leq \ell_j)}$  
such that the following conditions are satisfied.
\begin{enumerate}[$\bullet$]
\item $\place_i=\qplace_{h(i)}$.
Each token should have the same place as that
required by the corresponding element in $\region$.
\item
If $h(i)=(j,k)$ then
$j=m+1$ iff $x_i > \maxval$.
Tokens older than $\maxval$ should correspond to elements in multiset
$\mset_{m+1}$. The actual ages of these tokens are not
relevant.
\item
If $x_i\leq\maxval$ and $h(i)=(j,k)$ then
$\lfloor x_i\rfloor=y_{(j,k)}$.
The integral part of the age of tokens should agree with the natural number 
specified by the corresponding elements in $\word$.
\item
If $x_i\leq\maxval$ and $h(i)=(j,k)$ then
$\fract(x_i)=0$  iff $j=0$.
Tokens with zero fractional parts correspond to elements in multiset $\mset_0$.
\item
If $x_{i_1},x_{i_2}<\maxval$, $h(i_1)=(j_1,k_1)$ and 
$h(i_2)=(j_2,k_2)$
then $\fract(x_{i_1}) < \fract(x_{i_2})$ iff $j_1 < j_2$.
This condition implies $\fract(x_{i_1}) = \fract(x_{i_2})$ iff $j_1 = j_2$.
Thus, tokens with equal fractional parts correspond to
elements in the same multiset (unless they belong to $\mset_{m+1}$).
Furthermore, the ordering among the multisets inside $\region$ reflects the 
ordering among the fractional parts of the clock values (increasing from left to right).
\end{enumerate}
We sometimes identify a region $\region$ with the set of markings $\denotationof\region$ it 
represents (i.e., we write $\region$ instead of $\denotationof\region$). 
\end{defi}

\begin{figure}[htbp]
\begin{center}
\scalebox{1.2}{
\input{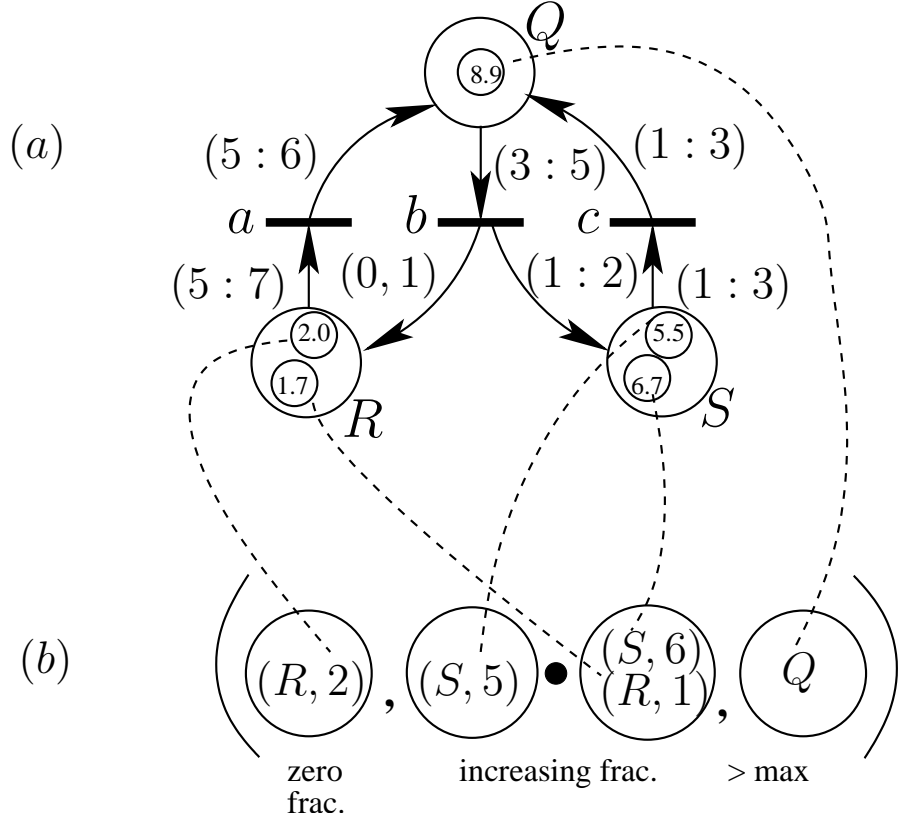}
}
\caption{Marking $\marking$ in (a) satisfies region $\region$ in (b).}
\label{fig:msatregion}
\end{center}
\end{figure}

\begin{exa}
Consider the TPN $N$ in Figure~\ref{fig:small:example} with $\maxval=7$.
Figure~\ref{fig:msatregion}(a) shows a marking $\marking=
\lst{(R, {2.0})\;,(S, {5.5}),\; 
(R, {1.7}),\;(S, {6.7}),\; (Q, {8.9})}$. 
Figure~\ref{fig:msatregion}(b) shows the unique region $\region=
(\lst{(R, 2)} ,\;\lst{(S, 5)} \wprod
\lst{(R, 1),\;(S, 6)},\;\lst{Q})$
such that $\marking\in\denotationof\region$.
(The symbol $\wprod$ stands for concatenation.)
In Figure~\ref{fig:msatregion}(b), each circle corresponds to a multiset
of tokens of $N$ with same fractional parts. 
Dotted lines show how the tokens of $\marking$ 
in TPN correspond to elements in the region $\region$.
\end{exa}

\medskip
\noindent{\bf Equivalence and orders.}
The region construction defines an equivalence relation $\equiv$
on the set of markings such that $\marking_1\equiv\marking_2$
if, for each region $\region$, it is the case that
$\marking_1\in\denotationof{\region}$ iff
$\marking_2\in\denotationof{\region}$.

It is well-known \cite{Alur:Dill:Region:Graph}
that $\equiv$ is a time-abstract bisimulation on the set
of markings. In other words,
if $\marking_1\trans{}\marking_2$ and 
$\marking_1\equiv\marking_3$ then there is an
$\marking_4$ such that $\marking_2\equiv\marking_4$
and $\marking_3\trans{}\marking_4$.

Notice that given a marking $\marking$, it is easy to compute the unique 
region $\region_\marking$ satisfied by $\marking$.

Next we define an order and a preorder on markings of TPN.
First, there is the usual order $\le$ on multisets (markings are multisets of
timed tokens). We have $\marking_1 \le \marking_2$ iff
$\forall p.\, \marking_1(p) \le \marking_2(p)$, i.e., $\marking_1$ can be
obtained from $\marking_2$ by removing some tokens.

The preorder $\morder$ abstracts from the precise values of the ages of the
tokens and considers only their relation to each other.
We define $\marking_1\morder\marking_2$ if there is an $\marking'_2$
with $\marking_1\equiv\marking'_2$ and $\marking'_2\mleq\marking_2$.
In other words, $\marking_1\morder\marking_2$ if
we can delete a number of tokens from $\marking_2$ and
as a result obtain a new marking which is $\equiv$ equivalent 
(but not necessarily $=$ equivalent) to $\marking_1$.
The relation $\morder$ is only a preorder on the set of markings, 
because it is not antisymmetric. However, it is an order on the
equivalence classes w.r.t. $\equiv$.

We let $\marking_1 \prec \marking_2$ denote that  $\marking_1 \morder \marking_2$ and 
$\marking_1\not\equiv \marking_2$.
Notice that 
 $\trans{}$ is {\it monotonic}
with respect to the preorder $\morder$, i.e, 
if $\marking_1\trans{}\marking_2$ and 
$\marking_1\morder\marking_3$ then there is an
$\marking_4$ such that $\marking_2\morder\marking_4$
and $\marking_3\trans{}\marking_4$.

Next we define a partial order $\morder$ on the set of regions.
 
\begin{defi}
\label{region:preceq:definition}
 Let $\region=\regiontuple$ and $\region'=\tuple{c_0,c_1\ldots c_l,c_{l+1}}$ be 
 regions. Then, 
 $\region \preceq \region'$ iff there is a strict monotone injection 
 $\funtype g {\set{0,\ldots,m+1}}{\set{0,\ldots,l+1}}$ 
 with $g(0) = 0$ and $g(m+1)=l+1$ and 
 $b_i\leq c_{g(i)}$ for each $i:0\leq i\leq m+1$.
We let $\region \prec \region'$ denote that $\region \preceq \region'$ 
and $\region \neq \region'$.
\end{defi}

The order $\preceq$ on regions agrees with the order $\morder$ on markings.

\begin{lem}\label{lem:monotone}
For regions $\region$ and $\region'$,
if $\region\preceq\region'$ then for each $\marking\in\denotationof{\region}, 
\marking'\in\denotationof{\region'}$,
 we have  $\marking \morder \marking'$.
\end{lem}
\proof
Directly from Def.~\ref{region:definition} and Def.~\ref{region:preceq:definition}.\qed

\begin{lem}\label{lem:uc_any_order}
Given a TPN and a region $\region$, the upward closure $\ucdenotationof\region$
w.r.t. $\le$ is the same as the upward-closure w.r.t. $\morder$.
Formally,
$\ucdenotationof\region := 
\{\marking\ |\ \exists \marking' \in \denotationof{\region}. \marking' \le
\marking\}
=
\{\marking\ |\ \exists \marking' \in \denotationof{\region}. \marking' \morder
\marking\}$
\end{lem}
\proof
The $\subseteq$ inclusion is trivial, since $\marking' \le \marking$
implies $\marking' \morder \marking$. 
To prove the $\supseteq$ inclusion let 
$\marking' \in \denotationof{\region}$ and $\marking' \morder \marking$.
Then, by definition of $\morder$ there exists some marking $\marking''$ s.t.
$\marking'' \le \marking$ and $\marking'' \equiv \marking'$. 
It follows from $\marking' \in \denotationof{\region}$ and the definition of
$\equiv$ that $\marking'' \in \denotationof{\region}$.
Thus $\marking$ is also in the first set.\qed

The following Lemma shows that the $\preceq$ preorder on regions of 
Def.~\ref{region:preceq:definition} is compatible with the $\morder$ preorder on
markings. Thus (sets of) regions can be used as a canonical representation of
upward-closed sets of markings, provided that they are closed under $\equiv$.
We define the upward closure of a region w.r.t. $\preceq$
by $\region\!\uparrow\, :=\, \{\region'\ |\ \region \preceq \region'\}$ and
generalize the definition of the denotation from regions to sets of regions
in the standard manner. So we define
$\eqdenotationof{\region\!\uparrow} := 
\bigcup_{\region \preceq \region'} \denotationof{\region'}$.

\begin{lem}\label{lem:compatible}
Consider a region $\region$ of a TPN and the preorder $\morder$ on markings and
regions as defined in Def.~\ref{region:preceq:definition}.
Then $\ucdenotationof\region = \eqdenotationof{\region\!\uparrow}$.
\end{lem}
\proof
If $\region$ is the empty region then the equivalence holds trivially.
For the rest assume that $\region$ is not empty.
If $\marking\in\ucdenotationof\region$ then there exists a marking
$\marking' \le \marking$ s.t. $\marking' \in \denotationof\region$, 
by Lemma~\ref{lem:uc_any_order}. It follows that 
$\region = \region_{M'} \preceq \region_{M} =: \region'$ and thus 
$M \in \denotationof{\region'} \subseteq \eqdenotationof{\region\!\uparrow}$.

If $\marking \in \eqdenotationof{\region\!\uparrow}$ then there exists some
region $\region'$ with $\region \morder \region'$ and $\marking \in
\denotationof{\region'}$.
Pick some marking $\marking' \in \denotationof{\region}$. 
By Lemma~\ref{lem:monotone} we get $\marking' \morder \marking$.
Thus we obtain $\marking \in \ucdenotationof\region$ by
Lemma~\ref{lem:uc_any_order}.\qed

One can symbolically represent certain upward-closed sets of markings 
as the upward closures of finite sets of regions.

\begin{defi}\label{def:MRUC}
A {\em Multi-region upward closure} (MRUC) $\alpha$ is represented as a finite
set of regions $\alpha := \{R_1, \dots, R_n\}$ where each $R_i$ is a region.
This represents an upward closed set of markings $\denotationof{\alpha}$
defined as follows.
\[
\denotationof{\alpha} := \bigcup_{i=1,\dots,n} \ucdenotationof{R_i}
\] 
\end{defi}

Note that, by Lemma~\ref{lem:compatible}, $\denotationof{\alpha} = 
\bigcup_{i=1,\dots,n} \denotationof{R_i\!\uparrow}$.

\begin{lem}
\label{regions:unionintersection}
Multi-region upward closures (MRUCs) are effectively closed under union and 
intersection.
\end{lem}
\proof
The union operation is trivial, since for MRUC $\alpha$, $\beta$ we have
$\denotationof{\alpha} \cup \denotationof{\beta} = \denotationof{\alpha \cup
\beta}$.

For the intersection operation
consider two MRUCs $\alpha := \{A_1, \dots, A_n\}$ and $\beta := \{B_1, \dots,
B_m\}$. Then 
\[
\denotationof{\alpha} \cap \denotationof{\beta} = 
\bigcup_{1 \le i \le n,\, 1 \le j \le m} \ucdenotationof{A_i} \cap \ucdenotationof{B_j}
\]
Thus it suffices to show that for any two regions $A,B$ one can construct
a MRUC ${\it inter}(A,B)$ s.t. 
$\denotationof{{\it inter}(A,B)} = \ucdenotationof{A} \cap
\ucdenotationof{B}$.
Given this, one can express the intersection as a new MRUC 
$\cup_{1 \le i \le n, 1 \le j \le m} {\it inter}(A_i, B_j)$, since
\[
\denotationof{\alpha} \cap \denotationof{\beta} = 
\bigdenotationof{\bigcup_{1 \le i \le n,\, 1 \le j \le m} {\it inter}(A_i, B_j)}
\]
We construct the MRUC ${\it inter}(A,B)$ for given regions $A,B$.
Let $A = (a_0, a_1 a_2 \dots a_n, a_\maxval)$ and
$B = (b_0, b_1 b_2 \dots b_m, b_\maxval)$. 

\medskip
\noindent{\bf Intuition:} For the multisets $a_0, b_0$ and
$a_\maxval, b_\maxval$ constructing the minimal requirements for the
intersection of their upward-closures is simple. It is just the maximum, i.e., 
the multiset union (see Def.~\ref{def:multiset} for multisets), and we have 
$a_0^\uparrow \cap b_0^\uparrow = (a_0 \cup b_0)^\uparrow$ (similarly for
$a_\maxval, b_\maxval$). 

The sequences of multisets $a_1 a_2 \dots a_n$ and $b_1 b_2 \dots b_m$
represent orderings of the fractional parts of the ages of tokens in those
multisets. However, the fractional part of $a_1$ could be smaller, equal to,
or larger than the fractional part of $b_1$, $b_2$, etc. All of these cases
must be considered. If two multisets
$a_i$, $b_j$ represent the same fractional part, then the minimal requirement
for markings in the upward-closure of the intersection is the maximum, i.e.,
the multiset union of $a_i$ and $b_j$. Otherwise they must appear individually
in the proper order of the fractional parts.

\vspace{1mm}

\noindent{\bf Construction:} Formally, let $F$ be the set of all injective, strictly monotone increasing
functions $f: \{1,\dots,n\} \rightarrow \{1,\dots,n+m\}$ and 
$G$ the set of all injective, strictly monotone increasing
functions $g: \{1,\dots,m\} \rightarrow \{1,\dots,n+m\}$.
(Note that $F$ and $G$ are finite.)
These functions are normally not surjective and we define
$R(f) := f(\{1,\dots,n\})$ and $R(g) := g(\{1,\dots,m\})$.
For any $f \in F$ and $g \in G$ we define a sequence of multisets 
\[
s(f,g) := c_1 c_2 \dots c_{n+m}
\]
such that for any $i \in \{1,\dots,n+m\}$
\begin{enumerate}[$\bullet$]
\item
If $i \in R(f) \cap R(g)$ then $\exists j,k.\, i=f(j)=g(k)$.
Let $c_i := a_j \cup b_k$.
\item
If $i \in R(f)$ and $i \notin R(g)$ then $\exists j=f^{-1}(i)$.
Let $c_i := a_j$.
\item
If $i \notin R(f)$ and $j \in R(g)$ then $\exists k=g^{-1}(i)$.
Let $c_i := b_k$.
\item
Else $c_i := \emptyset$.
\end{enumerate}
For each $f,g$, the sequence of multisets $s(f,g)$ describes a possible
interleaving/combination of the sequences $a_1\dots a_n$ and 
$b_1 \dots b_m$. However, $s(f,g)$ might contain some empty multisets, 
which must be removed in order to satisfy the requirements for 
regions (see Def.~\ref{region:definition}).
Given a sequence of multisets $x_1\dots x_k$, let $e(x_1 \dots x_k)$ be
the subsequence where all the empty multisets have been removed. 

We can now define the MRUC
\[
{\it inter}(A,B) := \bigcup_{f \in F,\, g \in G} \{(a_0 \cup b_0, e(s(f,g)),
a_\maxval \cup b_\maxval)\}
\]

\vspace{1mm}
\noindent{\bf Proof of correctness:} 
We show that this construction satisfies the required property
$\denotationof{{\it inter}(A,B)} = \ucdenotationof{A} \cap
\ucdenotationof{B}$.

\noindent
Let $M \in \denotationof{{\it inter}(A,B)}$. Then there exist $f\in F, g\in G$ 
s.t. $M \in \ucdenotationof{(a_0 \cup b_0, e(s(f,g)), a_\maxval \cup
b_\maxval)}$. 
Since $a_1,\dots,a_n$ is a subsequence of $e(s(f,g))$ and $a_0 \subseteq a_0 \cup
b_0$ and $a_\maxval \subseteq a_\maxval \cup b_\maxval$ we get
$\ucdenotationof{A} = \ucdenotationof{(a_0, a_1 a_2 \dots a_n, a_\maxval)} 
\supseteq \ucdenotationof{(a_0 \cup b_0, e(s(f,g)), a_\maxval \cup
b_\maxval)}$. Therefore, $M \in \ucdenotationof{A}$. By a symmetric argument
(with $a$ and $b$ interchanged) we obtain $M \in \ucdenotationof{B}$.
So finally we get $M \in \ucdenotationof{A} \cap \ucdenotationof{B}$.

Now we show the other inclusion. Let $M \in \ucdenotationof{A} \cap
\ucdenotationof{B}$. There exist markings $M_1 \le M$ and $M_2 \le M$
with $M_1 \in \denotationof{A}$ and $M_2 \in \denotationof{B}$.
Since $M_1, M_2$ are markings, they are multisets of (timed) tokens and we can
define a new marking $M'$ as their multiset union (see
Def.~\ref{def:multiset}) by
$M' := M_1 \cup M_2$ and obtain $M' \le M$.
Now there exist functions $f \in F$ and $g \in G$, expressing the relative
orders of the fractional parts in $M_1$ and $M_2$, s.t.
$M' \in \denotationof{(a_0 \cup b_0, e(s(f,g)), a_\maxval \cup b_\maxval)}$.
It follows that $M \in \ucdenotationof{(a_0 \cup b_0, e(s(f,g)), a_\maxval \cup
b_\maxval)}$
and thus $M \in \denotationof{{\it inter}(A,B)}$.\qed

We define functions $\pre$ and $\post$ on sets of markings $S$ such that
$\pre(S)$ and $\post(S)$ are the one-step predecessors and successors of
markings in $S$, respectively. Formally,  
$\pre(S) := \{M\ |\ \exists M' \in S.\, M \trans{} M'\}$ 
and
$\post(S) := \{M\ |\ \exists M' \in S.\, M' \trans{} M\}$.
By replacing the transition relation with its reflexive-transitive closure
we obtain the sets of all predecessors and successors, respectively.
Formally, 
$\pre^*(S) := \{M\ |\ \exists M' \in S.\, M \trans{*} M'\}$ 
and
$\post^*(S) := \{M\ |\ \exists M' \in S.\, M' \trans{*} M\}$.

The following lemmas show that for TPN and multi-region upward closures (MRUC)
$S$, one can effectively construct the sets $\post(S)$, $\pre(S)$ and
$\pre^*(S)$ as MRUC. 

\begin{lem}(\cite{ADMN:dlg})\label{post:lemma} 
Let $S$ be a set of markings which is represented as the upward-closure of a
finite set of regions, i.e., a MRUC. Then 
the set $\post(S)$ is effectively constructible as a MRUC.
\end{lem}

The construction for $\pre^*(S)$ is done by the classic 
technique of successive
construction of $\pre^{\le n}(S)$ for larger and larger $n$ (all of which are
upward closed and representable by MRUC) which eventually
converges to $\pre^*(S)$ by Higman's Lemma \cite{Higman:divisibility}, because 
$\morder$ is a well-founded preordering on regions. (The correctness is implied by
the compatibility of the preorder $\morder$ on regions with the order $\le$ on
markings, i.e., Lemma~\ref{lem:uc_any_order} and Lemma~\ref{lem:compatible}.)
A proof can be found in \cite{Parosh:Bengt:Timed:Networks} and a more
general result (for the more expressive formalism of `existential zones') has been
shown in \cite{Parosh:Aletta:bqoTPN}.

\begin{lem}\label{regions:prestar}
Let $S$ be a set of markings which is represented as the upward-closure of a
finite set of regions, i.e., a MRUC.
Then the sets $\pre(S)$ and $\pre^*(S)$ are effectively 
constructible as MRUC.
\end{lem}

Finally, it is known that, for TPN, the set $\post^*(S)$ cannot be effectively
constructed in any symbolic representation with a decidable membership
problem, since the reachability problem is undecidable \cite{Escrig:etal:TPN:nondecidability}.

\section{Zenoness}\label{sec:zeno}

\noindent
A zeno-computation of a timed Petri net is an infinite computation that
has a finite delay. 


\bigskip
\problemx{Zenoness-Problem}
{A timed Petri net $N$, and a marking $M$ of $N$.}
{Is there an infinite $M$-computation $\comp$ and a finite number $m$
  s.t. $\Delta(\comp) \le m$ ? }

\bigskip
We consider a timed Petri net $N$.
A marking $M$ is called a {\it zeno-marking} of $N$ iff the answer to 
the above problem is 'yes'.

\noindent
Note that the zeno-computation $\comp$ can be either a disc-computation or a
time-computation, depending on whether the first transition is discrete or timed.

We let ${\it ZENO}$ denote the set of all zeno-markings of $N$.
More generally, we define 
\[
{\it ZENO}^m := \{M\ |\ \exists\ \mbox{an infinite computation}\ \comp.\, M \dto{\comp}\ \wedge\ \Delta(\comp)
\le m\}
\]
Thus ${\it ZENO} = \bigcup_{m \ge 0} {\it ZENO}^m$.

The decidability of the zenoness-problem for timed Petri nets (i.e., the problem if 
$M \in {\it ZENO}$ for a given marking $M$, or, more generally, constructing
${\it ZENO}$) was mentioned in \cite{Escrig:etal:TPN} by Escrig, et.al. as
an open problem
for both discrete and dense-timed Petri nets. 
In this section, we show that for any TPN, 
a characterization of the
set ${\it ZENO}$ can be effectively computed.
We also show that
this implies the computability of ${\it ZENO}$ for discrete-timed Petri
nets.

The following outline explains the main steps of our proof.
{\parindent=0 pt
\begin{description}
\item [Step 1]
We translate the original
timed Petri net $N$ into an untimed {\it simultaneous-disjoint-transfer net} $N'$.
Simultaneous-disjoint-transfer nets are a subclass of 
{\it transfer Petri nets} \cite{Heinemann82,FS:TCS2001}
where all transfers
happen at the same time and do not affect each other (i.e., all sources 
and targets of all transfers are disjoint).
The computations of $N'$ represent, in a symbolic
way, the computations of $N$ that can be performed in time less than $1-\delta$
for some predefined $0< \delta < 1 $. 
\item [Step 2]
We consider the set ${\it INF}$ of markings of $N'$,
 from which an infinite computation is possible.
${\it INF}$ is upward-closed and can therefore 
be  characterized by the finite set ${\it INF}_{\it min}$ of 
its minimal elements.
While ${\it INF}_{\it min}$ is not computable for general transfer nets
\cite{DJS:ICALP99,Mayr:LCM:TCS}, it is computable for simultaneous-disjoint-transfer
nets, as shown in Lemma~\ref{lem:constr_infminprime}.
\item [Step 3]
We re-interpret the set ${\it INF}$ (resp. ${\it INF}_{\it min}$)
of $N'$ markings in the context of the timed
Petri net $N$ and construct from it a characterization of 
 the set ${\it ZENO}$, described by a multi-region upward closure (MRUC) 
(see Def.~\ref{def:MRUC}).
\end{description}}
To simplify the presentation, we first show Step 1 and Step 3. 
Then, we show how to perform Step 2.

\subsection{Step 1: Translating TPNs to Simultaneous-Disjoint-Transfer Nets}\label{subsec:SD-TN}
\ \\
First we define {\it simultaneous-disjoint-transfer nets}.

\begin{defi}\label{def:SD-TN}
Simultaneous-disjoint-transfer nets (short SD-TN) are a subclass of transfer
nets. A SD-TN $N$ is described by a tuple 
$(P,T, {\it Input}, {\it Output}, {\it Trans})$ where
\begin{enumerate}[$\bullet$]
\item $P$ is a set of places,
\item $T$ is a set of ordinary transitions, 
\item ${\it Input}, {\it Output}: T \to 2^P$ are functions that describe
the input and output places of every transition, respectively (as in ordinary 
Petri nets), and
\item ${\it Trans}$ describes the simultaneous and disjoint transfer
  transition. In order to emphasize the simultaneous operation of the
  transfers, we define ${\it Trans}$ as a single transition with many effects,
  rather than as a set of transitions.
We have ${\it Trans} = (I, O, \st)$ where 
$I \subseteq P$, $O \subseteq P$, and $\st \subseteq P \times P$.
${\it Trans}$ consists of two parts: (a)
$I$ and $O$ describe the 
 input and output places of the Petri net transition part;
(b) the pairs in $\st$ describe the source and target places of the
transfer part. 
Furthermore, the following restrictions on ${\it Trans}$ must be satisfied:
\begin{enumerate}[-]
\item If $(sr,tg), (sr',tg') \in \st$ then $sr,sr',tg,tg'$ are all different and
$\{sr,tg\} \cap (I \cup O) = \emptyset$.
\end{enumerate}
\end{enumerate}
Let $M: P \to \nat$ be a marking of $N$. 
We use $\mleq$ as the ordering on the set of markings (Section~\ref{defs:section}).
The firing of normal transitions
$t \in T$ is defined just as for ordinary Petri nets.
A transition $t \in T$ is enabled at marking $M$ iff
$\forall p \in {\it Input}(t).\, M(p) \ge 1$. Firing $t$ yields the new
marking $M'$ where
\[
\begin{array}{ll}
M'(p) = M(p)    & \mbox{if $p \in {\it Input}(t) \cap {\it Output}(t)$} \\
M'(p) = M(p)-1  & \mbox{if $p \in {\it Input}(t) - {\it Output}(t)$} \\
M'(p) = M(p)+1  & \mbox{if $p \in {\it Output}(t) - {\it Input}(t)$} \\
M'(p) = M(p)    & \mbox{otherwise}
\end{array}
\]
The transfer transition ${\it Trans}$ is enabled at $M$ iff 
$\forall p \in I.\, M(p) \ge 1$. Firing ${\it Trans}$ yields the new
marking $M'$ where 
\[
\begin{array}{ll}
M'(p) = M(p)    & \mbox{if $p \in I \cap O$} \\
M'(p) = M(p)-1  & \mbox{if $p \in I-O$}\\
M'(p) = M(p)+1  & \mbox{if $p \in O-I$} \\
M'(p) = 0       & \mbox{if $\exists p'.\, (p,p') \in \st$}\\
M'(p) = M(p)+M(p')  & \mbox{if $(p',p) \in \st$} \\
M'(p) = M(p)    & \mbox{otherwise}
\end{array}
\]
The restrictions above ensure that these cases are disjoint.
Note that after firing ${\it Trans}$ all source places of transfers are empty,
since, by the restrictions defined above, no place is both source and target
of a transfer. 

We use $M\trans{} M'$ to denote that $M'$ is reached from $M$ 
either by executing an ordinary Petri net transition $t\in T'$ or 
the transfer transition ${\it Trans}$. 
\end{defi}

In the following, sometimes  we use {\it transfer} transition to mean  
simultaneous-disjoint transfer transitions.

\subsubsection{Construction of SD-TN from a TPN}
\label{def:encoding}
\rm
For a given TPN $\tpn=\tpntuple$ we construct a SD-TN 
$N' = (P',T', {\it Input}, {\it Output}, {\it Trans})$. The intuition is
that $N'$ simulates symbolically all computations of $\tpn$ which can happen 
in time $< 1-\delta$ for some predefined $1 > \delta > 0$.
First we show how to construct the places of SD-TN.
Then we show how to simulate a discrete transition of $N$ by a set of 
transitions of $N'$.
Finally, we show how to simulate timed transitions of $N$ 
by simultaneous-disjoint-transfers and a set of normal discrete transitions as in ordinary PNs.

We let $\maxval$ be the maximal finite constant that appears in the arcs of the TPN.
We define a finite set of symbols 
$\symbols := \{k\ |\ k \in \nat, 0 \le k \le \maxval\} \cup 
\{k+\ |\ k \in \nat, 0 \le k \le \maxval\} \cup
\{k-\ |\ k \in \nat, 1 \le k \le \maxval\}$ and a total order on 
$\symbols$ by $k < k+ < (k+1)- < (k+1)$ for every $k$.

\subsubsection{Constructing places of SD-TN}
We let $P' = \{p(\sym)\ |\ p \in P, \sym \in \symbols\}$, i.e.,
for every place $p \in P$ of $\tpn$ we have a set
containing places of 
the form $p(\sym)$ such that $\sym \in \symbols$.
The set $P'$ is finite, since both $P$ and $\sym$ are finite.

A token in place $p(k)$ encodes a 
token of age exactly $k$ on place $p$.
A token in $p(k+)$ encodes a token in place $p$ of an age $\age$ which satisfies
$k < \age \le k+\delta$ for some a-priori defined $0 < \delta < 1$. This means that
the age of this token cannot reach $k+1$ in any computation taking time $< 1-\delta$.
A token in $p(k-)$ encodes a token in $p$ whose age $\age$ satisfies
$k-1+\delta < \age < k$ and which may or may not reach age $k$ during a computation
taking time $1-\delta$.
For instance, given $\delta=0.6$, a TPN token $(p,1.5)$ is encoded as $p(1+)$ 
while another TPN token $(p,2.7)$ is encoded as $p(3-)$.
The SD-TN tokens $p(k), p(k+)$ and $p(k-)$ are called {\em symbolic encodings}
of the corresponding TPN token $(p,a)$.

In particular, the age of a $p(k-)$ token could be chosen
arbitrarily close to $k$, such that its age could reach (or even exceed) $k$
in computations taking an arbitrarily small time. 

\subsubsection{Translating Discrete Transitions}
First we define a function ${\it enc}: {\it Intrv} \to 2^\symbols$ as follows.
\[
\begin{array}{lcl}
{\it enc}([x:y]) & := & \{\sym \in \symbols\ |\ x \le \sym \le y\} \\
{\it enc}((x:y]) & := & \{\sym \in \symbols\ |\ x < \sym \le y\}\\
{\it enc}([x:y)) & := & \{\sym \in \symbols\ |\ x \le \sym < y\} \\ 
{\it enc}((x:y)) & := & \{\sym \in \symbols\ |\ x < \sym < y\}
\end{array}
\]
For instance, ${\it enc}([1:2]) = \set{1,1+,2-,2}$ and
${\it enc}([1:2)) = \set{1,1+,2-}$.
We say that ${\it enc}(\interval)$ is the {\it encoding} of interval 
$\interval$. By the definition above, the bound $\infty$ is encoded as
$\maxval+$, i.e., ${\it enc}([1:\infty))=\set{1,1+,2-,2,\dots,\maxval,\maxval+}$. 

For every transition $t \in T$ in the TPN $\tpn$, we have a set $T'(t)$ of new
transitions in $N'$. 
The intuition is that the transitions in $T'(t)$ encode all possibilities of
the age intervals of input and output tokens.

\begin{figure}[htbp]
\begin{center}
\scalebox{0.9}{
\input{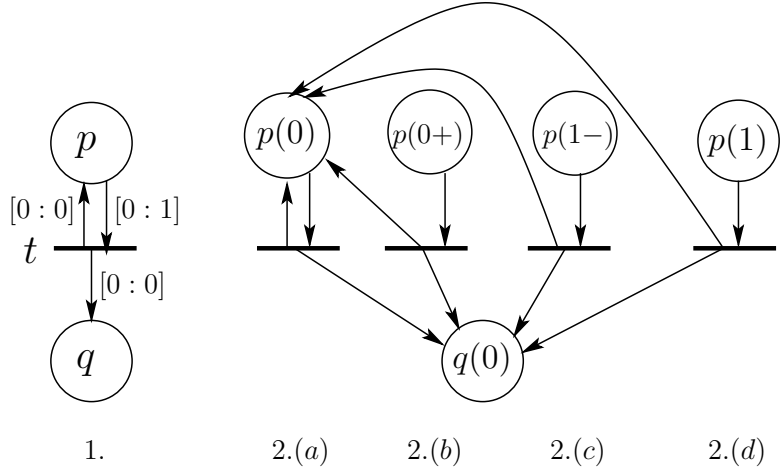}
}
\caption{Simulating ($1$) $t$ in TPN by ($2$) a set $T'(t)$ consisting 
of $4$ transitions in $2.(a)$, $2.(b)$, $2.(c)$ and $2.(d)$.}
\label{sdtrans:figure}
\end{center}
\end{figure}

\begin{exa}
Consider the TPN in Figure~\ref{sdtrans:figure}, part 1.
The only (discrete) transition $t$ has an input arc from place $p$ labeled
$[0:1]$ and two output arcs both labeled $[0:0]$ to places $p$ and $q$,
respectively. 
The translation of this transition into its corresponding SD-TN 
would yield 4 different transitions in $T'(t)$ with output arcs to both places
$p(0)$ and $q(0)$, and input arcs from
places $p(0), p(0+), p(1-)$ or $p(1)$, respectively, as shown in 
Figure~\ref{sdtrans:figure}, parts 2.(a), 2.(b), 2.(c), and 2.(d).
\end{exa}

\begin{figure}[htbp]
\begin{center}
\scalebox{0.9}{
\input{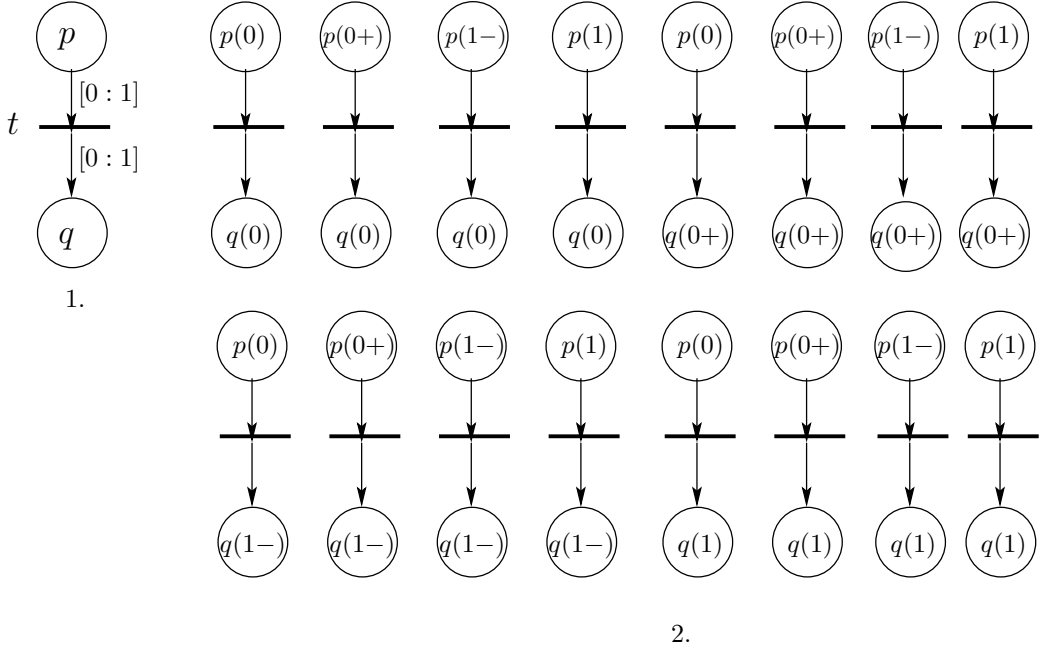}
}
\caption{Simulating (1) $t$ in TPN by (2) a set $T'(t)$ consisting 
of $16$ transitions. (For readability, these 16 transitions are listed
individually, rather than in a combined net.)}
\label{sdtrans2:figure}
\end{center}
\end{figure}

\begin{exa}
Consider the TPN in Figure~\ref{sdtrans2:figure}, part 1.
The only (discrete) transition $t$ has an input arc 
from place $p$  as in Figure~\ref{sdtrans:figure}, part 1., but the
 output arc to place $q$ is labeled by the interval $[0:1]$. 
This will yield the $16$ different transitions in $T'(t)$, 
shown in Figure~\ref{sdtrans2:figure}, part 2., since ${\it enc}([0:1])=\set{0,0+,1-,1}$.
\end{exa}

Each transition $t$ of TPN $N$ yields a set $T'(t)$ of transitions 
in the  corresponding SD-TN $N'$.
Each transition in the set $T'(t)$ is of the form $t'(A,B)$ where 
$A$ and $B$ are the set of input and output
places of $t'(A,B)$ respectively, i.e.,
${\it Input}(t'(A,B))=A$ and ${\it Output}(t'(A,B))=B$.
In the following, for each transition $t$ in TPN, 
 we compute  a set $\subInP(t)$ ($\subOutP(t)$) which contains the 
set of input (output)  places 
 for each 
 transition in $T'(t)$. 
  
For every $t \in T$,
consider the set of input arcs $\arcs_{in}(t)=\set{\place_1(\interval_1),\ldots,\place_m(\interval_m)}$ and 
the set of output  arcs $\arcs_{out}(t)=\set{\place'_1(\jinterval_1),\ldots,\place'_\ell(\jinterval_\ell)}$. 
Now, 
we define $\subInP(t) \subseteq 2^{P'}$ where each element in $\subInP(t)$ 
is a set $A$ of places and is given by 
\[A = \set{\place_1(\sym_1),\ldots,\place_m(\sym_m)}\]
where $\sym_i\in{\it enc}(\interval_i)$ for $i:1\leq i \leq m$.
Intuitively, each set $A$ in $\subInP(t)$  corresponds to a unique 
combination of encodings of input tokens of $t$ in $N$.

For every $t \in T$ we define $\subOutP(t) \subseteq 2^{P'}$ in a similar manner.
We define $\subOutP(t)$ where each element in $\subOutP(t)$ 
is a set $B$ of places and is given by 
\[B = \set{\place'_1(\sym'_1),\ldots,\place'_\ell(\sym'_\ell)}\]
where $\sym'_i\in{\it enc}(\jinterval_i)$ for $i:1\leq i \leq \ell$.
Similarly, each set $B$ in $\subOutP(t)$  corresponds to a unique 
combination of encodings of output tokens of $t$ in $N$.

We define
$
T'(t) := \{t'(A,B)\ |\ A \in \subInP(t), B \in \subOutP(t)\}
$
and finally $T' := \bigcup_{t \in T} T'(t)$.

\begin{exa}
Consider the example in Figure~\ref{sdtrans:figure}.
Here, $\inputs(t,p) = [0:1]$, $\outputs(t,p) = [0:0]$,
$\inputs(t,q)=\emptyset$ and $\outputs(t,q)=[0:0]$.
We have ${\it enc}([0:1])=\set{0,0+,1-,1}$ and
${\it enc}([0:0])=\set{0}$. 
Then 
$\subInP(t)=\set{\set{p(0)},\set{p(0+)},\set{p(1-)},\set{p(1)}}$ and 
$\subOutP(t)=\set{\set{q(0)}}$. 
The four transitions in Figure~\ref{sdtrans:figure}.2 
are given by $t'(\set{p(0)},\set{q(0)})$,
 $t'(\set{p(0+)},\set{q(0)})$,
 $t'(\set{p(1-)},\set{q(0)})$ and
 $t'(\set{p(1)},\set{q(0)})$, respectively.
$T'(t)$ consists of the above four transitions.
\end{exa}

\subsubsection{Translating Timed Transitions}\label{subsubsec:timedtrans}
So far, the transitions in $T'$ only encode the discrete transitions of $N$.
The passing of time will be encoded by a sequence of transitions, including
one use of the transfer transition.
Our construction must ensure the following properties.
\begin{enumerate}[$\bullet$]
\item
We need to keep discrete transitions and time-passing separate. 
Therefore, we must first
modify the net to obtain alternating discrete phases and time-passing phases.
\item
Time-passing phases must not directly follow each other. They must be
separated by at least one discrete transition.
\end{enumerate}
Our SD-TN is extended and modified in several steps.
\begin{enumerate}[(1)]
\item
First we add three extra places 
$p_{\it disc}$, $p_{\it time1}$ and $p_{\it time2}$ to 
$P'$ which act as control-states for the different phases. (The time-passing
phase has two sub-phases). The construction will ensure that at any time there
is exactly one token on exactly one of these places. 
\item
Normal transitions can fire if and only if $p_{\it disc}$ is marked.
Thus we modify all transitions $t \in T'$ by adding $p_{\it disc}$ to 
${\it Input}(t)$ and ${\it Output}(t)$. 
\item
We add an extra place $p_{\it count}$ to $P'$ which counts the number of fired
discrete transitions since the last time-passing phase. 
Thus we modify all transitions $t \in T'$ by adding $p_{\it count}$ to 
${\it Output}(t)$. This is needed to
ensure that time-passing phases are separated by at least one discrete
transition. A new time-passing phase can only start if $p_{\it count}$ is
non-empty, and $p_{\it count}$ will be cleared of tokens during the
time-passing phase.
\item
Now we add a new transition $t_{\it switch-time}$ which starts the
time-passing phase. We define
${\it Input}(t_{\it switch-time}) = \{p_{\it disc}, p_{\it count}\}$ and
${\it Output}(t_{\it switch-time}) = \{p_{\it time1}\}$. It can only fire if
$p_{\it count}$ is marked (thus time-passing phases cannot directly follow
each other) and moves the control-token from 
$p_{\it disc}$ to $p_{\it time1}$.
(Note that $p_{\it count}$ is not necessarily empty after this operation, since
it might have contained more than one token. The place $p_{\it count}$
will be cleared later by the transfer transition.)
\item
If the control-token is on $p_{\it time1}$ then the transfer transition 
${\it Trans}$ is the only enabled transition.
It encodes (in an abstract way)
the effect of the passing time on the ages of tokens.
After an arbitrarily small amount of time $<1$ passes, all tokens of age
$k$ have an age $>k$. This is encoded by the simultaneous-disjoint transfer
arc, which moves all tokens from places $p(k)$ to places $p(k+)$.
Furthermore, it will move the control-token from place $p_{\it time1}$
to place $p_{\it time2}$. Finally, it needs to clear the place $p_{\it count}$
of tokens. To do this, we add a new special place $p_{\it dump}$ (which is not
an input place of any transition; the number of tokens on $p_{\it dump}$ is 
semantically irrelevant) and transfer
all tokens from $p_{\it count}$ to $p_{\it dump}$.
Formally, ${\it Trans} := (I, O, \st)$ where 
$I := \{p_{\it time1}\}$, $O := \{p_{\it time2}\}$, and
$\st := \{(p(k), p(k+))\ |\ 0 \le k \le \maxval\} \cup \{(p_{\it count}, p_{\it dump})\}$.
Note that the transfer transition ${\it Trans}$ is enabled even if no tokens
are present on the places $p(k)$.
\item
Now the control-token is on place $p_{\it time2}$.
Next we add two new sets of transitions to $T'$, which encode what happens to tokens of age 
$k-$ when (a small amount $< 1$ of)
time passes. Their age might either stay below $k$, reach $k$ or exceed $k$.
Notice that we do not need to do anything in the first case.
\begin{enumerate}[$\bullet$]
\item For every $k \in \{1, \dots, \maxval\}$ we have a transition with input places
$p_{\it time2}$ and $p(k-)$ and output places $p_{\it time2}$ and $p(k)$. 
This encodes the second scenario.
\item Furthermore, for every $k \in \{1, \dots, \maxval\}$ we have a transition with input places
$p_{\it time2}$ and $p(k-)$ and output places $p_{\it time2}$ and $p(k+)$.
This encodes the third scenario.
\end{enumerate}
\item 
Finally, we add an extra transition $t_{\it switch-disc}$ with input place 
$p_{\it time2}$ and output place $p_{\it disc}$, which switches the net back to 
normal discrete mode.
\end{enumerate}
Note that after a time-passing phase the only tokens on places $p(k)$ are 
those which came from $p(k-)$, because all tokens on $p(k)$ were 
first transferred to $p(k+)$ by the transfer transition.
Furthermore, the place $p_{\it count}$ is empty after a time-passing phase,
and thus $t_{\it switch-time}$ is not immediately enabled. At least one
discrete transition must fire before the next time-passing phase. Therefore,
every infinite computation of the SD-TN $N'$ must contain infinitely many
discrete transitions.

\smallskip
\noindent{\bf Convention:} Since the number of tokens on place $p_{\it dump}$
is semantically irrelevant, we will ignore this place in the rest of
our proof. It was only introduced for technical reasons
to empty $p_{\it count}$ by the transfer, 
since we do not have reset-arcs, but only a transfer arc.

\begin{figure}[htbp]
\begin{center}
\scalebox{0.9}{
\input{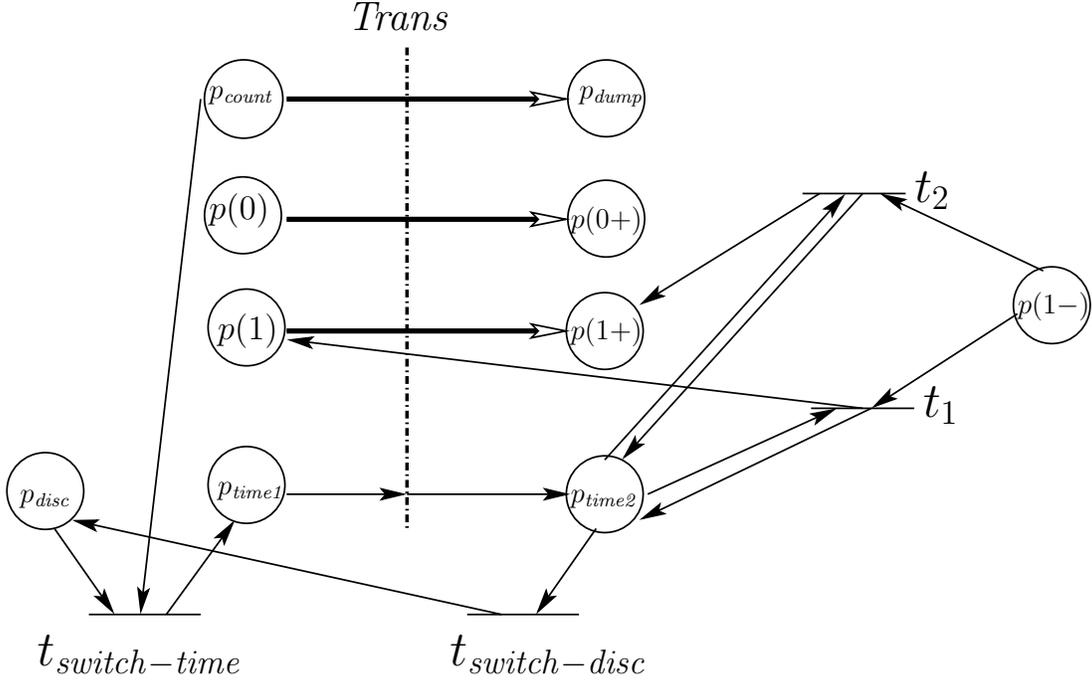}
}
\caption{Simulating a time-passing transition in a TPN for time $<1-\delta$,
  by the corresponding SD-TN.}
\label{transfer:figure}
\end{center}
\end{figure}

\begin{exa}
In Figure~\ref{transfer:figure}, we simulate the timed transitions 
of a TPN with a single place $p$ and $\maxval=1$. 
The transition $t_{\it switch-time}$ starts the time-passing phase by
moving the token from $p_{\it disc}$ to $p_{\it time1}$ and consumes one
token from $p_{\it count}$ (thus it cannot fire if $p_{\it count}$ is empty).
The transfer 
transition is described by the dotted line and the transfer arcs are shown as 
thick arrows from the source of the transfer to the target of the transfer, 
namely from $p(0)$ to $p(0+)$ and from $p(1)$ to $p(1+)$.
The place $p_{\it count}$ is cleared by moving all its tokens to the
(otherwise unused) place $p_{\it dump}$.
The Petri net part of a transfer (input from $p_{\it time1}$ and output to 
$p_{\it time2}$) is shown as ordinary arcs. The transitions
$t_1$ and $t_2$ move a token from $p(1-)$ to $p(1)$ and 
to $p(1+)$, respectively, if there is 
 a token in $p_{\it time2}$. Finally, $t_{\it switch-disc}$ moves the 
token from $p_{\it time2}$ back to $p_{\it disc}$ and ends the 
time-passing phase.
\end{exa}

\subsection{Step 3: Constructing ZENO}

In this section, we show how to compute the set ZENO as a MRUC.

\begin{defi}\label{def:INF}
Let $N$ be a TPN and 
$N' = (P',T', {\it Input}, {\it Output}, {\it Trans})$ the corresponding SD-TN, 
defined as in Subsection~\ref{subsec:SD-TN}.
\begin{enumerate}[$\bullet$]
\item
We say that a marking $M'$ of $N'$ is a {\em standard marking} if 
$M'(p_{\it disc})=1$ and $M'(p_{\it time1}) = M'(p_{\it time2})=0$
and $M'(p_{\it count})=0$. (It follows that a computation from a standard
marking cannot start directly with a time-passing phase.)
Let $\Omega$ be the set of all markings of $N'$ and $\Omega'$ 
the set of all standard markings of $N'$.
\item
We denote by ${\it INF}$ the set of all markings of $N'$ from which infinite
computations start. Since ${\it INF}$ is upward-closed in $\Omega$ with respect to $\mleq$ 
and $\mleq$ is a well-quasi-ordering, ${\it INF}$ can be characterized
by its finitely many minimal elements (see also Lemma~\ref{lem:Dickson}).
Let ${\it INF}_{\it min}$ be the set of minimal elements (markings). 
\item
Let ${\it INF}'$ and ${\it INF}_{\it min}'$ be
the restriction to standard markings of ${\it INF}$ and ${\it INF}_{\it min}$,
respectively. I.e., ${\it INF}' := {\it INF} \cap \Omega'$ and
${\it INF}_{\it min}' := {\it INF}_{\it min} \cap \Omega'$.
The set ${\it INF}'$ is not upward-closed in $\Omega$. However, by the
following Lemma~\ref{lem:standard_base}, 
${\it INF}'$ is the upward-closure of ${\it INF}_{\it min}'$ in $\Omega'$.
Thus ${\it INF}'$ can be characterized by the finite set ${\it INF}_{\it
  min}'$ of its minimal elements.
\end{enumerate}
\end{defi}

\begin{lem}\label{lem:standard_base}
${\it INF}'$ is the upward-closure of ${\it INF}_{\it min}'$ in $\Omega'$.
\end{lem}
\proof
Let $X := \{M' \in \Omega'\ |\ \exists M \in {\it INF}_{\it min}'.
\, M' \ge M\}$ be the upward-closure of ${\it INF}_{\it min}'$ in $\Omega'$.
We need to show that ${\it INF}' = X$. 

The inclusion $X \subseteq {\it INF}'$ holds trivially, by monotonicity of
SD-TN and the fact that all markings in $X$ are standard markings.

Now we show the other inclusion ${\it INF}' \subseteq X$. 
Let $M' \in {\it INF}' = {\it INF} \cap \Omega'$. Since $M' \in {\it INF}$, 
there exists some marking $M \in {\it INF}_{\it min}$ such that
$M \le M'$. Since $M \in {\it INF}$, it follows from the 
definition of ${\it INF}$ and the construction of the SD-TN $N'$ that
$M(p_{\it disc})+M(p_{\it time1})+M(p_{\it time2}) \ge 1$, i.e., 
at least one of these places must be marked or there cannot be an infinite
run. Since $M' \in \Omega'$ we have $M'(p_{\it disc})=1$ and
$M'(p_{\it time1}) = M'(p_{\it time2})=M'(p_{\it count})=0$. 
Therefore, by $M \le M'$, we have
that $M(p_{\it disc})=1$ and 
$M(p_{\it time1}) = M(p_{\it time2}) = M(p_{\it count}) = 0$
and thus $M \in \Omega'$. So we obtain 
$M \in {\it INF}_{\it min} \cap \Omega' = {\it INF}_{\it min}'$.
Since $M' \in \Omega'$ is a standard marking and $M' \ge M$, we finally obtain $M' \in X$
as required.\qed

The following definitions establish the connection between the markings of
the timed Petri net $N$ and the markings of the SD-TN $N'$.

\begin{defi}\label{def:int}
For every $\delta$ with $0 < \delta < 1$ 
we define a function ${\it int}_\delta: \msets{(\places\times\nnreals)} \to (P' \to \nat)$
that maps a marking $M$ of $N$ to its corresponding marking $M'$ in $N'$.
$M' := {\it int}_\delta(M)$ is defined as follows. Let 
\[
\begin{array}{lcll}
M'(p(k))    & := & M((p,k)) & \mbox{for $k \in \nat$, $0 \le k \le \maxval$.}\\
M'(p(k+))   & := & \sum_{k < x \le k+\delta} M((p,x)) & \mbox{for $k \in \nat$, $0 \le k \le \maxval-1$.}\\
M'(p(\maxval+))   & := & \sum_{\maxval < x} M((p,x)) \\
M'(p((k+1)-))   & := & \sum_{k+\delta < x < k+1} M((p,x)) & \mbox{for $k \in
  \nat$, $0 \le k \le \maxval-1$.}\\
M'(p_{\it disc})  & := & 1\\
M'(p_{\it time1}) & := & 0\\
M'(p_{\it time2}) & := & 0\\
M'(p_{\it count}) & := & 0
\end{array}
\]
Note that $M' = {\it int}_\delta(M)$ is a standard marking according to Def.~\ref{def:INF}.
\end{defi}

\noindent
For instance, for a 
TPN marking $M=\lst{(p,1),(p,0.5),(p,0.95),(p,1.9),(p,2.1),(p,3.9)}$ and 
$\maxval=2$, $\delta=0.8$ we obtain
 ${\it int}_\delta(M)=\lst{p(1),p(0+),p(1-),p(2-),p(\maxval+),p(\maxval+),p_{{\it disc}}}$.

The intuition is as follows. In an infinite computation $\comp$ starting at $M$
with $\Delta(\comp) < 1-\delta$, no TPN token $(p,x)$ with $k < x \le k+\delta$
can reach age $k+1$ by aging. This is reflected in $N'$ by the fact that 
$p(k+)$ tokens are not affected during the time-passing phase.
On the other hand, TPN tokens $(p,x)$ with $k+\delta < x < k+1$
can reach an age $\geq k+1$ by aging. This is reflected in $N'$ by the fact that 
$p((k+1)-)$ tokens can become $p(k+1)$ or $p((k+1)+)$ tokens during the 
time-passing phase.

The following lemma establishes a correspondence between fast
disc-computations of the TPN (i.e., starting with a discrete transition; 
see Section~\ref{defs:section}) and computations of the SD-TN.

\begin{lem}\label{lem:int}
Consider a TPN $N$ with marking $M_0$, the corresponding SD-TN $N'$ constructed as above,
and $0 < \delta < 1$. If there exists an infinite $M_0$-disc-computation $\comp$ such that  
$\Delta(\comp) < 1-\delta$ then  there exists an infinite
${\it int}_\delta(M_0)$-computation $\comp'$ in $N'$, i.e., ${\it int}_\delta(M_0) \in {\it INF}'$.
\end{lem}
\proof
We show that for every infinite $\marking_0$-disc-computation $\comp$ of the
form
\[
\begin{array}{l}
\marking_0 \disctrans \marking_0^1 \disctrans \marking_0^2 \disctrans\ldots \marking_0^{n_0}\\
\longrightarrow_{x_0} \\
\marking_1 \disctrans \marking_1^1 \disctrans \marking_1^2 \disctrans\ldots
\marking_1^{n_1}\\
\longrightarrow_{x_1} \\
\marking_2 \ldots
\end{array}
\]
with $n_i \ge 1$ and $\Delta(\comp) < 1-\delta$, there is a corresponding infinite computation in $N'$ of the
form
\[
\begin{array}{l}
{\it int}_{\delta_0}(M_0) 
\trans{} 
{\it int}_{\delta_0}(M_0^1)\!+\!\{p_{\it count}\}
\trans{}
{\it int}_{\delta_0}(M_0^2)\!+\!\{p_{\it count}^2\}
\trans{} \ldots 
{\it int}_{\delta_0}(M_0^{n_0})\!+\!\{p_{\it count}^{n_0}\}\\
\transclosure \\
{\it int}_{\delta_1}(M_1) 
\trans{} 
{\it int}_{\delta_1}(M_1^1)\!+\!\{p_{\it count}\}
\trans{}
{\it int}_{\delta_1}(M_1^2)\!+\!\{p_{\it count}^2\}
\trans{} \ldots 
{\it int}_{\delta_1}(M_1^{n_1})\!+\!\{p_{\it count}^{n_1}\}\\
\transclosure \\
{\it int}_{\delta_2}(M_2) \ldots
\end{array}
\]
with 
$\delta_0=\delta$ and
for all $i$,
$1 > \delta_{i+1} > \delta_i$. 
Let $\comp_i$ be the infinite suffix of $\comp$ starting at $M_i$.
The values of $\delta_i$ will be defined such that  $\Delta(\comp_i) < 1-\delta_i$.
(The condition $\delta_{i+1} > \delta_i$ is required, because 
$\Delta(\comp_{i+1}) < \Delta(\comp_i)$.)

For every discrete transition step $M_i^j \disctrans M_i^{j+1}$ there 
exists a transition step in $N'$ of the form 
${\it int}_{\delta_i}(M_i^j)+\{p_{\it count}^j\}
 \to_{dt} {\it int}_{\delta_{i}}(M_i^{j+1})+\{p_{\it count}^{j+1}\}$, where
 $dt\in T'(t)$ by the construction in Section~\ref{def:encoding}
and Def.~\ref{def:int}. Note that the functions ${\it int}_{\delta_i}$ always
return standard markings (with no tokens on place $p_{\it count}$). However,
in the computation of the SD-TN, the number of tokens on $p_{\it count}$
represents the number of steps since the last time-passing phase. 

For every timed transition step $M_i^{n_i} \xtimedtrans{x_i} M_{i+1}$ 
we have $\delta_{i+1} = \delta_i+x_i \le 1$.
By the construction in Section~\ref{def:encoding} and Def.~\ref{def:int} 
there is a sequence of transitions in $N'$
(the encoding of the time-passing phase) of the
form ${\it int}_{\delta_i}(M_i^{n_i}) + \{p_{\it count}^{n_i}\}
\transclosure {\it int}_{\delta_{i+1}}(M_{i+1})$.
The time-passing  phase can start at ${\it int}_{\delta_i}(M_i^{n_i}) +
\{p_{\it count}^{n_i}\}$, because $n_i \ge 1$, i.e., there is at least one
token on place $p_{\it count}$.
Note in particular that if some token $(p,x)$ with $k+\delta_i < x < k+1$
reaches an age equal to (or greater than) $k+1$ in the transition from $M_i^{n_i}$
to $M_{i+1}$ then its  encoding $p((k+1)-)$ can be transformed into a token
$p(k+1)$ or $p((k+1)+)$ in the time-passing phase of $N'$.
Furthermore, all tokens in $M_i^{n_i}$ with fractional part 0 are transformed into
tokens with a strictly positive fractional part in $M_{i+1}$, since $x_i > 0$.
In $N'$ this is encoded by the
fact that all $p(k)$ tokens become $p(k+)$ tokens in the time-passing phase.
Finally, all tokens are removed from $p_{\it count}$ in the time-passing phase.
Thus the resulting marking ${\it int}_{\delta_{i+1}}(M_{i+1})$ is a standard
marking again.\qed

The reverse implication of Lemma~\ref{lem:int} does not generally hold.
The fact that ${\it int}_\delta(M) \in {\it INF}'$ for some marking $M$ of a TPN $N$
does not imply that there is an infinite $M$-computation in the 
corresponding TPN. The infinite ${\it int}_\delta(M)$-computation
in $N'$ depends on the fact that the $p(k-)$ tokens do (or don't) become
$p(k)$ or $p(k+)$ tokens at the right step in the computation. 
For example, in an infinite computation taking time $0.5$, 
two different TPN tokens $(p,0.8)$ and $(p,0.9)$ are both
interpreted as $p(1-)$ in $N'$. However, $(p,0.8)$ cannot become $(p,1)$ by
aging unless $(p,0.9)$ becomes $(p,1.1)$, while their symbolic encodings $p(1-)$
can become $p(1)$ or $p(1+)$ in any order.

To establish a reverse correspondence between markings of $N'$ and markings of 
$N$ we need the following definitions.

\begin{defi}\label{def:permutation}
Consider a TPN $\tpn=\tpntuple$. Let $N'$ be the 
corresponding SD-TN with places 
$P' = \{p(\sym)\ |\ p \in P, \sym \in \symbols\} \cup \{p_{\it disc}, p_{\it
  time1}, p_{\it time2}, p_{\it count}\}$
and a standard marking $M': P' \to \nat$.  
Let $M'^-, M'^+$ be the sub-markings of $M'$ defined as follows.
\begin{enumerate}[$\bullet$]
\item $M'^-(p(k-))=M'(p(k-))$ for each place of the form $p(k-)$ in $P'$; 
 $M'^-(p(k+))=0$ and $M'^-(p(k))=0$ for each place of the form $p(k+)$ and
 $p(k)$ in $P'$, respectively.
$M'^-(p_x)=0$ for any $p_x \in \{p_{\it disc}, p_{\it time1}, p_{\it time2}, p_{\it count}\}$.
\item $M'^+(p(k+))=M'(p(k+))$ for each place of the form $p(k+)$ in $P'$. But 
$M'^+(p(k-))=0$ and $M'^+(p(k))=0$ for each place of the form $p(k-)$ and
  $p(k)$ in $P'$, respectively.
$M'^+(p_x)=0$ for any $p_x \in \{p_{\it disc}, p_{\it time1}, p_{\it time2}, p_{\it count}\}$.
\end{enumerate}
Let ${\it perm}(M'^-)$ be the set of all words
\[ 
\word_- = b_1 \wprod \dots \wprod b_n \in 
\left(\msets{(P \times \{0,\dots,{\it max}-1\})} - \{\emptyset\}\right)^*
\]
such that  for all $p$ and $k < {\it max}$ we have that
$
M'^-(p((k+1)-)) =  b_1((p,k)) + \ldots + b_n((p,k))
$.

Similarly, let ${\it perm}(M'^+)$ be the set of all words
\[
\word_+ = b_1 \wprod \dots \wprod b_n \in \left(\msets{(P \times
    \{0,\dots,{\it max}-1\})} - \{\emptyset\}\right)^*
\]
such that  for all $p$ and $k < {\it max}$, we have
$
M'^+(p((k)+)) =  b_1((p,k)) + \ldots + b_n((p,k))
$.
\end{defi}

Intuitively, ${\it perm}(M'^-)$ 
describes all possible permutations 
of the fractional parts of (the ages of) tokens in a TPN marking $M$ which are 
symbolically encoded as $p(k-)$ tokens in the corresponding SD-TN 
standard marking $M'$. Note that several different tokens can have the same
fractional part.
Similarly, the set ${\it perm}(M'^+)$ describes all possible permutations 
of the fractional parts of (the ages of) tokens in a TPN marking $M$ which are 
symbolically encoded as $p(k+)$ tokens in the corresponding SD-TN 
standard marking $M'$.

\begin{exa}
\label{permutation}
Let $\maxval=1$.
Consider $M'=\lst{p_{\it disc},p(1),q(1+),p(0+),q(1-),q(1-)}$.
Then 
${\it perm}(M'^-)=\set{\lst{(q,0)}\wprod\lst{(q,0)}\;,\;\lst{(q,0),(q,0)}}$ and
 ${\it perm}(M'^+)=\set{\lst{(p,0)}}$. Notice that 
$q(1+)$ does not belong to  ${\it perm}(M'^+)$, since $\maxval=1$.
\end{exa}

Every standard marking $M'$ of the SD-TN defines a set of 
TPN markings, depending on which permutation of the fractional parts of the ages
of the $p(k-)$-encoded tokens
and $p(k+)$-encoded tokens is chosen.

\begin{defi}\label{def:RXY}
Let $N'$ be a SD-TN. 
For every standard marking $M': P' \to \nat$ we define 
a multi-region upward closure (MRUC) $\regs(M')$ as follows.
The MRUC $\regs(M')$ contains all 
regions $\regs(M',\word_+,\word_-)$ of the form
$
(b_0, \word_+ \wprod \word_-, b_{\it max})
$,
where $b_0((p,k)) = M'(p(k))$ for all $p$ and all $k \le {\it max}$,
$\word_+ \in {\it perm}(M'^+)$, $\word_- \in {\it perm}(M'^-)$ and
$b_{\it max}(p) = M'(p({\it max}+))$ for all $p$.
\end{defi}

\begin{exa}
\label{convert}
Consider $M'=\!\lst{p_{\it disc},p(1),q(1+),p(0+),q(1-),q(1-)}$ and 
sets ${\it perm}(M'^+)$, ${\it perm}(M'^-)$ of Example~\ref{permutation}.
$\regs(M')$ consists of the $2$ regions shown in Figure~\ref{fig:reconstruct}.
\hide{
\begin{itemize}
\item 
$(\lst{p(1)}, \lst{p(0)}\wprod\lst{q(0)}\wprod\lst{q(0)}, \lst{q})$,
\item
$(\lst{p(1)}, \lst{p(0)}\wprod\lst{q(0),q(0)}, \lst{q})$.
\end{itemize}
}
\begin{figure}[htbp]
\begin{center}
\scalebox{0.8}{
\input{reconstruct.pstex_t}
}
\caption{$\regs(M')=\set{\region_1,\region_2}$}
\label{fig:reconstruct}
\end{center}
\end{figure}
\end{exa}

Next we show how an infinite disc-computation of the SD-TN corresponds to 
a zeno computation in the TPN which starts with a discrete transition.

\begin{lem}\label{lem:re_int}
Let $N$ be a TPN with corresponding SD-TN $N'$ and $M' \in {\it INF}'$. Then
\[
\exists \word_- \in {\it perm}(M'^-).\,\forall \word_+
\in {\it perm}(M'^+).\, \ucdenotationof{\regs(M',\word_+,\word_-)}
\subseteq \bigcup_{\delta > 0} {\it ZENO}^{1-\delta} \subseteq {\it ZENO}
\]
\end{lem}
\proof
Since $M' \in {\it INF}'$, 
there is an infinite $M'$-computation 
$\comp' = M' \to M_1' \to M_2' \to \dots$. The first transition in $\comp'$
is a discrete transition, since $M'$ is a standard marking. 
The computation $\comp'$ contains a 
(possibly infinite) number of time-passing phases (where the control-token
shifts to the place $p_{\it time1}$ and then $p_{\it time2}$) 
${\it tpp}_1, {\it tpp}_2, \dots$.
Now consider the {\em original} $p(k-)$ tokens in $M'$ which become $p(k)$ tokens
or $p(k+)$ tokens in the  $i$-th time-passing phase ${\it tpp}_i$. Other
tokens which were {\em newly created} during the computation $\comp'$ are not
considered here. (They will be treated differently; see below).
Let $\alpha_i$ be the multiset of $p(k-)$ tokens in $M'$ which 
become $p(k+)$ tokens in ${\it tpp}_i$ and $\beta_i$ the multiset of 
$p(k-)$ tokens in $M'$ which become $p(k)$ tokens in ${\it tpp}_i$.
(Note that this does not happen by the transfer
transition, but by normal transitions in second part of the time-passing
phase, where the control-token is on place $p_{\it time2}$.) 
We have $\alpha_i, \beta_i \mleq M'^-$, but not necessarily 
$\Sigma_{i \in \nat} (\alpha_i\mprod\beta_i) = M'^-$, because $p(k-)$ tokens can
also be used by normal transitions in the discrete phase or never become
$p(k)$ or $p(k+)$ tokens at all.
Let $\gamma := M'^- - \Sigma_{i \in \nat} (\alpha_i\mprod\beta_i)$.
Since $M'^-$ is finite, there exists a smallest number $m$ such that  
$\alpha_i \mprod \beta_i = \emptyset$ for all $i > m$.
It follows that there exists an 
infinite suffix $\comp''$ of $\comp'$ such that in $\comp''$ no {\em original}
$p(k-)$ token of $M'$ becomes a $p(k)$ or $p(k+)$ token. 

We define $\word_- \in {\it perm}(M'^-)$ by $\word_- :=
\gamma\bullet\beta_m\bullet\alpha_m\bullet\dots\bullet\beta_1\bullet\alpha_1$.

We need to prove that 
\[
\forall \word_+ \in {\it perm}(M'^+).\, \ucdenotationof{\regs(M',\word_+,\word_-)}
\subseteq \bigcup_{\delta > 0} {\it ZENO}^{1-\delta}
\]
For this it suffices to show that 
$\eqdenotationof{\regs(M',\word_+,\word_-)} 
\subseteq \bigcup_{\delta >0} {\it ZENO}^{1-\delta}$, 
because ${\it ZENO}^{1-\delta}$ is upward-closed. Now 
let $\word_+ \in {\it perm}(M'^+)$ and let $M \in
\eqdenotationof{\regs(M',\word_+,\word_-)}$. 
We need to show 
that $M \in {\it ZENO}^{1-\delta}$ for some $\delta > 0$, i.e., that there
exists an infinite $M$-computation $\comp$ with $\Delta(\comp) < 1-\delta$.

Since $M \in \eqdenotationof{\regs(M',\word_+,\word_-)}$ there exists a
$\delta$ with $0 < \delta < 1$ and ${\it int}_\delta(M) = M'$.
By our assumption above, $M' \in {\it INF}'$ is a standard marking where an
infinite computation $\comp'$ starts. The computation $\comp'$ 
begins with a normal transition (not a time-passing phase), since $M'$ is a
standard marking.  
Based on this $\comp'$, we now construct an infinite $M$-disc-computation $\comp$ with
$\Delta(\comp) < 1-\delta$.

A crucial feature of the construction of this particular $M$-disc-computation
$\comp$ is the order of the fractional parts of the ages of tokens.
While this order is given for the tokens already present in $M$, it can be
chosen conveniently (i.e., as needed) for those tokens which are newly created during $\comp$.
The main ideas for this construction are the following:
\begin{enumerate}[$\bullet$]
\item
Since $\Delta(\comp) < 1$, for any token it can happen at most once during
$\comp$ that it reaches the next higher integer age by aging.
In particular, initially present tokens which are interpreted as $p(k-)$ may
age to $p(k)$ or $p(k+)$, but not to $p((k+1)-)$ or higher during $\comp$.
\item
All time intervals on transition arcs in the timed Petri net have integer
bounds (see Section~\ref{defs:section}). Thus one can have intervals like
$(1:4]$ or $[2:7)$, but not $[1.3\, :\, 2.1]$.
This means that if a token is newly created during $\comp$ then the fractional
part of its age can be chosen nondeterministically arbitrarily closely to the next higher integer.
For example, if a token is created by an output arc labeled $[1:2)$ then its age could
be $1.7$, $1.9$, $1.99$, or $1.99999$, etc.
Consider an already existing token with an age whose fractional part 
is a nonzero value $x$. Now another token is newly created, and let $y$ be 
the fractional part of its age. Then all cases $y<x$, $y>x$ and $y=x$ are
possible, e.g., $y = x/2$ or $y=x+(1-x)/2$, or $y=x$.
This means that the newly created token could reach the next higher integer
age before, after, or at the same time as the old token, depending on which
value $y$ is chosen. For each of these
scenarios there is a computation in with the fractional part $y$ is chosen
to implement it.
In general, for any permutation of the orders of the fractional parts of the
ages of newly created tokens (w.r.t. already existing tokens and each other), 
there is some computation in which their ages are chosen
to create this order. Of course, this only applies to tokens which exist at the
same time in the net during the computation $\comp$, not those who are created 
(directly or indirectly) by each other.
\end{enumerate}

The computation $\comp$ 
has the form $M \disctrans M_{j_1} \to M_{j_2} \to \dots$ where 
the sequence $\{j_i\}_{i \in \nat}$ is a subsequence of $1,2,\dots$ 
(it skips the intermediate steps in the time-passing phases of $\comp'$)
and $M'_{j_i} = {\it int}_{\delta_{j_i}}(M_{j_i}) + \{p_{\it count}^n\}$ (for
some $n \ge 0$) and 
$\delta_{j_i} = \delta + \Delta(M\trans{} M_{j_1}\trans{} M_{j_2}\trans{}\dots\trans{} M_{j_i})$.
(The first transition in $\comp$ is a discrete transition, since also the
first transition in $\comp'$ is one.)

For every simulation of a discrete transition of $N$ in $\comp'$ (i.e., not in the time-passing phase)
of the form $M'_i \to M'_{i+1}$ where 
$M'_i = {\it int}_{\delta_i}(M_i) + \{p_{\it count}^n\}$ (for some $n \ge 0$) 
there is a corresponding discrete transition in $\comp$ of the form $M_i \disctrans M_{i+1}$ where
$\delta_{i+1} = \delta_i$ and
$M'_{i+1} = {\it int}_{\delta_{i+1}}(M_{i+1}) + \{p_{\it count}^{n+1}\}$. This follows directly
from Def.~\ref{def:SD-TN}. 
(Note that the extra parts with $\{p_{\it count}^n\}$ and 
$\{p_{\it count}^{n+1}\}$ are necessary. For technical reasons, 
the SD-TN counts the number of discrete transitions
since the last time-passing phase, while the functions ${\it int}_{\delta_i}$
always return standard markings without tokens on $p_{\it count}$.)

Now we consider the $i'$-th time-passing phase for $1 \le i' \le m$.
(Recall the definition above that $m$ is the index number of the last 
time-passing phase where {\em original} $p(k-)$ tokens of $M'$ change into $p(k)$
or $p(k+)$ tokens. The remaining case of $i' > m$ will be considered later.)
For every sequence of transitions $M'_i \transclosure M'_l$ in $\comp'$ representing
the $i'$-th time-passing phase there is a corresponding single time-transition
in $\comp$ of the form $M_i \xtimedtrans{\varepsilon_{i'}} M_l$,
where $M'_i={\it int}_{\delta_i}(M_i)+\{p_{\it count}^n\}$ 
(for some $n \ge 1$), 
$\delta_l = \delta_i + \varepsilon_{i'}$
and $M'_l = {\it int}_{\delta_l}(M_l)$. 
(Note that $M'_i$ must contain at least one token on $p_{\it count}$ for the
time-passing phase to start there and thus $n \ge 1$. On the other hand,
$M'_l$ is a standard marking, since it is reached at the end of a
time-passing phase and thus does not contain any tokens on $p_{\it count}$.)
The delay $\varepsilon_{i'}$
is chosen as $\varepsilon_{i'} := 1-f_{i'}$ where $f_{i'}$ is the 
fractional part of the age of those tokens in $M_i$
which are mapped to $\beta_{i'}$ by ${\it int}_{\delta_i}$. This ensures that 
in this timed transition the right tokens (of those originally present in $M$)
reach (those mapped to $\beta_{i'}$) or exceed (those mapped to $\alpha_{i'}$) 
the next higher integer age. For the other tokens of $M_i$, which were newly
created during $\comp$ we can arbitrarily choose the values of their
fractional parts, i.e., for every combination of these values there is a
possible computation which implements it. Thus one can assume that these fractional
parts are conveniently chosen such that they do (or don't) reach (or exceed)
the next higher integer age, just as required by the condition
${\it int}_{\delta_l}(M_l)=M'_l$.
Since ${\it int}_\delta(M)=M'$, only those tokens in $M$ with a fractional
part $>\delta$ were mapped to $p(k-)$ tokens in $M'$ and only those tokens
can reach (or exceed) age $k$ in $\comp$. Therefore it follows from our choice of
the $\varepsilon_{i'}$ for $i' \le m$ 
that $\sum_{i'=1}^m \varepsilon_{i'} < 1-\delta$. Thus we get 
$\lambda := (1-\delta) - \sum_{i'=1}^m \varepsilon_{i'} > 0$.
(The quantity $\lambda$ will be used to determine the $\varepsilon_{i'}$ for $i' >m$.)

Now we consider the $i'$-th time-passing phase for $i'>m$. These are 
the time-passing phases
in the infinite suffix $\comp''$ of $\comp'$ mentioned above. For them, it works
like the case above, except that the delays $\varepsilon_{i'}$ do no longer depend
on the initial marking $M$, because $\alpha_{i'} \mprod \beta_{i'} = \emptyset$ 
for $i'>m$. As shown above, none
of the original tokens of $M$ are involved in these $i'$-th time-passing phases
for $i' > m$. The
only tokens involved in this (reaching or exceeding the next higher integer
age in this phase) are tokens newly generated in $\comp$ (which have an age
greater than $\delta$ and are mapped to $p(k-)$). As explained above, 
the fractional parts of their ages can be chosen conveniently (i.e., as
needed) such that  they reach or exceed the next
higher integer age exactly as required for the correspondence with the
computation $\comp'$. In particular, their ages can be chosen arbitrarily
close to the next higher integer age such that  the required delays
$\varepsilon_{i'}$ (for $i' > m$)
can be made arbitrarily small. We choose 
$\varepsilon_{i'} := (\lambda/2)*2^{-i'}$ for $i' >m$. 

So we obtain $\Delta(\comp) = \sum_{i'\in \nat} \varepsilon_{i'} =
\sum_{1\le i'\le m} \varepsilon_{i'} + \sum_{i' > m} \varepsilon_{i'}
\le \sum_{1\le i'\le m} \varepsilon_{i'} + \lambda/2 < \sum_{1\le
i'\le m} \varepsilon_{i'} + \lambda = 1-\delta$.  Thus $\Delta(\comp)
< 1-\delta$ and $M \in {\it ZENO}^{1-\delta}$, as required.\qed

Now we describe the algorithm to compute the set 
${\it ZENO}$ as a multi-region upward closure. The algorithm computes a MRUC
$Z$, given by Definition~\ref{def:Z}, 
and we prove in Lemma~\ref{lem:Z_part_ZENO} and Lemma~\ref{lem:ZENO_part_Z} 
that $\denotationof{Z} = {\it ZENO}$.

\begin{defi}\label{def:Z}
Let $N$ be a TPN with corresponding SD-TN $N'$. 
\[
Z := \bigcup_{M' \in {\it INF}_{\it min}'}\ \bigcup_{\word_+ \in {\it perm}(M'^+)}
\ \bigcap_{\word_- \in {\it perm}(M'^-)}\ \pre^*(\{\regs(M',\word_+,\word_-)\})
\]  
\end{defi}

\subsection{Proof of Correctness}

We need to show that $Z$ is effectively constructible
and that $\denotationof{Z} = {\it ZENO}$.

The constructibility of $Z$ requires the following steps.
\begin{enumerate}[$\bullet$]
\item 
The set ${\it INF}'_{\it min}$ is finite and effectively
constructible. This will be shown in Subsection~\ref{subsec:step2}.
\item
For any $M' \in {\it INF}_{\it min}'$ the sets ${\it perm}(M'^+)$ and
${\it perm}(M'^-)$ are finite and effectively constructible.
This follows directly from Definition~\ref{def:permutation} and the finiteness
of $M'$.
\item
Since $\regs(M',\word_+,\word_-)$ is a region, we can interpret 
$\{\regs(M',\word_+,\word_-)\}$ as a MRUC. Then 
$\pre^*(\{\regs(M',\word_+,\word_-)\})$
can be effectively constructed as a MRUC by
Lemma~\ref{regions:prestar}.
(Note that $\pre^*$ is computed w.r.t. the relation 
$\longrightarrow \, =\, \timedtrans\,\cup\,\disctrans$ which includes both
timed- and discrete transitions. Thus the zeno-computations starting from
markings in $\denotationof{Z}$ may also start with a timed transition.)
\item
By Lemma~\ref{regions:unionintersection}, the finite union and intersection
operations on MRUC are computable and yield a MRUC $Z$.
\end{enumerate}

Now we show that $\denotationof{Z} = {\it ZENO}$.

\begin{lem}\label{lem:Z_part_ZENO}
$\denotationof {Z} \;\subseteq {\it ZENO}$.
\end{lem}
\proof
Let $M \in \denotationof Z$. 
Then there is an $M' \in {\it INF}_{\it min}'$ and a sequence
$\word_+ \in {\it perm}(M'^+)$ such that 
$M \in \denotationof{\bigcap_{\word_- \in {\it perm}(M'^-)}\pre^*(\{\regs(M',\word_+,\word_-)\})}$. 

We choose the sequence $\word_- \in {\it perm}(M'^-)$ according to
Lemma~\ref{lem:re_int} and so obtain $M \in \denotationof{\pre^*(\{\regs(M',\word_+,\word_-)\})}$
and
$\ucdenotationof{\regs(M',\word_+,\word_-)} \subseteq {\it ZENO}$. Thus $M \in {\it ZENO}$,
since $\pre^*({\it ZENO}) = {\it ZENO}$.\qed

\begin{lem}\label{lem:ZENO_part_Z}
${\it ZENO} \subseteq \denotationof {Z}.$
\end{lem}
\proof
Let $M \in {\it ZENO}$. By the definition of {\it zeno-marking},
there exists an
infinite $M$-computation $\comp$ and a finite number $m$ such that  
$\Delta(\comp) \le m$. 
It follows that there exists an infinite suffix of $\comp$ that takes 
only $< 1/2$ time. 
Thus there exists a marking $M_1$ such that  $M \transclosure M_1$ and
an infinite $M_1$-computation $\comp_1$ with $\Delta(\comp_1) < 1/2$.
Since $M_1$ contains finitely many tokens and $\comp_1$ is infinite, there
exists an infinite suffix of $\comp_1$ such that none of the original tokens of
$M_1$ is used in this infinite suffix (although some might still be present;
these are represented by $M_4$, see below). 
Since every infinite computation must contain infinitely
many discrete transitions (see Section~\ref{defs:section}), there exists an
infinite suffix of this infinite suffix of $\comp_1$ which starts with a 
discrete transition.

Thus there exist
markings $M_2$, $M_3$ and $M_4$ and a finite computation $\comp_2$ such that   
\begin{enumerate}[$\bullet$]
\item
$M_1 \dto{\comp_2} M_2 = M_3 + M_4$
\item
All tokens in $M_3$ were created during $\comp_2$.
\item
There is an infinite $M_3$-disc-computation $\comp_3$ with $\Delta(\comp_2\comp_3) < 1/2$,
and thus $\Delta(\comp_3) < 1/2$.
\end{enumerate}
Let $M_3' := {\it int}_{1/2}(M_3)$. Then we have 
$M_3' \in {\it INF}'$ by Lemma~\ref{lem:int}, since $\comp_3$ is 
an infinite disc-computation.
From Definition~\ref{def:RXY}, we have that
there are  permutations $\word_+ \in {\it perm}(M_3'^+)$ and 
$\word_- \in {\it perm}(M_3'^-)$ 
such that  $M_3 \in \denotationof{\regs(M_3',\word_+,\word_-)}$. 

Since $M_3' \in {\it INF}'$ and 
${\it INF'}$ is upward-closed (in $\Omega'$; see Def.~\ref{def:INF}),
there exists a marking $M_3'' \in {\it INF}_{\it min}'$ such that  $M_3''
\mleq M_3'$.
Therefore $M_3''^+ \mleq M_3'^+$, $M_3''^- \mleq M_3'^-$ and
 ${\it perm}(M_3''^+) \subseteq {\it perm}(M_3'^+)$ and ${\it perm}(M_3''^-) \subseteq {\it perm}(M_3'^-)$.

This means that 
there also exist permutations $\word_+' \in {\it perm}(M_3''^+)$ with 
$\word_+' \wleq \word_+$ and 
$\word_-' \in {\it perm}(M_3''^-)$ with $\word_-' \wleq \word_-$ (see
Def.~\ref{def:multiset}) and thus
$\ucdenotationof{\regs(M_3'',\word_+',\word_-')} \supseteq \ucdenotationof{\regs(M_3',\word_+,\word_-)}$.
It follows that
$M_3 \in \denotationof{\regs(M_3',\word_+,\word_-)} \subseteq
\ucdenotationof{\regs(M_3',\word_+,\word_-)}
\subseteq \ucdenotationof
{\regs(M_3'',\word_+',\word_-')}$.

Now consider all those tokens in $M_3$ which are mapped to $p(k-)$ tokens 
in $M_3'$, i.e., those with a fractional part of their age which is $> 1/2$.
These tokens (like all others in $M_3$) were all created during $\comp_2$ and none of them had an integer
age during $\comp_2$, because $\Delta(\comp_2) < 1/2$. Thus, the fractional
parts of their ages are totally independent and any permutation is possible, i.e.,
for any permutation there is a computation which implements it (for the
reasons explained in the proof of Lemma~\ref{lem:re_int}).

Therefore, for {\em every} $\word_- \in {\it perm}(M_3'^-)$ there is a marking $M_3^{\word_-}$ in $N$ such that 
\begin{enumerate}[$\bullet$]
\item
$M_1 \transclosure M_3^{\word_-} + M_4$
\item
$M_3^{\word_-} \in \denotationof{\regs(M_3',\word_+,\word_-)}$.
\end{enumerate}
Since $M_3'' \mleq  M_3'$ we have that for every 
$\word_-' \in {\it perm}(M_3''^-)$ there is a corresponding
$\word_- \in {\it perm}(M_3'^-)$ with $\word_-' \wleq \word_-$, i.e., 
$\word_-'$ is the restriction of $\word_-$ to $M_3''$.
It then follows from the property above that 
for {\em every} $\word_-' \in {\it perm}(M_3''^-)$
there is a marking $M_3^{\word_-'} := M_3^{\word_-}$ in $N$ s.t.
\begin{enumerate}[$\bullet$]
\item
$M_1 \transclosure M_3^{\word_-'} + M_4$
\item
$M_3^{\word_-'} \in \ucdenotationof{\regs(M_3'',\word_+',\word_-')}$.
\end{enumerate}
It follows that for {\em every} $\word_-' \in {\it perm}(M_3''^-)$
we have $M_3^{\word_-} + M_4 \in
\ucdenotationof{\regs(M_3'',\word_+',\word_-')}$
and thus 
$M_1 \in \pre^*(\{\regs(M_3'',\word_+',\word_-')\})$.
Since $M \in \pre^*(M_1)$ we finally obtain
\[
M \in \bigdenotationof{ \bigcap_{\word_-' \in {\it perm}(M_3''^-)}\ \pre^*(\{\regs(M_3'',\word_+',\word_-')\})}
\]
with $M_3'' \in {\it INF}_{\it min}'$ and $\word_+' \in {\it perm}(M_3''^+)$,
and thus $M \in \denotationof Z$.\qed

By Lemma~\ref{lem:Z_part_ZENO} and Lemma~\ref{lem:ZENO_part_Z} we have that 
${\it ZENO} = \denotationof{Z}$. It remains to show that 
${\it INF}'_{\it min}$ is effectively constructible.

\subsection{Step 2: Computing ${\it INF}'_{\it min}$}\label{subsec:step2}

Computability of the set {\it ZENO} (in the last section) requires 
that the minimal elements of 
any upward closed set is effectively constructible.
In this section, we show for any SD-TN, how to construct 
the set of minimal elements ${\it INF}_{\it min}$ of ${\it INF}$.
Then ${\it INF}'_{\it min}$ is obtained by just restricting 
${\it INF}_{\it min}$ to standard markings (see Def.~\ref{def:INF}).

For constructing ${\it INF}_{\it min}$, 
we use a result by Valk and Jantzen \cite{Valk_Jantzen:Acta85}. Our algorithm
depends on the concepts of 
 semi-linear languages, Presburger Arithmetic, Parikh's Theorem and Dickson's
 Lemma, described
in the following. Recall that we use $(v_1,\ldots,v_n)$ or $\vec v$ 
interchangeably to denote a vector of size $n$.

\begin{lem}\label{lem:Dickson}
{\bf (Dickson's Lemma \cite{Dickson:lemma})} \\
For every infinite sequence of vectors $\vecdef{x_1}, \vecdef{x_2}, \vecdef{x_3}, 
\dots$ in
$\nat^n$ there exists an infinite non-decreasing subsequence. 
In particular, there exist indices $i,j$ with 
$i < j$ s.t.~$\vecdef{x_i} \le \vecdef{x_j}$ ($\le$ taken component-wise).
\end{lem}

\subsubsection{Semilinear Sets}

First we define {\it linear} sets.

\begin{defi}
A set $\lang\subseteq\nat^n$ is called {\em linear},
if there exist vectors $\vecdef{\vect_0},\vecdef{\vect_1},\ldots,\vecdef {\vect_m}\in\nat^n$
such that
\[
L = \left\{\vecdef{\vect_0} + \sum_{i=1}^m k_i\vecdef{\vect_i}
\quad |\quad k_1,\ldots,k_m\in\nat\right\}
\]
We denote this linear set by $\lang=\lang(\vecdef{\vect_0}; \vecdef
{\vect_1},\ldots, \vecdef {\vect_m})$.
\end{defi}

\begin{exa}
$L((0,0);(0,2),(2,0)) = \setcomp{(0,0) + k_1(0,2) + k_2(2,0)}{k_1,k_2\in\nat}$ is linear.
\end{exa}

\begin{defi}
A subset of $\nat^n$ is called {\em semilinear} if it is a finite union 
of linear sets.
\end{defi}

\begin{thm}
\cite{Ginsburg:book}
Semilinear sets are closed under union, intersection, complementation and 
first-order quantification.
\end{thm}

Next we define the Parikh mapping $\parikh$.
Given a finite alphabet $\Sigma=\set{a_1,\ldots,a_n}$, 
$\parikh$ is a function from $\Sigma^*$ to $\nat^n$ , defined by 
$\parikh(\word) = (\#_{a_1}(\word),\ldots,\#_{a_n}(\word))$, where 
$\#_{a_i}(\word)$ is the number of occurrences of $a_i$ in $\word$.
Thus $\parikh(\epsilon)=(0,\dots,0)$ and $\parikh(\word_1\wprod\ldots\wprod \word_m)=\sum_{i=1}^m \parikh(\word_i)$.
Finally, given a language $\lang\subseteq \Sigma^*$,
$\parikh(\lang)=\setcomp{\parikh(\word)}{\word\in\lang}$.
If $\parikh(\lang)$ is semilinear for a language $\lang$, then $\lang$ is 
called a semilinear language.

\begin{thm}
\label{theorem:parikh}
(Parikh's Theorem) \cite{Parikh:JACM66} $\parikh(L)$ is effectively 
semilinear for each context-free 
language $\lang$.
\end{thm}

As a special case, Theorem~\ref{theorem:parikh} holds for regular languages, since 
every regular language is a context-free language \cite{Parikh:JACM66}.

\begin{exa}
Let $\Sigma=\set{a_1,a_2,a_3}$. \\
Then 
$\parikh(a_1a_2a_1a_3a_2a_3a_3)=(2,2,3)\in L((2,0,1);(0,1,1))$.\\
Also, $\parikh(ab^*ca)=\setcomp{(2,0,1) + n*(0,1,0)}{n\in\nat}$.
\end{exa}

\subsubsection{Presburger Arithmetic}

Presburger arithmetic is the first-order
theory of the integers with addition and the ordering relation 
over $\Z$, also denoted as $(\Z,\le,+)$. Formally,  Presburger arithmetic is
 the first-order theory over {\it atomic formulae} of the form 
\[
\displaystyle\sum_{1\leq i \leq n} \;\;a_ix_i \;\;\;\sim c
\]
where $a_i,c$ are integer constants, $x_i$-s are variables ranging 
over integers and $\sim$ is a comparison operator, where
$\sim\in\set{=,\neq,<,\leq,>,\geq}$. 
This means that
a {\it Presburger formula} $\rho$ is either an {\it atomic formula}, or 
 it is 
constructed from the Presburger formulae $\rho_1,\rho_2$ recursively as follows:
\[ 
\rho:= \neg \rho_1 \;\;| \;\;\rho_1 \wedge \rho_2 \;\;|\;\; \rho_1 \vee \rho_2
\;\;|\;\; \exists x_i. \rho_1(x_1,\ldots,x_n)
\]
where $\rho_1(x_1,\ldots,x_n)$ is a Presburger formula over free variables 
$x_1,\ldots,x_n$ and $1\leq i \leq n$.

\hide{
\item for all $d \geq 2$, we extend the language by predicates $"d|"$ 
for the divisibility relation.
This is definable in Presburger arithmetic. For instance,
the formula $\exists x. (x + \ldots +x) = y$ defines $d|y$ where $x$ is 
repeated $d$ times in $x + \ldots + x$ and $x$ does not appear in $y$.
\end{itemize}
The extended language allows quantifier elimination \cite{GS:PacMat66}. 
}

\begin{thm}
\label{Presburger}
{\bf (Presburger)} \cite{BenAri:logicforcs} Presburger arithmetic is decidable.
\end{thm}

As a shorthand notation, we work with $\Z_\omega = \Z \cup \{\omega\}$ instead
of the usual $\Z$, where $\omega$ is the first limit ordinal.
This is not a problem, since Presburger-arithmetic 
on $\Z_\omega$ can easily be
reduced to Presburger-arithmetic on $\Z$ as follows. For every variable $x$
one adds an extra variable $x'$ which is used in such a way that the original
state $x=k <\omega$ is represented by $(x,x') = (k,0)$ and the original state
$x=\omega$ is represented by $(x,x')=(0,1)$. It is easy to encode the usual 
properties like $\omega+k = \omega-k = \omega+\omega = \omega$.

\begin{thm}
\label{semilinear:presburger}
\cite{GS:PacMat66}
A subset of $\nat^n$ is semilinear iff it is definable in Presburger 
Arithmetic.
\end{thm}
\hide{
\proof
First we  construct a Presburger formula for 
a given semilinear set, following \cite{GS:PacMat66}.
It is enough to show that linear sets are Presburger definable.
This is due to the fact that if a linear set $S_i$ is 
definable by a formula $\rho_i$({\bf v}) (where {\bf v} 
is a n-tuple of variables over $\nat$ and {\bf v}(i) denotes 
the $i$th variable in {\bf v}, then 
 a semilinear set $S_1 \vee \ldots \vee S_k$ will be definable 
by $\bigvee_{i \leq k} \rho_i({\bf v})$.
Now, let $S=(\vecdef{\vect_0};\vecdef {u_0},\ldots,\vecdef {u_{m-1}})$ be a linear set.
Then put 
\[
\rho({\bf v})=(\exists k_0)(\exists k_1)\ldots(\exists k_{m-1})
\left(\bigwedge_{ i <  m} 0 \leq k_i \wedge \bigwedge_{i<n}\left(\vecdef {u_0}(i) + \sum_{j < m}\;\;k_i{\vecdef {u_j}}(i) = {\bf v}(i)\right)
\right)
\]

Next we show the other direction. Let $f(\vec x)$ be a Presburger formula. 
By Theorem~\ref{Presburger}, there exists a quantifier-free formula 
$f'(\vec x)$  which has the same set of solutions as $f(\vec x)$.
Now negation can be eliminated in the formula $f'(\vec x)$ in 
a straightforward way. For instance, $\neg (x = y)$ is equivalent to 
$ x < y \;\vee\; y < x$. Thus assume that $f'(\vec x)$ is a 
disjunction of conjunctions of atomic formulae.\qed
}
\hide{
Recently a popular way of deciding Presburger arithmetic is to use 
automata. 
An automata is given by a tuple $\automaton=\langle S, \Sigma, \eta, s_0, F\rangle$ where
$S$ is a finite set of states, $\Sigma$ is a finite alphabet, 
 $\funtype\eta{S \times \Sigma}S$ is the transition function,
$s_0\in S$ is the initial state and $F\subseteq S$ is the set of final states.
The {\it language} of $\automaton$, $\langof\automaton = \setcomp{\word\in\Sigma^*}{\hat\eta(s_0,\word)\in F}$ where $\hat\eta(s,\emptystring)=s$ and 
$\hat\eta(s,\word\wprod d) = \eta(\hat\eta(s,\word),d)$ for $s\in S, d \in \sigma$ and $\word \in \Sigma^*$.
The idea of automata-based approach is simple. We will 
use automata to recognize
 tuples of numbers by mapping words to tuples of 
numbers.
The automaton for a Presburger formula accepts precisely the 
words that represent the integers making the formula true.

 The encoding of integers is based on the 2's complement representation 
of integers, where most significant bit is the first digit.
For $d_n \wprod d_{n-1}\wprod \ldots d_0 \in \set{0,1}^+$ ($\set{0,1}^+$ is the 
set of non-empty words over $\set{0,1}$, 
we define $\langle d_n\wprod d_{n-1}\wprod \ldots \wprod d_0\rangle:=
-d_n2^n + \sum_{0\leq i < n} d_i2^i$.
For instance, given  the integer $5$, we have 
$\langle1\wprod0\wprod 1\rangle:=-1*2^2 + 0*2^1 + 1*2^0=3$ 
and $\langle \emptystring \rangle:=0$ for the empty word.
We extend this encoding to tuples of integers as follows.
A word $\word=\vecdef{d_n}\wprod\ldots\wprod\vecdef{d_0} \in
(\set{0,1}^r)^*$ (over vectors of length $r$) represents the 
tuple $\vecdef z = (z_1,\ldots,z_r) \in \Z^r$ of integers, where 
the $i$th "track" of the word $\word$ encodes the integer $z_i$.
That is for all $i:1 \leq i \leq r$, we have that $z_i= \langle d_{n,i},\ldots,d_{0,i}\rangle$ where $\vecdef{d_j} = (d_{j,1},\ldots,d_{j,r})$ 
for $0 \leq j \leq  n$.
Sets of tuples are interpreted as word languages over vectors as follows.
The set $A \subseteq \Z^r$ is represented by the language $L \subseteq (\set{0,1}^r)^*$ if for every $\vecdef z\in\Z^r$, it holds that $\vecdef z\in A$ iff all words (there can be more than one 
word representing same integer, e.g.,  $\langle d_n d_{n-1} \ldots d_0 \rangle = \langle d_n d_n d_{n-1} \ldots d_0\rangle$) 
that represent $\vecdef z$ are in $L$.
The automaton can be recursively constructed from the formula, where 
automata operations handle the logical connectives and quantifiers.
Specific algorithms for constructing automata for linear (in)equations 
have been developed in e.g., \cite{GBD:Presburger}.

We sketch the method of automata construction in the following.
Consider a formula $z < 2$. We use $z[\vecdef d]$ to get the 
integer by replacing the variables in $z$ by the bits in $\vecdef d$.
The set of states in the automaton for the above formula 
is $s_0 \;\cup\;\Z$ where $s_0$ is the initial state.
Note that we identify integers with states (so the set of states is infinite).
The idea is to keep track of the value of  $z$ as successive bits are read.
Thus, except for the special initial state, a state in $\Z$ represents the 
current value of $z$. The transition function
$\funtype{\eta}{(\set{s_0}\cup\Z) \times \set{0,1}^r}{\set{s_0}\cup\Z}$ is 
defined as follows for a letter $\vecdef d \in \set{0,1}^r$.
For $q\in \Z$, we define $\eta(q,\vecdef d)=2q + z[\vecdef d]$ and $\eta(s_0,\vecdef d)=-z[\vecdef d]$ for the initial state. 
Finally, for a word $\word \in (\set{0,1}^r)^*$, 
we define $\hat\eta(s,\word)=s2^n + z[\langle\word\rangle_\nat]$ 
and $\hat\eta(s_0,\word)=z[\langle\word\rangle_\nat]$,
where $q\in\Z$, 
$\word = \vecdef{d_n} \wprod\ldots\wprod\vecdef{d_0}$, and
$\langle\word\rangle_\nat$ is the binary interpretation of original 
integer given  by  $\sum_{0\leq i\leq n} \vecdef {d_i}$.

The automata construction is based on the observation that two states 
$s,s'\in \Z$ can be merged if, intuitively they are both small or 
both large with respect to the coefficients and constants in the formula.
Idea is that from a small value, one can only obtain smaller values and 
from a large value one can only obtain larger values using $\eta$.
See \cite{Klaedtke} for more details.
This yields a finite automaton for the formula.

However, the number of states generated in the automata-based approach 
is bounded by a doubly exponential lower and a triply exponential 
upper bound (with respect to the size of the formula).
}

\bigskip
\subsubsection{Result from Valk and Jantzen}
\ \\

We recall a result from \cite{Valk_Jantzen:Acta85}.

\begin{thm}\label{thm:Valk_Jantzen}(Valk \& Jantzen
  \cite{Valk_Jantzen:Acta85})\quad
Given an upward-closed set $V \subseteq \nat^k$, the finite set 
$V_{\it min}$ of minimal elements of $V$ is
effectively computable iff for any vector $\vecdef u \in \nat_\omega^k$ the
predicate $\vecdef u\!\downarrow \cap \;V \neq \emptyset$ is decidable.
\end{thm}
\proof
Assume that the minimal elements of $V$, denoted by $V_{\it min}$ can be computed.
Then $V=V_{\it min} + \nat^k$ gives a semilinear representation of $V$.
Since $\vecdef u\!\downarrow$ is also a semilinear set, a representation of which 
can be found effectively, the 
predicate $\vecdef u\!\downarrow \cap \;V \neq \emptyset$ is decidable.

On the other hand, assume that the predicate is decidable
 for any vector $\vecdef u \in \nat_\omega^k$. The following method then
  effectively constructs  $V_{\it min}$.
  First start with a singleton set of vectors $\vset_0:=\set{(\omega,\ldots,\omega)}$
  with $k$ $\omega$-s. Let $\vset_i$ be the set of vectors that we need 
  to consider in the $i$-th iteration 
  and $\vmin_i$ the set of minimal elements found 
  for $V_{\it min}$ in the $i$-th iteration. Initially $\vmin_0:=\emptyset$.
  We let $\pred_V(\vecdef u)$ denote $\vecdef u\downarrow\cap\; V\neq\emptyset$.
  We repeat the following. 

\smallskip 
{\bf Stage 1:} In this stage, we perform the following two loops 
 sequentially. 

\begin{description}
\item [Loop 1]
  We choose some vector $\vecdef u$ from $\vset_i$ 
  and compute $\pred_V(\vecdef u)$. If the value is false, then we remove $u$ from $\vset_i$.
  We get out of this loop  if $\pred_V(\vecdef u)$ is true or $\vset_i=\emptyset$.
  
  After exiting from the above loop if $\vset_i=\emptyset$, then $V_{\it min}=\vmin_i$ and we 
  stop the algorithm.
  Otherwise, $\pred_V(\vecdef u)$ is true; $\vecdef u\downarrow$ 
  contains at least one element of  $V_{\it min}$ and one such element will 
  be found in the next loop.
 \item  [Loop 2]
  We repeat the following until all coordinates of $\vecdef u$ are considered.
  Choose some coordinate $\vecdef u(i)$ of $\vecdef u$ which has not yet been considered
  and replace $\vecdef u(i)$ in $\vecdef u$ by the smallest natural number such that 
  $\pred_V(\vecdef u)$ for this new vector is still true.

  The above computed new vector will then be an element of $V_{\it min}$. So,
  we update $\vmin_{i+1}=\vmin_i \;\cup\;\set{\vecdef u}$.
\end{description}

\smallskip
 {\bf Stage 2:}
  Let the new found vector be $\vecdef u=(z_1,\ldots,z_k)$. 
 In this stage,  we try to find other vectors in $V_{\it min}$.
  We let 
\[
W'_i=\setcomp{(z'_1,\ldots,z'_k)\in\nat^k_\omega\,}{\,\exists j:1 \leq j \leq k:
  z'_j = z_j-1 
\ \wedge\ \forall m \neq j.\, z'_m=\omega}.
\]
  We update $\vset_{i+1}:= {\it min}(\vset_i,\vset'_i)$ where 
${\it min}(\vset,\vset')=\setcomp{{\it min}(\vecdef u,\vecdef {u'})}{\vecdef u\in
  \vset , \vecdef {u'}\in \vset'}$ 
and ${\it min}$ of two vectors are evaluated component-wise.
  Then we increment the iterator by $i:=i+1$ and go back to Loop 1.\qed

\subsubsection{Computing ${\it INF}_{\it min}$ for a Petri net}



While a marking of a normal untimed Petri net (or a SD-TN) is a mapping
$M: P \to \nat$ (see Def.~\ref{def:SD-TN}), an $\omega$-marking is
defined as a mapping $M: P \to \nat_\omega$, where $\nat_\omega = \nat \cup
\{\omega\}$.
In the following we work with $\omega$-markings, i.e., when we speak of
markings these may be $\omega$-markings.

For any Petri net $N$ let ${\it INF}$ be the set of markings where infinite
runs start, and ${\it INF}_{\it min}$ the finite set of minimal elements of
${\it INF}$, similarly as for SD-TN in Def.~\ref{def:INF}.
We use the result of Valk and Jantzen to compute ${\it INF}_{\it min}$ for a Petri net.
To apply this algorithm, we require the computability of the predicate
$M\downarrow\;\cap\;{\it INF}\neq \emptyset$  
 ($\pred_{\it INF}(\vecdef \marking)$) for
any $\omega$-marking $\marking$. The decidability of this predicate 
was first shown in \cite{BM:STACS99}. We include a description 
of this construction here (adapted to our notation), because the
more general construction for SD-TN in the next section is based on it and
would be hard to understand without it.

\begin{defi}\label{def:cover_g} {\bf (Coverability graph)} \cite{KaMi:schemata}\\
Given a Petri net $N$ (with $k$ places) with initial $\omega$-marking $M_0$, the Karp-Miller coverability
graph is a finite directed graph $\covergraph = (G,\to)$ 
with $G \subseteq \nat_\omega^k$ whose vertices are labeled with
$\omega$-markings of $N$. It is constructed as follows. 

Starting from $M_0$, one begins to construct the (generally infinite) 
computation graph of $N$, i.e., the graph of reachable markings, connected by
arcs representing fired transitions.
However, if one encounters a marking $M_2$ which is strictly bigger than a previously 
encountered marking $M_1$ (i.e., $M_2 \ge M_1$ and $M_2 \neq M_1$) then
one replaces $M_2$ by $M_2 + \omega (M_2 - M_1)$. This describes the effect
that by repeating the sequence of transitions between $M_1$ and $M_2$ one
could reach markings with arbitrarily many tokens on those places $p$ 
where $M_2(p) > M_1(p)$. (Note that such sequences can be repeated because
Petri nets are monotonic.)
If one encounters the same $\omega$-marking as previously, then one creates a
loop. 

It follows from Dickson's Lemma (see Lemma~\ref{lem:Dickson}) that the
generated graph is finite and the construction terminates.
\end{defi}

The following properties of the coverability graph follow directly from the
construction (see \cite{KaMi:schemata}).

\begin{lem}\label{lem:prop_covergraph}\ 
\begin{enumerate}[\em(1)]
\item
For every marking $M$, reachable from the initial marking $M_0$, 
there is an $\omega$-marking $M_\covergraph$ 
in the coverability graph such that $M \mleq M_\covergraph$.
\item
For every $\omega$-marking $\marking_\covergraph$ 
in $\covergraph$, there are markings $M$ reachable from $M$ which 
contain arbitrarily large numbers of tokens in the places with $\omega$ in
$\marking_\covergraph$.
\item
The arcs in the coverabiliy graph are induced by the transitions in the Petri
net. If it is possible to fire some sequence of transitions from a marking
$M_\covergraph$ in the coverability graph, leading to a marking
$M_\covergraph'$, then there is a reachable marking
$M \mleq M_\covergraph$ in the Petri net which can fire the same sequence of
transitions, leading to a marking $M' \mleq M_\covergraph'$.
\end{enumerate}
\end{lem}

\begin{defi}\label{def:eff_vec} {\bf (Effect Vector)}
To every transition $t$ in a normal untimed Petri net with $k$ places one can associate 
a vector $\vecdef {v_t} \in \Z^k$ which describes the effect of the transition on the 
markings of the net, i.e., the change in the marking caused by firing the
transition.
This means that if $M_1 \dto{t} M_2$, then $M_2 = M_1 + \vecdef {v_t}$. 
We call $\vecdef {v_t}$ the {\em effect-vector} of transition $t$.
\end{defi}

\begin{lem}
\label{lemma:pn}
\cite{BM:STACS99} 
Given a Petri net $N$ with $k$ places and an $\omega$-marking 
$\marking_0\in\nat^k_\omega$ where $\nat_\omega=\nat \cup \set{\omega}$ 
 and $\omega$ denotes the first limit ordinal 
(satisfying $z + \omega = z - \omega = \omega$ for $x\in \nat$), 
it is decidable if $\marking_0\downarrow \cap {\it INF} \neq \emptyset$.
\end{lem}
\proof
We show that if ${M_0\!\downarrow} \cap {\it INF} \neq \emptyset$ then this
condition will be detected by the following construction. Furthermore, we
prove that the construction does not yield any false positives.

\smallskip
\noindent{\bf Construction:}

Let $\covergraph = (G,\to)$ with $G \subseteq \nat_\omega^k$ 
be the coverability graph of $N$ from the initial marking $\marking_0$, 
which is computable (see Def.~\ref{def:cover_g} and \cite{KaMi:schemata}).

The main idea is to analyze the coverability graph $\covergraph$ and look for a cycle
s.t. the transitions fired in this cycle have a combined positive effect 
on the marking (and will thus be repeatable).
It will be shown that such a cycle in $\covergraph$ exists if and only if 
${M_0\!\downarrow} \cap {\it INF} \neq \emptyset$.

First, for every $\omega$-marking $\marking$ in the coverability graph $\covergraph$, we compute 
a finite-state automaton $\automaton_\marking$ as follows.
\begin{enumerate}[$\bullet$]
\item
The transition graph of $\automaton_\marking$ is the largest strongly
connected subgraph of $\covergraph$ containing
$\marking$.
\item
The initial state of $\automaton_\marking$ is $\marking$.
\item
$\automaton_\marking$ has only one final state, which is also $\marking$.
\item
Let $l$ be the number of edges in $\automaton_\marking$.
We label every arc in $\automaton_\marking$
with a unique symbol $\Lambda_i$ for $i:1\leq i \leq l$. 
To every symbol $\Lambda_i$, we associate the 
{\it effect-vector} (see Def.~\ref{def:eff_vec}) 
$\vecdef{\zeta_i}\in {\integers}^k$ 
that describes the effect of the transition that was fired in 
the step from one node to the other.
\end{enumerate}
Let $\langof{\automaton_\marking}$ be the regular language 
(over alphabet $\{\Lambda_i\,|\, 1 \leq i \leq l\}$)
recognized by $\automaton_\marking$.
The aim is to find a cyclic path in $\automaton_\marking$ from a marking
$\marking$ back to $\marking$ 
where the sum of all the effect-vectors of all traversed arcs is 
$\geq \vecdef{0}$. This cyclic path is not necessarily a simple cycle.
The effect-vector of an arc that is traversed 
$j$ times is counted $j$ times. Such a cyclic path with positive 
overall effect is repeatable infinitely often and thus 
corresponds to a possible infinite computation of the system $N$.

Given the automaton $\automaton_\marking$ with
 $\marking$ as its initial and the only final state,
every word 
in $\langof{\automaton_\marking}$ corresponds to a cyclic path from $\marking$ to $\marking$. 
For any word $\word$, let $\sizeof{\word}_{\Lambda_i}$ be the 
number of occurrences of $\Lambda_i$ in $\word$. The question now is 
if there is a word $\word \in \langof{\automaton_\marking}$ such that 
\[
\sum_{1\leq i \leq l} \sizeof{\word}_{\Lambda_i}\vecdef{\zeta_i} \geq
\vecdef{0}
\]
Such words characterize loops starting and ending in the same 
node of the coverability graph.
We show how to answer the above question in the following.
\begin{enumerate}[$\bullet$]
\item 
First we compute the Parikh image of $\langof{\automaton_\marking}$, i.e., 
the set $\setcomp{(\sizeof{\word}_{\Lambda_1},\ldots,\sizeof{\word}_{\Lambda_l})}{\word\!\in\!\langof{\automaton_\marking}}$. This set  is 
effectively semilinear by 
Parikh's Theorem.
\item By Theorem~\ref{semilinear:presburger}, we compute a Presburger 
formula  $\presburger(x_1,\ldots,x_l)$ from the semilinear set computed 
above. The variables $x_1,\ldots,x_l$ count the number of times each edge 
$\Lambda_i$ appears in a word $\word\in\langof{\automaton_\marking}$.
\item
Finally, to decide if
$\sum_{1\leq i \leq l} \sizeof{\word}_{\Lambda_i}\vecdef{\zeta_i} \geq \vecdef{0}$, 
we check the satisfiability of $\presburger_\automaton=\presburger(x_1,\ldots,x_l)\;\wedge\; \sum_{1\leq i \leq l} x_i\vecdef{\zeta_i} \geq \vecdef{0}$, which is 
again a Presburger formula.
By Theorem~\ref{Presburger}, we can decide whether this formula 
is satisfiable.
\end{enumerate}
For every marking $\marking$ in the coverability graph $\covergraph$
(these are finitely many) we check this condition for the 
automaton $\automaton_\marking$ and we say that 
$M_0\downarrow \cap {\it INF} \neq \emptyset$ is true if and only if 
the condition holds for at least one automaton $\automaton_\marking$.

\smallskip
{\bf Correctness:}
Now we show the correctness of the above construction.
If ${M_0\!\downarrow} \cap {\it INF} \neq \emptyset$ then there exists a
marking $M \in \nat^k$ with $M \mleq M_0$ and $M \in {\it INF}$. Thus there 
exists an infinite $M$-computation $\comp$.
By Dickson's lemma, there are markings $M',M''$ and a sequence of transitions
${\it Seq}$ 
such that $M \transclosure M' \xrultrans{Seq} M''$ and $M' \mleq M''$.
Thus the total effect of ${\it Seq}$ is non-negative.

Now, from Lemma~\ref{lem:prop_covergraph}, we 
know that
there is a $\omega$-marking $M_\covergraph$ in the coverability graph such that $M'' \mleq M_\covergraph$. 
Due to monotonicity of the transition relation, there is a path 
labeled with transitions in ${\it Seq}$ and which leads us 
from $M_\covergraph$ to a $\omega$-marking larger than $M_\covergraph$.
Repeating this process from the larger node will finally lead us to 
 a node which is largest of all $\omega$-markings larger than
$M_\covergraph$. We will reach such a node
 $\marking_\covergraph^{\maxval}$, since the graph is finite.
 This means that we can fire transitions
  in $Seq$ from $\marking_\covergraph^{\maxval}$ and we get back to 
   $\marking_\covergraph^{\maxval}$ itself (since there are no $\omega$-marking 
   larger than $\marking_\covergraph^{\maxval}$ in
   $\covergraph$ and by monotonicity ${\it Seq}$ leads to a larger or equal 
   node in $\covergraph$).  
So, $M_\covergraph^{\maxval} \xrultrans{Seq} M_\covergraph^{\maxval}$, i.e.,
there are $\omega$-markings $M_1,\ldots,M_n$ such that 
$M_\covergraph^{\maxval} \trans{} M_1 \trans{} \ldots \trans{} M_n = M_\covergraph^{\maxval}$ with effect-vectors 
$\vecdef{\zeta_1},\ldots,\vecdef{\zeta_{n}}$ such that $\sum_{1\leq i \leq n} \;\;\vecdef{\zeta_i} \geq \vecdef 0$.
This is the condition checked in our construction.

To prove the other direction, 
suppose that there is a word 
$\word\in\langof{\automaton_{\marking_\covergraph}}$ for some $\omega$-marking $\marking_\covergraph$
in the coverability graph such that 
$\sum_{1\leq i \leq l} \sizeof{\word}_{\Lambda_i}\vecdef{\zeta_i} \geq \vecdef{0}$.
This means that there 
is a $\omega$-marking $M_\covergraph$ from which there is a path (through a 
sequence $Seq$ of transitions) back to itself 
with non-negative effect.
From Lemma~\ref{lem:prop_covergraph} we know that
there are markings $M'$ reachable from $M_0$ which 
agree with $M_\covergraph$ in its finite coordinates, and can be made 
arbitrarily large in the coordinates where $M_\covergraph$ is $\omega$.
We can choose one such marking $M'$ such that it contains enough tokens 
in those coordinates where $M_\covergraph$ is $\omega$ to be able to perform 
one iteration of ${\it Seq}$.
Now,  ${\it Seq}$ has a non-negative effect.
This means that one can 
repeatedly execute ${\it Seq}$ starting from $M'$. 
The reachability of such an $M'$ from $M_0$ and a non-negative loop from $M'$ 
implies the existence of an infinite $M_0$-computation. 
This means that $M_0\downarrow \;\cap\; {\it INF}\; \neq \emptyset$.\qed

\begin{figure}[htbp]
\begin{center}
\scalebox{0.9}{
\input{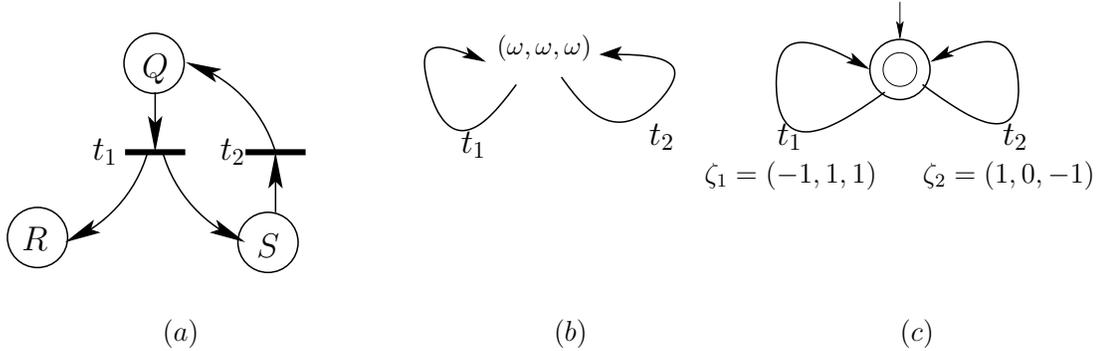}
}
\caption{(a). A small Petri net, (b). Coverability graph for this net from $(\omega,\omega,\omega)$. (c) 
Automaton $A_{(\omega,\omega,\omega)}$.}
\label{pred:zeno}
\end{center}
\end{figure}

\begin{exa}
\label{example:pred}
Consider the Petri net in Figure~\ref{pred:zeno}(a) and 
 the coverability graph (Figure~\ref{pred:zeno}(b)) of the above Petri net from  
a $\omega$-marking $\marking=(\omega,\omega,\omega)$ \footnote{Markings of a Petri net are written as multisets over places 
and vectors over the set of natural numbers interchangeably.} where
$\marking(Q)=\marking(R)=\marking(S)=\omega$.
We show that $\marking\downarrow\;\cap\;{\it INF}\neq \emptyset$.
The automaton produced for the single node in the coverability graph is shown in Figure~\ref{pred:zeno}(c). Notice that $\Lambda_1=t_1$ and $\Lambda_2=t_2$. Also, the effect-vectors
$\vecdef{\zeta_1}$ and $\vecdef{\zeta_2}$ show the effect of firing $t_1$ and $t_2$ 
respectively. Notice that $\langof{A_{(\omega,\omega,\omega)}}=\setcomp{\word}{\word\in\wstar{t_1,t_2}}$. This means that $\parikh(\langof{A_{(\omega,\omega,\omega)}})=L((0,0);(1,0),(0,1))$. Finally, we compute 
a Presburger formula $\presburger(x_1,x_2)$ for the above linear set 
and from it, construct the formula $\presburger(x_1,x_2)\:\wedge\;x_1\vecdef{\zeta_1} + x_2\vecdef{\zeta_2}\geq \vecdef 0$. 
One of the  solutions of this formula is given by $x_1=x_2=k$ for any natural number $k$.
This means $\marking\downarrow\;\cap\;{\it INF} \neq \emptyset$.
\end{exa}

\begin{figure}[htbp]
\begin{center}
\scalebox{1.0}{
\input{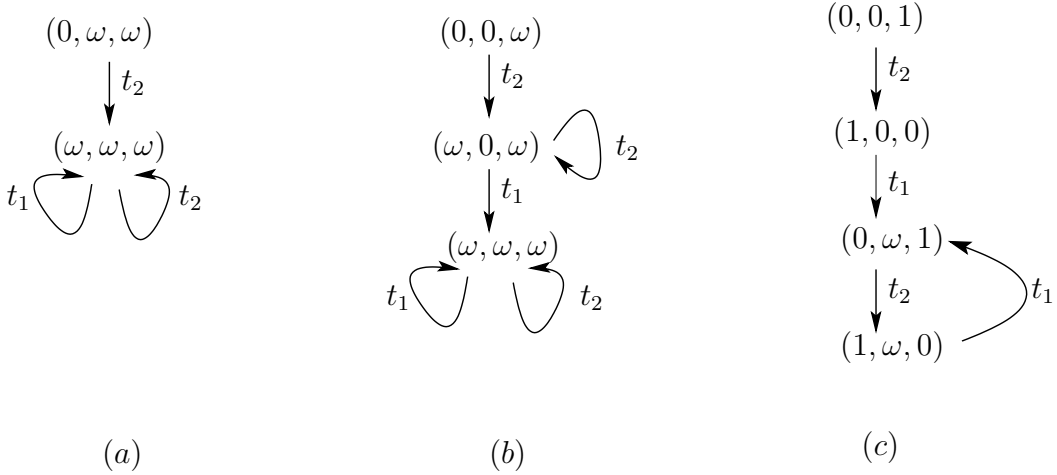}
}
\caption{$(a)$. Coverability graph from $(0,\omega,\omega)$. $(b)$ Coverability graph from $(0,0,\omega)$. $(c)$ 
Coverability graph from  $(0,0,1)$.}
\label{fig:pred}
\end{center}
\end{figure}
\begin{exa}
In the above, we show an example for computing $\pred_{{\it INF}}(\marking)$ for 
an $\omega$-marking $\marking$.
Now we show how to compute ${\it INF}_{\it min}$ for the same Petri net using Valk and Jantzen's algorithm. We start with a single marking $(\omega,\omega,\omega)$. 
Immediately, we get out of 
Loop 1, since $\pred_{{\it INF}}((\omega,\omega,\omega))$ is true (as shown in 
Example~\ref{example:pred}).
In Loop 2, one finds a minimal element in ${\it INF}_{\it min}$.
This is done by first reducing the first coordinate for $Q$ in
 $(\omega,\omega,\omega)$ to $0$. In Figure~\ref{fig:pred}(a), 
we show the coverability graph from $(0,\omega,\omega)$. $\pred_{{\it INF}}((0,\omega,\omega))$ is true, since we reach a node 
$(\omega,\omega,\omega)$ in the coverability graph  from $(0,\omega,\omega)$
and $\pred_{{\it INF}}((\omega,\omega,\omega))$ is already shown to be true in the 
previous example. Then we replace the $\omega$ in place $R$ to $0$ and 
compute the coverability graph for $(0,0,\omega)$ in
Figure~\ref{fig:pred}(b). 
$\pred_{{\it INF}}((0,0,\omega))$ is true again  by the same reasoning.
Notice that $\pred_{{\it INF}}((0,0,0))$ is false.
So, finally we show the coverability graph from marking  $(0,0,1)$  in 
Figure~\ref{fig:pred}(c) and 
$\pred_{{\it INF}}((0,0,1))$ is true.  Thus $(0,0,1)$ is included in 
${\it INF}_{\it min}$.

\begin{figure}[htbp]
\begin{center}
\scalebox{1.0}{
\input{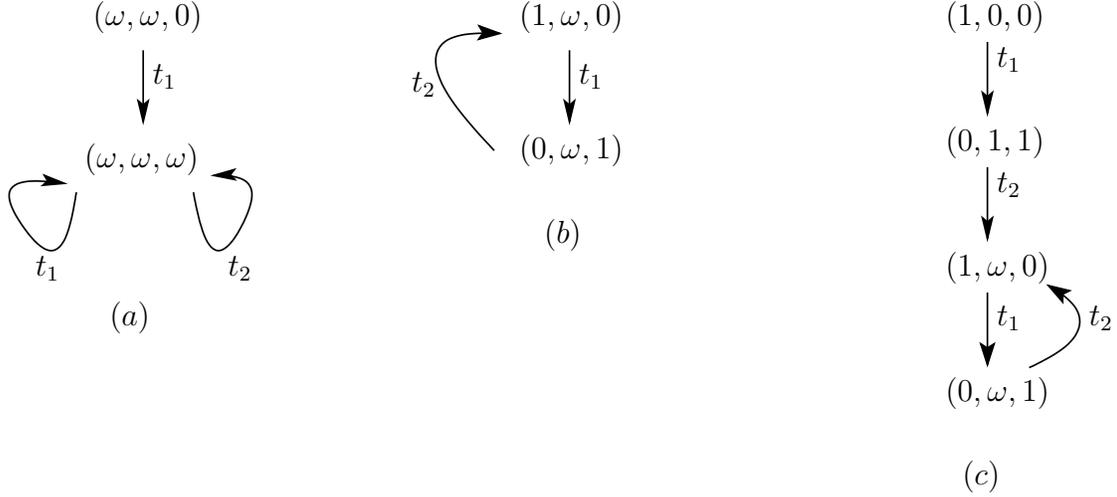}
}
\caption{(a). Coverability graph from $(\omega,\omega,0)$ (b). Coverability graph from $(1,\omega,0)$. (c) 
Coverability graph from  $(1,0,0)$.}
\label{fig:pred:2}
\end{center}
\end{figure}
In Stage 2, we have $W'_0=\set{(\omega,\omega,0)}$ and 
$W_1={\it min}((\omega,\omega,\omega),(\omega,\omega,0))=\set{(\omega,\omega,0)}$.

Now we go to Loop 1 again.
From Figure~\ref{fig:pred:2}(a), it is evident that 
$\pred_{{\it INF}}((\omega,\omega,0))$ is true. Now, we again perform Loop 2.
We find that  $\pred_{{\it INF}}((0,\omega,0))$  is false, but 
  $\pred_{{\it INF}}((1,\omega,0))$  is true (the coverability 
graph from $(1,\omega,0)$ is shown in Figure~\ref{fig:pred:2}(b)). 
We show the coverability from $(1,0,0)$ in Figure~\ref{fig:pred:2}(c)
and it follows that 
$\pred_{{\it INF}}((1,0,0))$ is true.
Thus $(1,0,0)$ is another member of ${\it INF}_{\it min}$.

In Stage 2, we have $W'_1=(0,\omega,\omega)$ and $W_2={\it min}((0,\omega,\omega),(\omega,\omega,0))=(0,\omega,0)$.
Now $\pred_{{\it INF}}((0,\omega,0))$ is false and $W_2=\emptyset$ and the
construction terminates.
Thus ${\it INF}_{\it min}=\set{(0,0,1),(1,0,0)}$. 
\end{exa}

\subsubsection{Computing ${\it INF}_{\it min}$ for SD-TNs}
\ \\

To compute ${\it INF}_{\it min}$ for SD-TNs, we will use 
Valk and Jantzen's Theorem~\ref{thm:Valk_Jantzen} again.
This algorithm requires a decision procedure for the predicate
  $\marking_0\downarrow \cap {\it INF} \neq \emptyset$ 
for any given $\omega$-marking $\marking_0\in\nat^k_\omega$ for an SD-TN.
First we construct a 
coverability graph for a given SD-TN. We need the following
definitions and notational conventions.

\begin{defi}
By Def.~\ref{def:SD-TN} of SD-TN, the source
places and target places of transfers are disjoint and thus after a
simultaneous transfer all source places are empty. We call a marking 
an `after transfer marking' (AT-marking) if it is reached just after firing 
${\it Trans}$. We represent markings as vectors in $\nat^k$ of the form
(transfer source places, other places). So AT-markings have the form
$(\overrightarrow{0}, \overrightarrow{v})$ with $\overrightarrow{0} \in \nat^{k'}$ and $\overrightarrow{v} \in \nat^{k''}$
with $k=k'+k''$ where $k'$ is the number of transfer source places. 
The corresponding markings in 
the coverability graph $\covergraph$ are called $\omega$-AT-markings and have the form
$(\overrightarrow{0}, \overrightarrow{v})$ with $\overrightarrow{v} \in \nat_\omega^{k''}$.
\end{defi}


First we show that the coverability graph for SD-TN can be effectively
constructed (Lemma~\ref{lem:constr_coverability}), then we prove that this
graph satisfies the required properties (Lemma~\ref{lemma:cover:sdtn})
and finally we give an example.

\begin{lem}\label{lem:constr_coverability}
For any SD-TN $N$ with initial marking $M_0$, the coverability graph 
can be effectively constructed.
\end{lem}
\proof
We use $\omega$-markings from $\nat_\omega^k$ (where $k$ is the number of places). 
One proceeds from $M_0$ similarly as in the Karp-Miller construction
(\cite{KaMi:schemata}; see also Def.~\ref{def:cover_g}) 
except for the transfer arc. The detection of loops is done 
slightly differently in the two cases (with and without the transfer arc).
\begin{enumerate}[(1)]
\item {\it Loop without transfer arc:}
If one encounters the case $M_1 \longrightarrow_{\it Seq} M_2$ with 
\begin{enumerate}[$\bullet$]
\item $M_1 < M_2$, 
\item {\it Seq} is a sequence of transitions of $N$ such that
the transfer arc was not used in {\it Seq},
\end{enumerate}
then we replace 
$M_2$ by $M_2 + \omega(M_2-M_1)$ as in the case of Petri nets. 
Notice that  $\omega\marking = \marking'$ such that $\marking'(p)=\omega$ for all place 
$p$ with $\marking(p) > 0$.
Obviously, {\it Seq} can be repeated
arbitrarily often to yield an arbitrarily high number of tokens on the places
where $M_2$ is strictly larger than $M_1$.
\item {\it Loop containing transfer arc:}
Let $M_1$ and $M_2$ be two markings {\it reached just after transfers}, i.e.,
$\longrightarrow_{\it Trans} M_1 \longrightarrow_{\it Seq} M_1' \longrightarrow_{\it Trans} M_2$ (where {\it Seq} may 
contain other transfers). We call such markings $\omega$-AT-markings (AT for
`after transfer'). 
If $M_1 < M_2$  then 
we replace $M_2$ by $M_2 + \omega(M_2-M_1)$.
The sequence of transitions $\longrightarrow_{\it Seq}\longrightarrow_{\it Trans}$ can be repeated 
arbitrarily often to yield arbitrarily high numbers of tokens on the places
where $M_2$ is strictly bigger than $M_1$. This is possible, because in SD-TN 
the set of places which are sources of transfers and the set of places which 
are targets of transfers are disjoint by Def.~\ref{def:SD-TN}.
Thus the transfers in $\longrightarrow_{\it Seq}\longrightarrow_{\it Trans}$ do not negatively affect
those places $p$ where $M_1(p) < M_2(p)$. 
This point does not carry over to general transfer nets.
In particular, all transfer-target places, once marked by $\omega$ in this construction, 
will stay $\omega$ in the future. Furthermore, all transfer source places
are empty after the transfer, since all transfers are simultaneous. 
\item
If one reaches an $\omega$-marking encountered before, then one creates a loop.
\end{enumerate}
It is easy to show that the so-generated coverability graph is finite.
Assume the contrary, i.e., that there is an infinite sequence $M_0,M_1,\ldots$ 
of {\em different} nodes in the coverability graph.
Now, there are two cases.
\begin{enumerate}[$\bullet$]
\item In this infinite sequence, there is only 
a finite number of occurrences of the transfer 
transition ${\it Trans}$. Suppose $M_r$ was the last marking produced by transfer 
transition. Consider the sequence $M_{r+1},M_{r+2},\ldots$. This sequence is
 still infinite. By Dickson's lemma (Lemma~\ref{lem:Dickson}), any such infinite 
sequence of markings of the SD-TN contains an infinite non-decreasing
subsequence. Since, by our assumption above, all markings $M_i$ are different,
this subsequence must be strictly increasing.
Thus, in our construction above, it would happen infinitely often that a place
is marked by $\omega$ which previously had only held a finite number.
However, since the infinite suffix $M_{r+1},M_{r+2},\ldots$ does not contain
any transfer, all places marked $\omega$ stay at $\omega$. 
This yields a contradiction, since there are only finitely many places in the net.
\item There is an infinite number of 
markings produced by the transfer transition ${\it Trans}$, which appear
 in the sequence $M_0,M_1,\ldots$. We take the subsequence 
$M'_0,M'_1,\ldots$ of $M_0,M_1,\ldots$ such that each marking $M'_i$ 
for $i \geq 0$ is a marking produced by the transfer transition
(i.e., an $\omega$-AT-marking).
Since there are infinitely many transfer transitions in 
the sequence $M_0,M_1,\ldots$, the sequence $M'_0,M'_1,\ldots$ is 
also infinite. Now, like the previous case,
we will always find a strictly increasing subsequence of $M'_0,M'_1,\ldots$.
Thus, by the construction above, we would infinitely often introduce the
 number $\omega$ into some places of the net. However, this could only happen
to places which are {\em not} sources of transfers, since all source-places
of transfers are marked zero in $\omega$-AT-markings. Since those places
marked by $\omega$ are not sources of any transfers, 
they will always remain marked $\omega$. 
(Here we require the specific property from SD-TN. This does not hold for
general transfer nets, where a target place of one transfer could be 
the source place of another.)
This yields a contradiction, because 
there are only finitely many places in the net and $\omega$ could not be 
introduced infinitely often as required above.
\end{enumerate}
Since our assumption above led to a contradiction in both cases, the
opposite must be true, i.e., the generated coverability graph is
finite.\qed


\smallskip
\noindent
{\bf Remark:} Notice that if a place $\place$ is a source of a transfer transition,
then $\marking_1(p) < \marking_2(p)$ does not in general imply that $p$ 
may eventually contain an arbitrarily high number of tokens. 
This is due to the fact that the loop may contain 
a transfer transition which will remove all tokens from $p$.

\begin{lem}\label{lemma:cover:sdtn}\ 
\begin{enumerate}[\em(1)]
\item
For every reachable marking $M$ from the initial marking $M_0$ in an SD-TN, 
there is an $\omega$-marking $M_\covergraph$ 
in the coverability graph such that  $M \mleq M_\covergraph$.
\item
For every $\omega$-marking $\marking_\covergraph$ 
in $\covergraph$, there are markings $M$ reachable from $M_0$ which 
contain arbitrarily large numbers of tokens in the places with $\omega$ in
$\marking_\covergraph$.
\item
The arcs in the coverability graph are induced by the transitions in the SD-TN.
If some sequence of transitions if possible to fire from a marking
$M_\covergraph$ in the coverability graph, leading to a marking
$M_\covergraph'$, then there is a reachable marking
$M \mleq M_\covergraph$ in the SD-TN which can fire the same sequence of
transitions, leading to marking $M' \mleq M_\covergraph'$.
\end{enumerate}
\end{lem}
\proof
The proof is similar to the correctness proof 
of the Karp-Miller algorithm for ordinary Petri nets \cite{KaMi:schemata}.
\begin{enumerate}[(1)]
\item
First, for every computation path staring at $M_0$ in the SD-TN there is a
corresponding path in the coverability graph constructed in
Lemma~\ref{lem:constr_coverability}. Furthermore, markings are only replaced
by larger $\omega$-markings in the coverability graph. By the monotonicity of
SD-TN, the first result follows.
\item
By the construction of the coverability graph for SD-TN in
Lemma~\ref{lem:constr_coverability},
values $\omega$ can be introduced in two ways: by encountering an increasing
loop without transfer arcs or an increasing loop with transfer arcs.

In the first case, the loop can simply be repeated arbitrarily often to yield 
arbitrarily high numbers of tokens on the increasing places 
(marked by $\omega$ in the coverability graph), because of the monotonicity of
the net, just as for ordinary Petri nets.

In the second case, new $\omega$ are only introduced for increasing loops
between $\omega$-AT-markings, i.e., loops of the form 
$\longrightarrow_{\it Trans} (\vec{0},\vec{v}) \longrightarrow_{\it Seq} M_1
\longrightarrow_{\it Trans} (\vec{0},\vec{v}')$
where $\vec{v}' > \vec{v}$. Since the source places
of transfers are all marked $0$ in these markings, no $\omega$s are introduced
to them here. (However, source places of transfers may aquire $\omega$ 
(either permanently or just temporarily until the next transfer) by
ordinary Petri nets loops in the first case described above.)
By the special restrictions on transfers in SD-TN (unlike in general transfer
nets) the places marked by vectors $\vec{v}, \vec{v}'$ which may aquire
$\omega$ here are never the source of {\em any} transfer. Thus the loop
$\longrightarrow_{\it Seq} \longrightarrow_{\it Trans}$ can be repeated 
arbitrarily often to yield markings with arbitrarily high numbers of tokens on 
those places where $\vec{v}'$ is strictly larger than $\vec{v}$.
\item
The third property follows directly from the definition of the coverability graph.\qed
\end{enumerate}

\begin{rem}
It follows directly from Lemma~\ref{lem:constr_coverability}
and Lemma~\ref{lemma:cover:sdtn} that place-bounded\-ness is
decidable for simultaneous-disjoint transfer nets, while it is undecidable for
general transfer nets \cite{DJS:ICALP99,Mayr:LCM:TCS}.
\end{rem}

\begin{figure}[htbp]
\begin{center}
\scalebox{0.9}{
\input{sdtn_pred.pstex_t}
}
\caption{(a) A small SD-TN. (b) Coverability graph $\covergraph$ for this net.}
\label{sdtn_pred:zeno}
\end{center}
\end{figure}

\begin{exa}
Consider a small SD-TN shown in Figure~\ref{sdtn_pred:zeno}(a).
In Figure~\ref{sdtn_pred:zeno}(b),  we show the coverability graph
$\covergraph$ from a marking $\marking=(2,0,0)$ of SD-TN where 
$\marking(p_1)=2, \marking(p_2)=0$ and $\marking(p_3)=0$. 
We omit the transfer arcs in the  coverability graph if the source place 
of transfer does not contain a token. Notice that 
${\it Trans}=(\emptyset,\emptyset,(p_1,p_3))$ and $(0,0,2)$ and $(0,\omega,\omega)$ are the only 
$\omega$-AT-markings in $\covergraph$.
\end{exa}

\subsubsection{Computing $\pred_{\it INF}$ for SD-TNs.}
\ \\

Now that we can compute the coverability graph for SD-TN, we continue to
develop the algorithm for deciding the predicate $\pred_{\it INF}$, i.e., 
deciding if $\marking_0\downarrow \cap {\it INF} \neq \emptyset$ 
for any given $\omega$-marking $\marking_0\in\nat^k_\omega$ for an SD-TN.

\begin{lem}\label{lem:find_infrun}
Given an SD-TN $N$ with $k$ places and an $\omega$-marking $M_0 \in \nat_\omega^k$, it is decidable if
${M_0\!\downarrow} \cap {\it INF} \neq \emptyset$. 
\end{lem}
\proof
First we give an algorithm to detect the non-emptiness of the intersection 
 ${M_0\!\downarrow} \cap {\it INF}$. 
Let $\covergraph = (G,\to)$ with $G \subseteq \nat_\omega^k$ 
be the coverability graph of $N$ from initial marking $M_0$. 
An infinite computation $\comp$ from a marking $M$ in $M_0\!\downarrow$ 
is detected as follows. There are two cases.
Either there are finitely many or infinitely many transfers in such 
an infinite computation.
\begin{enumerate}[$\bullet$]
\item In the first case, the transfer transition ${\it Trans}$ is 
used only finitely often and $\comp$ has an infinite suffix $\comp'$
which starts at some marking $M'$ and 
only normal Petri net transitions are used in $\comp'$. 
Since $M \transclosure M'$, there is a node
$M_\covergraph$ in $\covergraph$ such that $M' \mleq M_\covergraph$. 
To find out whether there is a positive effect of such cycles consisting 
of ordinary Petri net transitions,
we let $N'$ be the ordinary Petri net obtained from $N$ by removing 
the transfer transition ${\it Trans}$. 
So $\comp'$ is an infinite $M'$-computation of
$N'$. 
Let ${\it INF}_{N'} \subseteq \nat^k$ be the (upward-closed) set of markings from which infinite
computations of $N'$ start. So we have $M_\covergraph\downarrow \cap \;{\it INF}_{N'} \neq \emptyset$.
In fact, we consider each $\omega$-marking $\marking_\covergraph\in G$ and 
detect the presence of an infinite computation with just ordinary Petri net 
transitions if the following condition (Cond1) holds.
\[ 
\mbox{(Cond1)} \quad\quad \exists M_\covergraph \in G.\, M_\covergraph\!\downarrow \cap \;{\it INF}_{N'} \neq \emptyset
\] 
This is a problem about ordinary Petri nets and it has already been shown to be decidable
 (Lemma~\ref{lemma:pn}). Deciding (Cond1) requires only finitely many calls to
 the decision procedure in Lemma~\ref{lemma:pn}, because $G$ is finite.
\hide{
\begin{figure}[!ht]
\begin{center}
\epsfig{figure=sdtn_pred1.eps, width=.7\textwidth}
\caption{(a). Petri net $N'$ derived from $N$ in 
Figure~\ref{sdtn_pred:zeno}(a). (b) Coverability graph $\covergraph$ 
from $(0,0,2)$.}
\label{sdtn_pred1:zeno}
\end{center}
\end{figure}
In Figure~\ref{sdtn_pred1:zeno}(a), we show a Petri net 
$N'$ derived from $N$ by removing the transfer transition and 
in Figure~\ref{sdtn_pred1:zeno}(b), we show the coverability 
graph from $M_\covergraph=(0,0,2)$. Notice that there are other choices for $M_\covergraph$.
Evidently there are self loops in the automata $\automaton_{(0,\omega,2)}$ and 
$\automaton_{(\omega,\omega,2)}$ (obtained as in Lemma~\ref{lemma:pn}) 
and this lets the condition Cond1 to 
be satisfiable. 
}
\item
In the second case, the transfer transition ${\it Trans}$ is used
infinitely often in $\comp$. 
Recall that in Lemma~\ref{lemma:pn}, we construct automata from the coverability graph,
for each of its nodes and associate an effect-vector with each edge of such an automaton.
In this case, the presence of transfer transitions in the cycles of SD-TNs does not let us follow 
such a procedure directly. This is due to the fact that the effect of the transfer 
depends on the amount of tokens in the source places of the transfer and that is 
not a constant number.

In this case, first we compute the effect-vectors between two $\omega$-AT-markings 
$\qmarking,\qmarking'$ in the coverability graph such that $\qmarking'$ is reachable from 
$\qmarking$.
\hide{
In this construction we use semilinear languages, Presburger arithmetic
and Parikh's theorem. Semilinear sets are finite unions of linear subsets
of $\Z^n$ (for arbitrary $n$). 
It is known that a set is semilinear iff it is definable in 
Presburger arithmetic \cite{GS:PacMat66}. Presburger arithmetic is the first-order
theory of the integers with addition and order, also denoted as
$(\Z,+,\le)$. Presburger arithmetic is decidable; see, e.g., \cite{Berman:TCS80}.
Parikh's theorem \cite{Parikh:JACM66} shows that the Parikh-image (the set of vectors,
each representing the numbers of the different letters in a word) of regular (and even
context-free) languages is effectively semilinear.
}
For any pair of $\omega$-AT-markings $\qmarking,\qmarking' \in G$ we can effectively construct 
a semilinear set $\effect(\qmarking,\qmarking') \subseteq \Z^k$ which represents all possible
effects of sequence of transitions of the form {\it Seq}.${\it Trans}$ 
with $\qmarking \longrightarrow_{{\it Seq}}\longrightarrow_{\it Trans}
\qmarking'$ 
where {\it Seq} is a sequence of transitions which 
does not contain ${\it Trans}$. This is done as
follows. 
\hide{
\begin{figure}[!ht]
\begin{center}
\epsfig{figure=sdtn_pred_aut.eps, width=.7\textwidth}
\caption{(a). 
Automaton $\automaton$ with 
$\marking=(0,0,2)$ as the initial state and 
$X=(\omega,\omega,2)$ as the final state.
(b) Automaton $\automaton'$ with 
$\marking=(0,\omega,\omega)$ as the initial state and 
$X=(\omega,\omega,\omega)$ as the final state.}
\label{sdtn_pred_aut:zeno}
\end{center}
\end{figure}
}
First, we compute the semilinear sets 
$\effect'(\qmarking,X) \subseteq \Z^k$ for all $X \in G$ such that  $X \longrightarrow_{\it Trans} \qmarking'$
in the coverability graph $\covergraph$ and $\qmarking \transclosure X$ without using ${\it Trans}$.
The sets $\effect'(\qmarking,X)$ are semilinear and effectively constructible, 
by computability of Presburger-arithmetic and its equivalence with semilinear
languages (Theorem~\ref{semilinear:presburger}).
This is due to the fact that
 $\covergraph$ is a finite graph whose arcs are labelled with constant vectors 
in $\Z^k$ and 
the Parikh-image of regular languages is effectively semilinear. This means that one can consider $\qmarking$ as the
initial- and $X$ as the final state of a finite automaton $\automaton$.
Each edge in $\automaton$ is labelled by a unique symbol $\Lambda$ and 
there is an associated effect-vector $\zeta$ for the effect of the 
transition by that edge. 
Let $\presburger(x_1,\ldots,x_l)$ be the Presburger 
formula for the Parikh-image of $\langof\automaton$ where 
$l$ is the number of edges in the coverability graph.
A valuation of the variable $x_i$ for $i:1\leq i \leq l$ gives how many 
times the symbol $\Lambda_i$ appears in a word in $\langof\automaton$.  
Given $k$ as the number of places in SD-TN, 
we have $\effect'(\qmarking,X)$  given by a Presburger formula
\[
 \presburger_X(y_1,\ldots,y_k)=\exists x_1\ldots,x_l.\, 
\presburger(x_1,\ldots,x_l) \wedge \bigwedge_{1\leq i \leq k} y_i = \sum_{1\leq j \leq l} x_j\zeta_j(i)
\]
\hide{
We illustrate the computation for $\effect'(\qmarking,X)$ with an example.
Consider the $\omega$-AT-markings $\qmarking=(0,0,2)$ and $\qmarking'=(0,\omega,\omega)$ 
in the coverability graph of Figure~\ref{sdtn_pred:zeno}. In this case, 
$X\in\set{(\omega,\omega,2)}$, since $(0,0,2) \trans{*} (\omega,\omega,2) 
\longrightarrow_{{\it Trans}} (0,\omega,\omega)$.
Consider an automaton $\automaton$ (Figure~\ref{sdtn_pred_aut:zeno}(a))
with initial marking $\qmarking$ and final marking $X$ and the edges 
from the coverability graph of Figure~\ref{sdtn_pred:zeno} except those 
marked as ${\it Trans}$.
The language of this automaton is expressed by the regular expression 
$\Lambda_1\wstar{\Lambda_2}\Lambda_3\wstar{\Lambda_4,\Lambda_5}$, 
where the alphabet 
of $\automaton$ is given by the set $\set{\Lambda_1,\ldots,\Lambda_5}$.
Now $\parikh(\langof \automaton)=L((1,0,1,0,0);(0,1,0,0,0),(0,0,0,1,0),(0,0,0,0,1))$. 
From $\parikh(\langof \automaton))$, we compute $\effect'(\qmarking,X)$ as 
follows. 
We know that the effect-vectors for $\Lambda_1,\Lambda_2,\Lambda_4$ are the 
same as 
$(0,1,0)$ (corresponding to $t_1$) 
and those for $\Lambda_3,\Lambda_5$ are given by $(1,-1,0)$ (corresponding to $t_2$).
This means that 
 $\effect'(\qmarking,(\omega,\omega,2))=L(1*(0,1,0)+1*(1,-1,0);\;1*(0,1,0),\;1*(0,1,0),\;1*(1,-1,0))=L((1,0,0);(0,1,0),(1,-1,0))$.
Furthermore, consider $\qmarking=\qmarking'=(0,\omega,\omega)$.
In this case, $X=(\omega,\omega,\omega)$. Consider 
the automaton $\automaton'$ (Figure~\ref{sdtn_pred_aut:zeno}(b)) 
with $\qmarking$ 
as initial state and $X$ with final state and it has 
only allowed transitions  $t_1$ and $t_2$ as before. 
The four edges in the coverability graph are labelled by $\Lambda_1,\ldots,\Lambda_4$.
Now 
$\wstar{\Lambda_1}\Lambda_2\wstar{\Lambda_3,\Lambda_4}$ gives the language of $\automaton'$ and effect-vectors for $\Lambda_1,\Lambda_3$ are $(0,1,0)$ and 
and those for $\Lambda_2,\Lambda_4$ are $(1,-1,0)$.
$\parikh(\langof {\automaton'})
=L((0,1,0,0);(1,0,0,0),(0,0,1,0),(0,0,0,1))$ and 
 $\effect'(\qmarking,X)=L((1,-1,0);(0,1,0),(1,-1,0))$.
}

Secondly, we 
obtain $\effect(\qmarking,\qmarking')$ as a 
 Presburger formula by introducing the effect of transfers 
(${\it Trans}=(I,O,\st)$) as follows.
Consider the set ${\bf X}$ containing $\omega$-markings
X 
such that $\qmarking \trans{*} X \xrultrans{{\it Trans}} \qmarking'$.
For each $X\in{\bf X}$,
 we compute 
 a Presburger formula  
\[
\presburger_X''(z_1,\ldots,z_k) = \exists y_1,\ldots,y_k.\;
(\presburger_X(y_1,\ldots,y_k)
\;\wedge\;\presburger'_X(y_1,\ldots,y_k,z_1,\ldots,z_k))
\]
where $\presburger_X'(y_1,\ldots,y_k,z_1,\ldots,z_k)$ is a conjunction of 
the following formulae.
\begin{enumerate}[$\bullet$]
\item $\forall j,j':(p_{j'},p_j) \in \st.\, z_j = y_j + y_{j'} \;\wedge\;
 z_{j'} = 0$.
Here, $\st$ is from Def.~\ref{def:SD-TN}. 
This corresponds to  a transfer from place $p_{j'}$ to place $p_j$
 whenever $(p_{j'},p_j) \in \st$. 
\item $\forall p_j\in I.\, z_j = y_j - 1
\;\wedge\; \forall p_j \in O.\, z_j = y_j + 1$.
This corresponds to Petri net part of transfers, since $I$ contains 
places from which there is an input arc to the transfer transition and $O$ 
contains places from which there is an output arc to the transfer transitions.  
\item $\forall j. (p_j\not\in\st\, \wedge\, p_j\not \in I \cup O)\,\Rightarrow\, z_j = y_j$.
Here $p_j\not\in\st$ is used to mean that there are no pairs $(p,q)\in\st$, 
such that $p_j=p$ or $p_j=q$. 
This means that there is no change in the number of tokens at the other places.
 \end{enumerate} 
Finally the effect $\effect(\qmarking,\qmarking')=\bigvee_{X\in{\bf X}}\presburger_X''(z_1,\ldots,z_k)$. 
%
By Theorem~\ref{Presburger}, we can compute a semilinear set from the 
Presburger formula given above for $\effect(\qmarking,\qmarking')$.
\hide{
We illustrate the above computation of $\effect(\qmarking,\qmarking')$ 
in the following.
Consider $\qmarking=(0,0,2)$ and $\qmarking'=(0,\omega,\omega)$ 
in the coverability graph of Figure~\ref{sdtn_pred:zeno}. 
Recall that 
 $\effect'(\qmarking,X)=L((1,0,0);(0,1,0),(1,-1,0))$.
Then $\effect(\qmarking,\qmarking')$
can be written as $L((0,0,1);(0,1,0),(0,-1,1))$ (moving all tokens 
from $p_1$ to $p_3$, since $(p_1,p_3)\in\st$).
In the second case, 
$\qmarking=\qmarking'=(0,\omega,\omega)$ and $X=(\omega,\omega,\omega)$.
Recall that in this case $\effect'(\qmarking,X)
=L((1,-1,0);(0,1,0),(1,-1,0))$.
This means that here
$\effect(\qmarking,\qmarking')=L((0,-1,1);(0,1,0),(0,-1,1))$.
}

Now we construct a new finite graph $\covergraph' = (G',\to)$ as follows. $G' \subseteq
G$ is the set of $\omega$-AT-markings in $G$. For $\qmarking,\qmarking' \in G'$ we have
$\qmarking \to \qmarking'$ in $\covergraph'$ iff 
$\qmarking \longrightarrow_{{\it Seq}''}\longrightarrow_{\it Trans} \qmarking'$
in $\covergraph$ where 
${\it Seq}''$ does
not contain ${\it Trans}$. The arc between $\qmarking$ and $\qmarking'$ is labeled with 
(a symbolic Presburger-arithmetic representation of the semilinear set) $\effect(\qmarking,\qmarking')$.

We check the following condition (Cond2).
\[
\begin{array}{ll}
\mbox{(Cond2)} &
\exists n \in \nat.\, \qmarking_0,\dots,\qmarking_n \in G'.\, \qmarking_0 \to
\qmarking_1 \to \dots \to \qmarking_n = \qmarking_0.\\
& \displaystyle
\exists \overrightarrow{v_i} \in \effect(\qmarking_i,\qmarking_{i+1}). 
\sum_{i=0}^{n-1} \overrightarrow{v_i} \ge \overrightarrow{0}
\end{array}
\]
Note that the $\qmarking_i$ above do not need to be disjoint.

Now we show how to check the condition (Cond2). We transform the graph $\covergraph'$,
whose arcs are labeled with semilinear sets $\effect(\qmarking,\qmarking')$ 
into
a new equivalent graph $\covergraph''$  
whose arcs are labeled with constant vectors.
Since $\effect(\qmarking,\qmarking')$ is effectively semilinear, it can be represented as 
a finite union of
linear sets of the form
$\lang(\vecdef{u_i};\vecdef{w_i^1},\ldots,\vecdef{w_i^{n_i}})$ where $i: 1\leq i \leq m$ and $m \geq 1$.
$\covergraph''$ contains the nodes of $\covergraph'$ and some additional 
nodes:
\begin{enumerate}[$\bullet$]
\item if there is an edge between two nodes $\qmarking,\qmarking'$ 
labeled by $\effect(\qmarking,\qmarking')$ (of the above form)
in $\covergraph'$,
 we add 
 new nodes $\qmarking'_i$ for $i:1\leq i \leq m$ in $\covergraph''$.
\end{enumerate}
Also, for any pair of nodes $\qmarking,\qmarking'$ in $C'$, labeled by
$\bigcup_{1\leq i \leq m}\lang(\vecdef{u_i};\vecdef{w_i^1},\ldots,\vecdef{w_i^{n_i}})$,
we have the following arcs in $\covergraph''$.
For each $i:1\leq i \leq m$, we have
\begin{enumerate}[$\bullet$]
\item an edge from $\qmarking$ to $\qmarking'_i$, labeled by $\vecdef{u_i}$.
\item edges from $\qmarking'_i$ to $\qmarking'_i$, labeled by $\vecdef{w_i^j}$ 
for $j:1\leq j \leq n_i$.
\item an edge from $\qmarking'_i$ to $\qmarking'$, labeled by $\vecdef{0}$.
\end{enumerate}
Let $\covergraph''=(G'',\to)$ be the graph obtained in this way. 
We get immediately that the
following condition (Cond3) holds for $C''$ iff (Cond2) holds for $C'$.
\[
\begin{array}{ll}
\mbox{(Cond3)} & \exists n \in \nat.\, \qmarking_0,\dots,\qmarking_n \in G''.\\
& \displaystyle
(\qmarking_0 \dto{v_0} \qmarking_1 \dto{v_1} \dots \dto{v_{n-1}} \qmarking_n = \qmarking_0)\;\;\wedge\;\;
\sum_{i=0}^{n-1} \overrightarrow{v_i} \ge \overrightarrow{0}
\end{array}
\]

The condition (Cond3) is decidable, since $C''$ is a finite graph and by 
Parikh's theorem \cite{Parikh:JACM66} the Parikh-image
of regular languages is effectively semilinear. 
(Just interpret $C''$ as a finite automaton and try out any $\qmarking_0 \in G''$ as initial 
and final state.)
Then we proceed as in Lemma~\ref{lemma:pn}.
Thus (Cond2) is decidable.
\end{enumerate}

\begin{figure}[htbp]
\begin{center}
\scalebox{0.9}{
\input{sdtn_pred3.pstex_t}
}
\caption{(a). Graph $\covergraph'$ derived from $\covergraph$ in 
Figure~\ref{sdtn_pred:zeno}(b). (b) Graph $\covergraph''$ derived from 
$\covergraph'$.}
\label{sdtn_pred2:zeno}
\end{center}
\end{figure}

\begin{exa}
In Figure~\ref{sdtn_pred2:zeno}(a)
we show $\covergraph'$ obtained from 
$\covergraph$ of Figure~\ref{sdtn_pred:zeno}(b) with edges labeled by 
their Presburger-arithmetic representation. 
We have ${\it Effect}((0,0,2),(0,\omega,\omega)) = 
\{(0,1,0) + k_1 (0,1,0) + k_2 (0,-1,1)\ |\ k_1,k_2 \in \nat\}$ 
and 
${\it Effect}((0,\omega,\omega),(0,\omega,\omega)) = 
\{(0,-1,1) + k_1 (0,1,0) + k_2 (0,-1,1)\ |\ k_1,k_2 \in \nat\}$. (Note that
the transfer moves all tokens from the first component to the third component.)
In Figure~\ref{sdtn_pred2:zeno}(b), finally we show the graph $\covergraph''$ 
obtained from $\covergraph'$ in Figure~\ref{sdtn_pred2:zeno}(a).  
\end{exa}

\medskip
\noindent{\bf Correctness of the above constructions:}
Now we show the correctness of the above two constructions (by using
Lemma~\ref{lemma:cover:sdtn}).

\begin{enumerate}[$\bullet$]
\item Firstly, we show that (Cond1) is sufficient and necessary for the existence of an 
infinite $M$-computation $\comp$ with finitely many transfers for some $M \mleq M_0$.

Suppose there is an  infinite $M$-computation $\comp$ with finitely 
many transfers.
Then $\comp$ has an infinite suffix $\comp'$,
starting at some marking $\marking'$ which uses only ordinary Petri net 
transitions. Since $N'$ is obtained by removing transfer transitions,
 $\comp'$ is an infinite $M'$-computation of $N'$.
This implies that Cond1 holds for $N'$ (Lemma~\ref{lemma:pn}). 
Since the coverability 
graph for $N'$ is a subgraph of that for $N$, Cond1 also holds for $N$.
On the other hand, from Lemma~\ref{lemma:pn}, we have
that if Cond1 holds for $N'$, then there is an
 infinite $M'$-computation. Since $M \transclosure M'$, 
 we have an infinite $M$-computation in $N$.

\item Secondly, we show that (Cond2) is sufficient and necessary for the existence of an 
infinite $M$-computation with infinitely many transfers for some $M \mleq M_0$.

If Cond2 is satisfied (i.e., there is a sequence $Seq$ of transitions 
with non-negative effect), then there exist markings 
$M \mleq M_0$ where $M_0\in\covergraph$ and $\marking' \le M_0$ 
such that  $M \transclosure \marking'$ 
(by definition of $\covergraph,\covergraph',\covergraph''$ 
and Lemma~\ref{lemma:cover:sdtn}) such that 
$\marking'$ is large enough to
perform $Seq$ once from $\marking'$. Now, $Seq$ has a non-negative effect, 
therefore one can keep on repeating $Seq$ resulting into an infinite 
$M'$-computation. This implies that there is an infinite $M$-computation.
 
Now we show the other direction. Assume that there is some
$M \in \nat^k$ with $M \le M_0$ and $M \in {\it INF}$ and some infinite
$M$-computation $\comp$ which uses ${\it Trans}$ infinitely often.
Thus it contains infinitely many AT-markings. 
Thus, by Dickson's Lemma (Lemma~\ref{lem:Dickson}, \cite{Dickson:lemma}),
there is a computation (possibly containing several transfers)
where $M \transclosure (\overrightarrow{0},\overrightarrow{x_1}) \longrightarrow_{\it Seq} (\overrightarrow{0},\overrightarrow{x_2})$ with
$\overrightarrow{x_2} \ge \overrightarrow{x_1}$. Thus the total effect 
of the sequence {\it Seq} is
non-negative. 
From Lemma~\ref{lemma:cover:sdtn}, 
it follows that there exists an $\omega$-AT-marking
$\qmarking_0 \in G$ with $\qmarking_0 \ge (\overrightarrow{0},\overrightarrow{x_2})$. 
In fact there exists a largest such
$\qmarking_0$ (as in case of Petri nets, see  Lemma~\ref{lemma:pn}) 
such that  we have $\qmarking_0 \longrightarrow_{\it Seq} \qmarking_0$ in $C$.
So, $\effect(\qmarking_0,\qmarking_0) \ge \overrightarrow{0}$.
The sequence {\it Seq} can be decomposed into 
${\it Seq} = {\it Seq}_1{\it Seq}_2\dots{\it Seq}_n$ with 
$\qmarking_i \longrightarrow_{{\it Seq}_i} \qmarking_{i+1}$
for $1 \le i \le n-1$ and $\qmarking_n = \qmarking_0$. Here
$\{\qmarking_0,\dots,\qmarking_n\}$ is the set of $\omega$-AT markings visited in {\it Seq}. In other words, each {\it Seq}$_i$ contains the transfer transition only once at the end.
It follows that $\qmarking_0 \to \qmarking_1 \dots \to \qmarking_n = \qmarking_0$ is a cyclic
path in $C'$ and $\overrightarrow{v_i} \in \effect(\qmarking_i,\qmarking_{i+1})$ and
$\sum_{i=0}^{n-1} \overrightarrow{v_i} = \effect(\qmarking_0,\qmarking_n) \ge \overrightarrow{0}$. Therefore the condition 
(Cond2) is satisfied.
\end{enumerate}
Altogether we obtain that ${M_0\!\downarrow}\; \cap\; {\it INF} \neq \emptyset$
iff (Cond1) or (Cond2) is satisfied. (It is possible that both (Cond1) and
(Cond2) are true.) Since both conditions are decidable, we
obtain decidability of ${M_0\!\downarrow} \;\cap\; {\it INF} \neq \emptyset$.\qed

\begin{lem}\label{lem:constr_infminprime}
For any SD-TN $N'$ the set ${\it INF}'_{\it min}$ can be effectively constructed.
\end{lem}
\proof
Since ${\it INF}$ is upward-closed, we can, by Lemma~\ref{lem:find_infrun}
and Theorem~\ref{thm:Valk_Jantzen}, construct the minimal elements of 
the set ${\it INF}$, i.e., 
the set ${\it INF}_{\it min}$. We obtain ${\it INF}'_{\it min}$ by the 
restriction of ${\it INF}_{\it min}$ to standard markings.\qed

\subsection{Characterizing ZENO}

\begin{thm}\label{thm:main_zeno}
Let $N$ be a TPN. The set ${\it ZENO}$ is effectively constructible as a MRUC.
\end{thm}
\proof
We first construct the SD-TN $N'$ corresponding to $N$, according to
Section~\ref{def:encoding}. Then we consider the MRUC $Z$ from Def.~\ref{def:Z}.

We have ${\it ZENO} = \denotationof Z$ by Lemma~\ref{lem:Z_part_ZENO} and 
Lemma~\ref{lem:ZENO_part_Z}. The MRUC $Z$ is 
effectively constructible by Lemma~\ref{lem:constr_infminprime},
Definition~\ref{def:Z}, Lemma~\ref{regions:prestar} and
Lemma~\ref{regions:unionintersection}.\qed

\section{The Zenoness-Problem for Discrete-timed Petri Nets}

\noindent
In this section, we discuss how to characterize the set ${\it ZENO}$ 
for discrete-timed Petri nets, thus solving 
the open problem from \cite{Escrig:etal:TPN}. First
we describe how the semantics of a discrete-timed Petri net is 
different from that of a dense-timed Petri net. 

\begin{enumerate}[$\bullet$]
\item Firstly, the ages of the 
token are natural numbers rather than real numbers. 
\item Secondly, the timed transition takes only discrete steps. 
\end{enumerate}

A direct solution for discrete-timed nets is to simply 
modify the construction of the SD-TN $N'$ in Section~\ref{def:encoding}
by removing the time-passing phase in
Subsubsection~\ref{subsubsec:timedtrans}. The resulting net $N'$ is then a
normal Petri net, since it does not contain a transfer arc.
This modified construction would yield ${\it ZENO}$ for the discrete-time
case, because (unlike in the dense-time case) every infinite zeno-computation in a discrete-time net has an 
infinite suffix taking no time at all.

\begin{figure}[htbp]
\begin{center}
\scalebox{0.9}{
\input{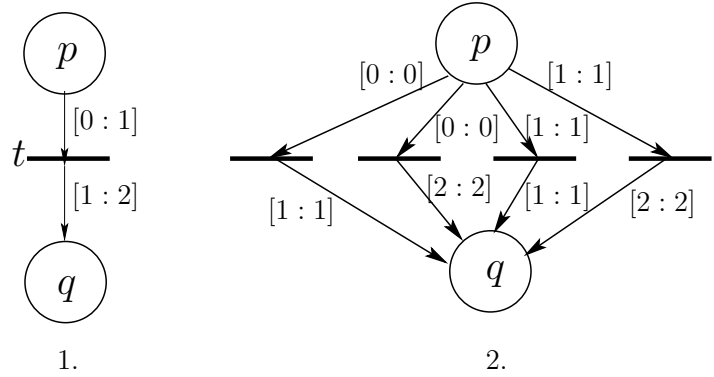}
}
\caption{Simulating (1) $t$ in TPN by (2) a set consisting 
of $4$ transitions in 2.}
\label{disc:encoding:zeno}
\end{center}
\end{figure}

In the special case where all time intervals on transitions are bounded
(i.e., $\infty$ does not appear) there is another solution.
Here one can encode discrete-timed nets into dense-timed nets, as shown in
Figure~\ref{disc:encoding:zeno}.
The trick is to split the intervals on the input (output) arcs to 
several point intervals on a number of transitions.

\section{Arbitrarily Fast Computations}\label{sec:all_zeno}

\noindent
If $M_0 \in {\it ZENO}$ then, by definition, there exists an infinite
$M_0$-computation that requires only finite time, i.e., 
$\exists m, \comp.\, \Delta(\comp) \le m$.
It follows that for any smaller number $m'$ with $0 < m' \le m$ there exists
some marking $M'$ with $M_0 \dto{*} M'$ and
an infinite suffix $\comp'$ of $\comp$ s.t. $\comp'$ is an infinite
$M'$-computation with $\Delta(\comp') \le m'$.
Thus, there exist more and more markings
with faster and faster computations. Formally, 

\begin{equation}\label{eq:zeno}
\forall \epsilon > 0.\,\exists
M_\epsilon \in {\it Post}^*(M_0),\,\mbox{an infinite}\ \comp_\epsilon.\ M_\epsilon \dto{\comp_\epsilon}\ \wedge\ \Delta(\comp_\epsilon) \le \epsilon
\end{equation}

\smallskip
\noindent
However, this does {\em not} imply that there exists some
fixed reachable marking $M$ where arbitrarily fast computations start, because each 
$M_\epsilon$ could be different. 
The existence of arbitrarily fast computations from a fixed reachable marking
is a stronger condition than
zenoness, defined as follows.

\begin{equation}\label{eq:allzeno}
\exists M \in {\it Post}^*(M_0).\,\forall \epsilon > 0.\,\exists\ \mbox{an infinite}
\ \comp_\epsilon.\ M \dto{\comp_\epsilon}\ \wedge\ \Delta(\comp_\epsilon) \le \epsilon
\end{equation}

\smallskip
\noindent
In general, condition~(\ref{eq:zeno}) does {\em not} imply
condition~(\ref{eq:allzeno}), as will be shown by Lemma~\ref{lem:zeno_allzeno}. 

\bigskip
\problemx{All-Zenoness-Problem}
{A timed Petri net $N$, and a marking $M$ of $N$.}
{For all $\epsilon >0$ does there exist an infinite $M$-computation $\comp_\epsilon$ s.t. $\Delta(\comp_\epsilon) \le \epsilon$ ?}

\bigskip
\noindent
A marking $M$ is called an {\it allzeno-marking} of $N$ iff the answer to 
the above problem is 'yes'.

We consider a timed Petri net $N$.
We let ${\it ALLZENO}$ denote the set of the allzeno-markings of $N$.

\begin{figure}[htbp]
\begin{center}
\scalebox{0.9}{
\input{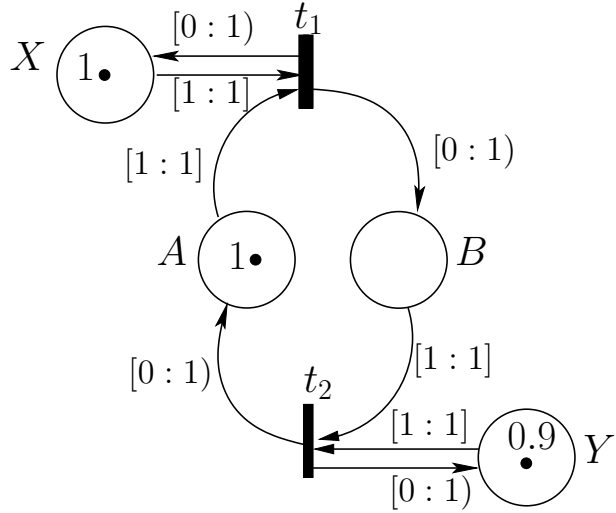}
}
\caption{A TPN with initial marking 
$M_0 := [(X,1), (A,1), (Y,0.9)] \in {\it ZENO}$.
No reachable marking is in ${\it ALLZENO}$, but
allzeno markings exist, e.g., $[(X,1),(Y,1),(A,1),(B,1)]$.
Note the half-open intervals $[0:1)$ which do not include $1$.}\label{fig:nasty}
\end{center}
\end{figure}

\begin{lem}\label{lem:zeno_allzeno}
For all TPN we have ${\it Pre}^*({\it ALLZENO}) \subseteq {\it ZENO}$.
There exist TPN (e.g., the TPN in Figure~\ref{fig:nasty}) where the inclusion
is strict.
\end{lem}
\proof
The inclusion ${\it ALLZENO} \subseteq {\it ZENO}$ follows directly from the
definitions (let, e.g., $\epsilon := 1$). Since 
${\it Pre}^*$ is monotonous, we get 
${\it Pre}^*({\it ALLZENO}) \subseteq {\it Pre}^*({\it ZENO}) = {\it ZENO}$.

Now we consider the example TPN in Figure~\ref{fig:nasty} with initial marking 
\[
M_0 := [(X,1), (A,1), (Y,0.9)]
\] 
There is a zeno run $\comp$ from $M_0$ of
the following form: Transitions $t_1$ and $t_2$ alternate and the length of
the delays between them drops exponentially. 

Formally, 
$\comp = (\rightarrow_{t_1} \rightarrow_{\delta_i} 
\rightarrow_{t_2} \rightarrow_{\delta_{i+1}})_{i=0,2,4,\dots}$
with $\delta_i = (0.1)*2^{-i}$ and thus $\Delta(\pi) \le 0.2$. 
Therefore $M_0 \in {\it ZENO}$. 

Now we show 
that $M_0 \notin {\it Pre}^*({\it ALLZENO})$.

In every reachable marking $M \in {\it Post}^*(M_0)$ there is one
token on place $X$, one token on place $Y$ and either one token on
place $A$ or one token on place $B$. Without restriction we consider 
the case where there is a token on place $A$; the other case is symmetric.
So we have $M = [(X,\chi), (A,\alpha), (Y,\psi)]$. If $\chi >1$, $\alpha >1$ or $\psi>1$
or $\chi \neq \alpha$ then there is no infinite run at all. Otherwise,
if $\chi < 1$ then for $\epsilon := (1-\chi)/2 >0$ there is no run $\comp_\epsilon$
from $M$ with $\Delta(\comp_\epsilon) \le \epsilon$, and 
thus $M \notin {\it ALLZENO}$. There remains the case where $\chi=\alpha=1$.
Then transition $t_1$ must fire immediately, because otherwise the tokens become too
old (i.e., $>1$) and there is no infinite run. Let the resulting marking be 
$M' = [(X,\chi'), (Y,\psi), (B,\beta)]$. By construction of the net, we have 
$\beta < 1$. If $\psi \neq \beta$ then there is no infinite run. So we must
have $\psi = \beta < 1$. Then, for $\epsilon := (1-\psi)/2 >0$ there is no
infinite run $\comp_\epsilon$ from either $M'$ or $M$ with $\Delta(\comp_\epsilon) \le
\epsilon$. Thus $M \notin {\it ALLZENO}$.
So we have shown that no reachable 
$M \in {\it Post}^*(M_0)$ is in ${\it ALLZENO}$, i.e., 
${\it Post}^*(M_0) \cap {\it ALLZENO} = \emptyset$. Therefore, 
$M_0 \notin {\it Pre}^*({\it ALLZENO})$.\qed

Now we show that the All-Zenoness-Problem for TPN is decidable.
In fact, the set ${\it ALLZENO}$ is effectively constructible as a MRUC.

\medskip
\noindent{\bf Intuition:} The construction of ${\it ALLZENO}$ 
is similar to the construction of ${\it ZENO}$ in Section~\ref{sec:zeno}.
The main differences can be understood with the following observations.
\begin{enumerate}[$\bullet$]
\item
In arbitrarily fast runs (unlike in zeno-runs) no tokens of the {\em initial}
marking can reach the next higher integer age by aging. For example,
a token of age $1-\epsilon$ for $\epsilon > 0$ cannot reach age 1 in a run
$\comp$ with $\Delta(\comp) \le \epsilon/2$.
On the other hand, tokens which are newly created during the run can reach the
next higher integer age by aging, since their ages may be 
chosen (nondeterministically) arbitrarily
close to the next higher integer. This is because all the bounds of the time
intervals on transition arcs in the TPN are integers. 
\item
If it were not for the initial marking, we
would have the following situation: If there is a run $\comp$ with
$\Delta(\comp) = \epsilon$ where $0 < \epsilon < 1$ then there also exists a run $\comp'$ with
$\Delta(\comp')=\epsilon/2$. One just replaces any delay of length $\delta$ 
in $\comp$ by a shorter delay $\delta/2$ in $\comp'$ and any token of age $x$ which is newly
created in $\comp$ is replaced in $\comp'$ by a newly created token (on the
same place) of age $x+(\lceil{x}\rceil-x)/2$.
Furthermore, a token with an integer age $i$ will always have a non-integer
age $i+\delta$ after some delay $\delta$ for any $0 < \delta < 1$, i.e.,
regardless of how small $\delta$ is.
\item
How to treat the tokens of the initial marking? Since none of them can age to
the next higher integer in arbitrarily fast computations, they cannot be
encoded as $p(k-)$ tokens in the corresponding SD-TN. Instead they are all
encoded as $p(k)$ tokens (if they have an integer age) or as 
$p(k+)$ tokens (if they have a non-integer age).
\item
Finally, there is the problem that arbitrarily fast computations can be either
disc-com\-pu\-tations or time-computations, depending on whether their first
transition is discrete or timed.
In the construction of the set ${\it ZENO}$ this was elegantly solved, because
this construction included the $\pre^*$ operation which is taken w.r.t. {\em all}
transitions (both discrete and timed).
However, since of construction of ${\it ALLZENO}$ does not include $\pre^*$,
this difference must be addressed explicitly here.
\item
Given this, one can encode arbitrarily fast computations of TPN into
computations of SD-TN, in a similar way as for zeno-computations (with delay
$<1$) in Section~\ref{sec:zeno}. 
\end{enumerate}

\medskip
\noindent{\bf Construction of ${\it ALLZENO}$:} Given a TPN $\tpn$, we
first construct a SD-TN $N'$ in the same way as in
Subsection~\ref{subsec:SD-TN}.  Then we define a mapping ${\it int}$
from markings of $\tpn$ to markings of $N'$, similarly as in
Definition~\ref{def:int}.

\begin{defi}\label{def:allzeno_int}
We define a function ${\it int}: \msets{(\places\times\nnreals)} \to (P' \to \nat)$
that maps a marking $M$ of $N$ to its corresponding marking $M'$ in $N'$.
$M' := {\it int}(M)$ is defined as follows.
\[
\begin{array}{lcll}
M'(p(k))    & := & M((p,k)) & \mbox{for $k \in \nat$, $0 \le k \le \maxval$.}\\
M'(p(k+))   & := & \sum_{k < x < k+1} M((p,x)) & \mbox{for $k \in \nat$, $0 \le k \le \maxval-1$.}\\
M'(p(\maxval+))   & := & \sum_{\maxval < x} M((p,x)) \\
M'(p((k+1)-))   & := & 0  & \mbox{for $k \in \nat$, $0 \le k \le \maxval-1$.}\\
M'(p_{\it disc})  & := & 1\\
M'(p_{\it time1}) & := & 0\\
M'(p_{\it time2}) & := & 0\\
M'(p_{\it count}) & := & 0
\end{array}
\]
Note that $M' = {\it int}(M)$ is a standard marking according to
Def.~\ref{def:INF},
and $M'$ does not contain any $p(k-)$ tokens.

Next we define an operation $\tau$ which encodes the effect of passing 
an arbitrarily small, but non-zero, amount of time.
No tokens can age to the next higher integer age in arbitrarily short time,
but all tokens of an integer age $k$ will have an age $> k$ afterwards.
Given a standard marking $M \in \Omega'$ (recall Def.~\ref{def:INF})
of the SD-TN $N'$, we define $M' := \tau(M)$ as follows.
\[
\begin{array}{lcll}
M'(p(k))    & := & 0  & \mbox{for $k \in \nat$, $0 \le k \le \maxval$.}\\
M'(p(k+))   & := & M(p(k+)) + M(p(k)) & \mbox{for $k \in \nat$, $0 \le k \le \maxval-1$.}\\
M'(p(\maxval+))   & := & M(p(\maxval+)) + M(p(\maxval)) \\
M'(p((k+1)-))   & := & M(p((k+1)-))  & \mbox{for $k \in \nat$, $0 \le k \le \maxval-1$.}\\
M'(p_{\it disc})  & := & M(p_{\it disc})\\
M'(p_{\it time1}) & := & M(p_{\it time1})\\
M'(p_{\it time2}) & := & M(p_{\it time2})\\
M'(p_{\it count}) & := & M(p_{\it count})
\end{array}
\]
Note that the operation $\tau$ is only defined on standard markings and its
result is also a standard marking.
\end{defi}

Unlike in Section~\ref{sec:zeno}, there is a more direct correspondence between
the computations of a marking $M$ and the computations of ${\it int}(M)$ and
$\tau({\it int}(M))$. (Recall the Def.~\ref{def:INF} of ${\it INF}'$.)

\begin{lem}\label{lem:allzeno_int1}
Consider a TPN $N$ with marking $M_0$ and the corresponding SD-TN $N'$. \\
$M_0 \in {\it ALLZENO} \Longrightarrow 
\left(
{\it int}(M_0) \in {\it INF}' \,\vee\,\tau({\it int}(M_0)) \in {\it INF}'
\right)
$.
\end{lem}
\proof 
Let $M_0 \in {\it ALLZENO}$. 
Then there exist arbitrarily fast computations from $M_0$. It follows that
there are either arbitrarily fast disc-computations from $M_0$, or
arbitrarily fast time-computations from $M_0$ (or both). Let 
\[
D := \{(\lceil x\rceil -x)\ |\ \exists p.\, M_0((p,x)) > 0\ \wedge\ (\lceil x\rceil -x)>0\}
\]
\begin{enumerate}[(1)]
\item
First we consider the case that there are arbitrarily fast disc-computations from $M_0$. 
There are two cases.
\begin{enumerate}[(a)]
\item
If $D = \emptyset$ then all tokens in $M_0$ have integer ages. It follows
that ${\it int}(M_0)$ does not contain any $p(k+)$ or $p(k-)$ tokens.
We let $\delta := 1/2$ and obtain ${\it int}_\delta(M_0) 
= {\it int}_{1/2}(M_0) = {\it int}(M_0)$.
By our assumption there are arbitrarily fast disc-computa\-tions from $M_0$
and thus there exists an infinite $M_0$-disc-computation
$\comp$ with $\Delta(\comp)  < 1/2 = 1-\delta$. Therefore,
by Lemma~\ref{lem:int}, ${\it int}(M_0) = {\it int}_\delta(M_0) \in {\it INF}'$.
\item
If $D \neq \emptyset$ then we define $\epsilon > 0$ as the minimal non-zero
distance of the age of any token in $M_0$ from the next higher integer.
\[
\epsilon := \min(D) > 0
\]
Let $\delta := 1-\epsilon/2$. Then ${\it int}_\delta(M_0) = {\it int}(M_0)$.
By our assumption there are arbitrarily fast disc-computations from $M_0$
and thus there exists an infinite $M_0$-disc-computation
$\comp$ with $\Delta(\comp) \le \epsilon/3 < 1-\delta$. Therefore,
by Lemma~\ref{lem:int}, 
${\it int}(M_0) = {\it int}_\delta(M_0) \in {\it INF}'$.
\end{enumerate}
\item
Now we consider the case that there are arbitrarily fast time-computations
from $M_0$. Again there are two cases.
\begin{enumerate}[(a)]
\item
Assume $D = \emptyset$, i.e., all tokens in $M_0$ have integer ages.
Since there are arbitrarily fast time-computations from $M_0$, there
exists a marking $M_1$ such that $M_0 \to_\lambda M_1$ 
with $0 < \lambda < 1/3$ and an infinite disc-computation $\comp$ from $M_1$
with $\Delta(\comp) < 1/3$. It follows that 
$\tau({\it int}(M_0)) = {\it int}(M_1)$.
We let $\delta := 1/2$ and obtain ${\it int}_\delta(M_1) 
= {\it int}_{1/2}(M_1) = {\it int}(M_1) = \tau({\it int}(M_0))$.
Since $\comp$ is an infinite $M_1$-disc-computation with 
$\Delta(\comp) < 1/3 < 1/2 = 1-\delta$, Lemma~\ref{lem:int} yields
${\it int}_\delta(M_1) \in {\it INF}'$. 
Therefore $\tau({\it int}(M_0)) = {\it int}_\delta(M_1) \in {\it INF}'$.
\item
Now assume $D \neq \emptyset$.
As before, we define $\epsilon := \min(D) > 0$
and $\delta := 1-\epsilon/2$. 
Since there are arbitrarily fast time-computations from $M_0$, there
exists a marking $M_1$ such that $M_0 \to_\lambda M_1$ 
with $0 < \lambda < \epsilon/3$ and an infinite disc-computation $\comp$ from $M_1$
with $\Delta(\comp) < \epsilon/3$.
It follows that $\tau({\it int}(M_0)) = {\it int}(M_1)$, 
because $\lambda < \epsilon$.
Furthermore, ${\it int}_\delta(M_1) = {\it int}(M_1)$, 
because $\lambda < \epsilon/3 < \epsilon/2 = 1-\delta$.
Thus $\tau({\it int}(M_0)) = {\it int}_\delta(M_1)$. 
Since $\comp$ is an infinite $M_1$-disc-computation with 
$\Delta(\comp) < \epsilon/3 < \epsilon/2 = 1-\delta$, Lemma~\ref{lem:int} yields
${\it int}_\delta(M_1) \in {\it INF}'$. 
Therefore $\tau({\it int}(M_0)) = {\it int}_\delta(M_1) \in {\it INF}'$.\qed
\end{enumerate}
\end{enumerate}

\begin{lem}\label{lem:allzeno_int2}
Consider a TPN $N$ with marking $M_0$ and the corresponding SD-TN $N'$. \\
${\it int}(M_0) \in {\it INF}' \Longrightarrow M_0 \in {\it ALLZENO}$.
\end{lem}
\proof
Let $M' := {\it int}(M_0) \in {\it INF}'$. Then, by
Lemma~\ref{lem:re_int}, we have  
\[
\exists \word_- \in {\it perm}(M'^-).\,\forall \word_+
\in {\it perm}(M'^+).\, \ucdenotationof{\regs(M',\word_+,\word_-)}
\subseteq \bigcup_{\delta > 0} {\it ZENO}^{1-\delta}
\]
From the definition of the function ${\it int}$ we know that 
$M'^-$ is empty and thus $\word_- = \epsilon$, i.e., the empty sequence.
Thus, $\forall \word_+ \in {\it perm}(M'^+).\, \ucdenotationof{\regs(M',\word_+,\epsilon)}
\subseteq \bigcup_{\delta > 0} {\it ZENO}^{1-\delta}$, and therefore 
$M_0 \in \bigcup_{\delta > 0} {\it ZENO}^{1-\delta}$.
It follows that there exists some fixed $\delta > 0$ such that 
$M_0 \in {\it ZENO}^{1-\delta}$.
Let $\epsilon := 1-\delta < 1$. Then there exists some $M_0$-computation
$\comp_\epsilon$ s.t. $\Delta(\comp_\epsilon) \le \epsilon < 1$.
This $M_0$-computation $\comp_\epsilon$ in $N$ corresponds to an $M'$-computation in $N'$.
Therefore, in $\comp_\epsilon$, no original tokens in $M_0$ reach the next higher integer age by aging, 
because $M' := {\it int}(M_0)$, i.e., because there are no $p(k-)$ tokens in $M'$.

We now show that there exist arbitrarily fast $M_0$-computations
$\comp_{\epsilon/n}$ with $\Delta(\comp_{\epsilon/n}) \le \epsilon/n$ for any
$n \ge 1$.
For any $n\ge 1$ we obtain $\comp_{\epsilon/n}$ by modifying $\comp_\epsilon$
as follows. Every timed transition $\rightarrow_{\delta_i}$ 
in $\comp_\epsilon$ is replaced by a timed transition
$\rightarrow_{\delta_i/n}$ in $\comp_{\epsilon/n}$. In order to ensure 
that in $\comp_{\epsilon/n}$ the same tokens do (or don't) reach/exceed the
next higher integer age during the same timed transition as in $\comp_\epsilon$,
we modify the ages of the newly created tokens. Any token of age $x$ which is newly
created in $\comp_\epsilon$ is replaced in $\comp_{\epsilon/n}$ by a newly
created token (on the same place) of age $x+(n-1)(\lceil{x}\rceil-x)/n$.
This is possible, because all bounds of the time intervals on transition arcs
in the TPN are integers.
Since no {\em original} tokens in $M_0$ age to the next higher integer age
in those runs, this suffices to make $\comp_{\epsilon/n}$ a feasible run from
$M_0$. So we obtain that $\comp_{\epsilon/n}$ is a $M_0$-computation and
$\Delta(\comp_{\epsilon/n}) = \Delta(\comp_\epsilon)/n \le \epsilon/n$.
Therefore, $M_0 \in {\it ALLZENO}$.\qed

\begin{lem}\label{lem:allzeno_int}
Consider a TPN $N$ with marking $M_0$ and the corresponding SD-TN $N'$. \\
$M_0 \in {\it ALLZENO} \iff 
\left(
{\it int}(M_0) \in {\it INF}' \,\vee\,\tau({\it int}(M_0)) \in {\it INF}'
\right)$.
\end{lem}
\proof
The ``$\Rightarrow$'' implication holds by Lemma~\ref{lem:allzeno_int1}. For
the ``$\Leftarrow$'' implication there are two cases.
\begin{enumerate}[(1)]
\item
${\it int}(M_0) \in {\it INF}'$. Then $M_0 \in {\it ALLZENO}$ by
Lemma~\ref{lem:allzeno_int2}.
\item
$\tau({\it int}(M_0)) \in {\it INF}'$. Let 
\[
D := \{(\lceil x\rceil -x)\ |\ \exists p.\, M_0((p,x)) > 0\ \wedge\ (\lceil x\rceil -x)>0\}
\]
If $D \neq \emptyset$ then let $\epsilon := \min(D)/2 > 0$ else let $\epsilon :=
1/2$. Let $\epsilon_i := \epsilon/i$ for $i \ge 1$.
Let $M_i$ be the marking that is reached from $M_0$ after $\epsilon_i$ time
passes, i.e., $M_0 \xtimedtrans{\epsilon_i} M_i$. 
Since $\epsilon_i < \min(D)$ (or $\epsilon_i < 1$ if $D=\emptyset$), 
we have ${\it int}(M_i) = \tau({\it int}(M_0))$ 
and thus ${\it int}(M_i) \in {\it INF}'$ for all $i \ge 1$.
It follows from Lemma~\ref{lem:allzeno_int2} that $M_i \in {\it ALLZENO}$.
Therefore there exist arbitrarily fast time-computations from $M_0$ and thus
$M_0 \in {\it ALLZENO}$.\qed
\end{enumerate}

Similarly as in Section~\ref{sec:zeno}, we compute the set 
${\it ALLZENO}$ as a multi-region upward closure.
We compute a MRUC ${\it AZ}$ and prove that $\denotationof{{\it AZ}} = {\it ALLZENO}$.

\begin{defi}\label{def:AZ}
Let $N$ be a TPN with corresponding SD-TN $N'$, as in
Subsection~\ref{subsec:SD-TN},
and ${\it INF}_{\it min}'$ from Def.~\ref{def:INF}.
Let ${\it INF}_{\it min}''$ be the restriction of ${\it INF}_{\it min}'$
to markings without tokens on $p(k-)$ places. Let
\[
{\it INF}_{\it min}'' := \{M \in {\it INF}_{\it min}'\ |\ \forall p,k.\ M(p(k-))=0\}
\]
and 
\[
\Gamma := \{M' \in \Omega'\ |\ M' \in {\it INF}_{\it min}'' \vee 
\tau(M') \in {\it INF}_{\it min}''\}
\] 
and
\[
{\it AZ} := \bigcup_{M' \in \Gamma} 
\ \bigcup_{\word_+ \in {\it perm}(M'^+)}
\{\regs(M',\word_+,\epsilon)\}
\]  
\end{defi}

Note that it follows from the definition of the function $\tau$ and the
finiteness of ${\it INF}_{\it min}''$ that $\Gamma$ is finite.

\begin{lem}\label{lem:AZ_is_ALLZENO}
$\denotationof{{\it AZ}} = {\it ALLZENO}$.
\end{lem}
\proof
Let $M \in \denotationof{{\it AZ}}$. 
Then there is an $M' \in \Gamma$ and a
$\word_+ \in {\it perm}(M'^+)$ such that 
$M \in \ucdenotationof{\regs(M',\word_+,\epsilon)}$.
Thus there exists some marking $M'' \le M$ s.t. $M'' \in \denotationof{\regs(M',\word_+,\epsilon)}$.
Therefore ${\it int}(M'') = M' \in \Gamma$.
Since ${\it INF}_{\it min}'' \subseteq {\it INF}'$, it follows that
${\it int}(M'') \in {\it INF}' \,\vee\, \tau({\it int}(M'')) \in {\it INF}'$.
By Lemma~\ref{lem:allzeno_int} we have $M'' \in {\it ALLZENO}$ and 
thus $M \in {\it ALLZENO}^{\uparrow} = {\it ALLZENO}$.

To prove the reverse inclusion, let $M \in {\it ALLZENO}$.
Then, by Lemma~\ref{lem:allzeno_int}, ${\it int}(M) \in {\it INF}'$
or $\tau({\it int}(M)) \in {\it INF}'$.
\begin{enumerate}[$\bullet$]
\item
Consider the case where ${\it int}(M) \in {\it INF}'$.
From the definition of the function ${\it int}$ (Def.~\ref{def:allzeno_int})
it follows that ${\it int}(M)$ does not contain any tokens on $p(k-)$ places.
Therefore, there exists some marking $M'' \in {\it INF}_{\it min}''$ s.t.
${\it int}(M) \ge M'' \in \Gamma$.
\item
Consider the case where $\tau({\it int}(M)) \in {\it INF}'$.
From the definition of the functions ${\it int}$ and $\tau$ (Def.~\ref{def:allzeno_int})
it follows that $\tau({\it int}(M))$ does not contain any tokens on $p(k-)$ places.
Therefore, there exists some marking $M' \in {\it INF}_{\it min}''$ s.t.
$\tau({\it int}(M)) \ge M'$. It follows from the definition of the functions
${\it int}$ and $\tau$ and the
fact that $M' \in {\it INF}_{\it min}''$ that there exists some marking 
$M'' \le {\it int}(M)$ s.t. $\tau(M'')=M'$. 
Since $M' \in {\it INF}_{\it min}''$, we have $M'' \in \Gamma$.
Therefore there exists some marking $M'' \in \Gamma$ s.t. ${\it int}(M) \ge M''$.
\end{enumerate}
Thus in both cases there is some marking $M'' \in \Gamma$ 
s.t. ${\it int}(M) \ge M''$.

It follows that there exists some 
$\word_+ \in {\it perm}(M''^+)$ such that
$M \in \ucdenotationof{\regs(M'',\word_+,\epsilon)} \subseteq
\denotationof{{\it AZ}}$.\qed

\begin{thm}\label{thm:main_allzeno}
Let $N$ be a TPN. The set ${\it ALLZENO}$ is effectively constructible as a MRUC.
\end{thm}
\proof
We first construct the SD-TN $N'$ corresponding to $N$, according to
Subsection~\ref{subsec:SD-TN}. Then we consider the MRUC ${\it AZ}$ from Def.~\ref{def:AZ}.
We have ${\it ALLZENO} = \denotationof{{\it AZ}}$ by
Lemma~\ref{lem:AZ_is_ALLZENO}.
The MRUC ${\it AZ}$ is 
effectively constructible by Lemma~\ref{lem:constr_infminprime},
Definition~\ref{def:AZ}, and
Lemma~\ref{regions:unionintersection}.\qed

Finally, we consider the problem whether, for a given marking, there exists an
infinite computation which takes no time at all.

\bigskip
\problemx{Zerotime-Problem}
{A timed Petri net $N$, and a marking $M$ of $N$.}
{Does there exist an infinite $M$-computation $\comp$ such that
  $\Delta(\comp)=0$ ?}

\bigskip
A marking $M$ is called a {\it zerotime-marking} of $N$ iff the answer to 
the above problem is 'yes'.

For a timed Petri net $N$, we 
let ${\it ZEROTIME}$ denote the set of its zerotime-markings.

The construction of the set ${\it ZEROTIME}$ as a MRUC is similar to the construction of
${\it ALLZENO}$. The differences are that in the construction of the SD-TN 
$N'$ the transitions which encode the
time-passing phase (i.e., Subsubsection~\ref{subsubsec:timedtrans}) are left
out. (Thus $N'$ is a normal Petri net.) 
Furthermore, the function $\tau$ is not needed, since all
zerotime-computations are disc-computations.

\begin{lem}\label{lem:zerotime_int}
Consider a TPN $N$ with marking $M_0$ and the corresponding Petri net $N'$
as in Subsection~\ref{subsec:SD-TN} (without Subsubsection~\ref{subsubsec:timedtrans}). Then
$M_0 \in {\it ZEROTIME} \iff {\it int}(M_0) \in {\it INF}'$.
\end{lem}
\proof
If $M_0 \in {\it ZEROTIME}$ then it has an infinite disc-computation $\comp$
with $\Delta(\comp)=0$. Thus ${\it int}(M_0) \in {\it INF}'$ 
by the proof of Lemma~\ref{lem:allzeno_int1}.
If ${\it int}(M_0) \in {\it INF}'$ then $M_0 \in {\it ZEROTIME}$, because
there are no time-passing phases in the Petri net $N'$.\qed

The definition of the needed MRUC ${\it ZT}$ is a simplification of
Definition~\ref{def:AZ}.

\begin{defi}\label{def:ZT}
Let $N$ be a TPN with corresponding Petri net $N'$, as in
Subsection~\ref{subsec:SD-TN} (without Subsubsection~\ref{subsubsec:timedtrans}),
and ${\it INF}_{\it min}'$ from Def.~\ref{def:INF}.
Let ${\it INF}_{\it min}''$ be the restriction of ${\it INF}_{\it min}'$
to markings without tokens on $p(k-)$ places. Let
\[
{\it INF}_{\it min}'' := \{M \in {\it INF}_{\it min}'\ |\ \forall p,k.\ M(p(k-))=0\}
\]
and
\[
{\it ZT} := \bigcup_{M' \in {\it INF}_{\it min}''} 
\ \bigcup_{\word_+ \in {\it perm}(M'^+)}
\{\regs(M',\word_+,\epsilon)\}
\]  
\end{defi}

\begin{lem}\label{lem:ZT_is_ZEROTIME}
$\denotationof{{\it ZT}} = {\it ZEROTIME}$.
\end{lem}
\proof
This follows directly from the definitions and Lemma~\ref{lem:zerotime_int},
similarly as in Lemma~\ref{lem:AZ_is_ALLZENO}.\qed

\begin{thm}\label{thm:main_zerotime}
Let $N$ be a TPN. The set ${\it ZEROTIME}$ is effectively constructible as a MRUC.
\end{thm}
\proof
We first construct the Petri net $N'$ corresponding to $N$, according to
Subsection~\ref{subsec:SD-TN} (without
Subsubsection~\ref{subsubsec:timedtrans}). 
Then we consider the MRUC ${\it ZT}$ from Def.~\ref{def:ZT}.
We have ${\it ZEROTIME} = \denotationof{{\it ZT}}$ by
Lemma~\ref{lem:ZT_is_ZEROTIME}.
The MRUC ${\it ZT}$ is 
effectively constructible by Lemma~\ref{lem:constr_infminprime},
Definition~\ref{def:ZT}, and
Lemma~\ref{regions:unionintersection}.\qed

\section{Universal Zenoness}\label{sec:non_zeno}

The zenoness problem in Section~\ref{sec:zeno} can be seen as existential
zenoness, i.e., the question whether there exists an infinite zeno computation,
and it is decidable by Theorem~\ref{thm:main_zeno}.

Here we consider the {\em universal} zenoness problem, i.e., the question 
whether {\em all} infinite computations from a given marking are zeno (i.e.,
take only finite time).

\bigskip
\problemx{Universal Zenoness Problem}
{A timed Petri net $N$ and a marking $M$.}
{Is it the case that for every infinite $M$-computation $\comp$,
there exists a finite number $m$ s.t. $\Delta(\comp) \le m$ ?}

\bigskip
We will prove the undecidability of the universal zenoness problem by a 
reduction from an undecidable problem for lossy counter machines
\cite{Mayr:LCM:TCS}.
To simplify the presentation, we no not consider the universal zenoness
problem directly, but its negation.

\bigskip
\problemx{Non-Zenoness-Problem}
{A timed Petri net $N$ and a marking $M$.}
{Does there exist an infinite $M$-computation $\comp$,
 such that $\Delta(\comp)=\infty$ ?}

\bigskip
Obviously, a Petri net $N$ with marking $M$ is a positive instance of
the Universal Zenoness Problem if and only if it is a negative instance of 
the Non-Zenoness-Problem.

A marking $M$ is called a {\it nonzeno-marking} of $N$ iff the answer to 
the Non-Zenoness-Problem problem is 'yes'.

We consider a timed Petri net $N$.
We let ${\it NONZENO}$ denote the set of the non-zeno-markings of $N$.
The set ${\it NONZENO}$ is not the complement of the set ${\it ZENO}$. A marking of
a TPN can have infinite zeno runs or infinite nonzeno runs or both or
neither.

\smallskip
In the following, we show that the Non-Zenoness-Problem is undecidable,
which implies the undecidability of the Universal Zenoness Problem.
The proof is done 
by reducing the universal termination problem for {\it lossy counter machines}
to the Non-Zenoness-Problem for TPN.

\subsection{Lossy Counter machines}

Lossy counter machines (LCM) \cite{Mayr:LCM:TCS} are Minsky-counter machines where the
values in the counters can spontaneously decrease (i.e., part of the counter
value is lost). Different versions of LCM are
defined by the way in which this decrease can happen (e.g., just 1 lower, any
lower value, or a reset to zero), which is formally expressed by so-called
lossiness relations \cite{Mayr:LCM:TCS}. Here we consider the classic variant
of LCM where counters can spontaneously change to any lower value.
In this model, any test for zero of a counter could always be successful by
a spontaneous reset to zero. Thus classic LCM are equivalent to 
the following model.

A {\em lossy counter machine} is a tuple $\lcm=\lcmtuple$, where
$\states$ is a finite set of {\em states}, $\state_0\in\states$ is the initial state, $\counters$ is a finite set
of counters and $\ctransitions$ is a finite set of {\em instructions}.
An instruction is a triple of the form
$\tuple{\state,\instr,\state'}$, where
$\state,\state'\in\states$ and $\instr$ is 
either 
an {\it increment} (of the form $\inc\counter$);
a {\it decrement} (of the form $\dec\counter$); or 
a {\it reset} (of the form $\counter:=0$) for a counter $\counter$ in 
$\counters$.

A {\em configuration $\conf$} of $\lcm$ is of the form
$\tuple{\state,\cvalue}$, where $\state\in\states$ and $\cvalue$ is
a mapping from the set $\counters$ of counters to the set $\nat$ of
natural numbers.
We define a transition relation $\cmtrans$ on the set of configurations
such that $\tuple{\state,\cvalue}\cmtrans\tuple{\state',\cvalue'}$
iff one of the following conditions is satisfied:
\begin{enumerate}
\item
$\tuple{\state,\inc{\counter},\state'}\in\ctransitions$,
$\cvalue'(\counter)=\cvalue(\counter)+1$ and
$\cvalue'(\counter')=\cvalue(\counter')$ if
$\counter'\neq\counter$.
\item
$\tuple{\state,\dec{\counter},\state'}\in\ctransitions$,
$\cvalue(\counter)>0$, $\cvalue'(\counter)=\cvalue(\counter)-1$
and $\cvalue'(\counter')=\cvalue(\counter')$ if
$\counter'\neq\counter$.
\item
$\tuple{\state,\counter:=0,\state'}\in\ctransitions$,
$\cvalue'(\counter)=0$ and $\cvalue'(\counter')=\cvalue(\counter')$ if 
 $\counter'\neq\counter$.
\item
$\state'=\state$, $\cvalue'(\counter)=\cvalue(\counter)-1$
for some $\counter\in\counters$, and 
$\cvalue'(\counter')=\cvalue(\counter')$ if $\counter'\neq\counter$.
\end{enumerate}
We use $\cmtransclosure$  for denoting the reflexive, transitive closure of $\cmtrans$.
For a configuration $\conf$, a {\em $\conf$-computation $\comp$} 
of $\lcm$ is a sequence of configurations
$\conf_0,\conf_1,\conf_2,\ldots$, where $\conf_0=\conf$ and 
$\conf_i\cmtrans\conf_{i+1}$, for $i\geq 0$.

The {\em universal termination } problem for LCMs is defined as
follows (see \cite{Mayr:LCM:TCS}).

\bigskip
\problemx{$\exists n.\, {\it LCM}^\omega$}
{A LCM $\lcm$ with $4$ counters and a control-state $q_0$.}
{Does there exist a finite number $n$ such that there is an infinite
  computation of $\lcm$ from the configuration $\conf_0=(q_0,n,0,0,0)$?}
\bigskip

\begin{thm}\cite{Mayr:LCM:TCS}
\label{lcm:theorem}
$\exists n.\, {\it LCM}^\omega$ is undecidable.
\end{thm}
\smallskip

\subsection{Undecidability}
\label{sec:proof:undec}

We show the undecidability of the non-zenoness problem 
for TPNs through a reduction from $\exists n.\, {\it LCM}^\omega$.

Given an instance of $\exists n.\, {\it LCM}^\omega$, i.e.,
an LCM $\lcm$ and  a state $\state_0$ of $\lcm$, 
we construct an equivalent instance of the non-zenoness problem, i.e.,
we derive a TPN  $\tpn$ and a marking $\marking$ of $\tpn$,
such that non-zenoness problem for TPNs has a positive
answer if and only if $\exists n.\, {\it LCM}^\omega$ has a positive answer.

The idea is as follows. First the TPN performs a loop, taking zero time,
which puts a number $n$ of tokens on a certain place. This encodes guessing
the number $n$. Then the TPN faithfully simulates the computation of the LCM
from configuration $(q_0,n,0,0,0)$ in such a way that every single step takes
at least one time unit. This simulation of the LCM is the only possible 
infinite non-zeno run of the TPN since the initial guessing-loop takes zero time.
Thus the TPN has an infinite non-zeno run iff there exists a number $n$ s.t.
the LCM has an infinite run from $(q_0,n,0,0,0)$.

The following encoding of LCM into TPN is similar to the constructions in
\cite{Escrig:etal:TPN,Parosh:Aletta:infinity}, except that we enforce that
every simulation step takes at least one time unit. This delay is crucial for 
our proof.

Consider the LCM $\lcm=\lcmtuple$. We construct a corresponding timed Petri
net (TPN) $\tpn=\tpntuple$ as follows.
For each state $\state\in\states$ there is a place in $\places$
which we call place $\state$.
We use $\places_{\states}$ to denote the set of places of $\tpn$
corresponding to the states $\states$.
Also, for each counter $\counter\in\counters$ there is a place in $\places$
which we call place $\counter$.
We use $\places_{\counters}$ to denote to the set of places corresponding
to counters. There are also six intermediate places for simulating each 
increment and decrement instructions and five such places for 
simulating each reset instruction of the LCM. 

A configuration $\conf$ of $\lcm$ is encoded by a marking $\marking$ 
in $\tpn$ when the following conditions are satisfied.
\begin{enumerate}[$\bullet$]
\item
 The state of $\conf$ is defined in $\tpn$ by the element 
of $\places_\states$ which contains a token.
(The TPN $\tpn$ satisfies the invariant that there is at most
one place in $\places_{\states}$ which contains a token).
\item
The value of a counter $\counter$ in $\conf$ is defined in $\marking$ by
the number of tokens in place $\counter$ which have ages equal to 
$0$. (Tokens which have ages greater than $0$ are considered to have been lost
and do not affect the value of the counter).
\end{enumerate}
Losses in $\lcm$ are simulated either by making the age of the token 
strictly greater than $0$, or by firing a special ${\it loss}_\counter$
 transition
which can always remove tokens from the place $\counter$
 in $\places_{\counters}$.
%
%
Transitions in $\lcm$ are encoded by functions
 $\inputs$ and $\outputs$ in $\tpn$ reflecting  the above 
properties and are defined as follows.

\begin{figure}[htbp]
\begin{center}
\scalebox{0.9}{
\input{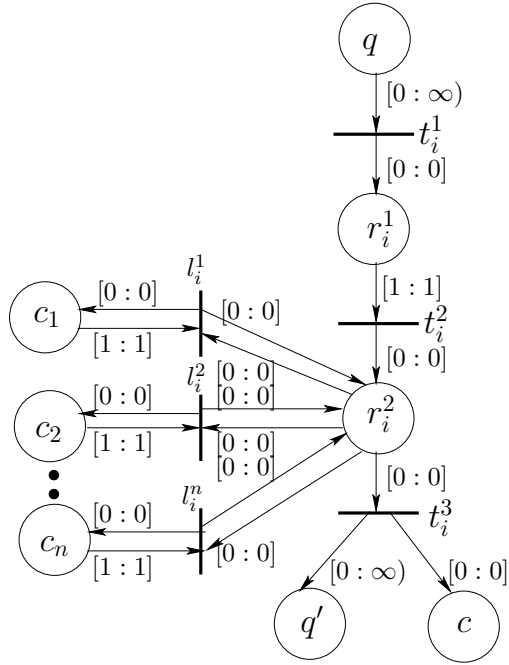}
}
\caption{Simulating the operation of increasing the counter $c$.}
\label{constrinc:fig}
\end{center}
\end{figure}

\begin{figure}
\begin{center}
\scalebox{0.9}{
\input{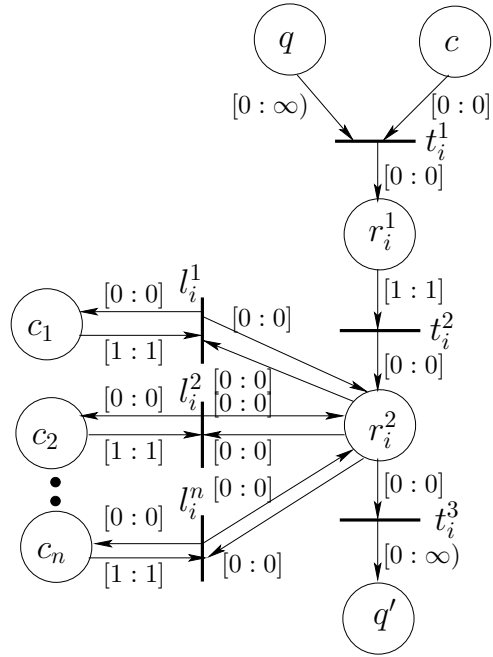}
}
\caption{Simulating the operation of decreasing the counter $c$.}
\label{constrdec:fig}
\end{center}
\end{figure}
\begin{enumerate}[$\bullet$]
\item
An {\it increment} $\imath=\tuple{\state,\inc{c},\state'}$ in $\ctransitions$
is simulated by a set of transitions  in $\transitions$ which are of the form in
Figure~\ref{constrinc:fig}. 
These transitions effectively move a token from place $\state$ to place $\state'$
and adds a token of age $0$ to place $\counter$. However, 
we let at least one time
 unit pass during these transitions. 
To achieve this, 
we use two intermediate places $r_{\imath}^{1}$ and $r_{\imath}^{2}$ for 
each increment instruction $\imath$.
The transition $t_\imath^1$ is fired by 
moving a token from place $\state$ to place $r_{\imath}^{1}$ and resets 
its age to $0$.
The token in $r_{\imath}^{1}$ has to stay there for a 
time equal to  $1$ and then the transition $t_\imath^2$ is fired. 
If more time passes, then this token in $r_{\imath}^{1}$ 
will forever stay in place $r_{\imath}^{1}$ after which no tokens
will ever reside in any place in $\places_{\states}$ and 
thus the net will deadlock.
The idea is that the TPN should not have any zeno-run during the simulation of 
any operation of the LCM. 
So, during the simulation of the increment-operation, we need to wait at least for 
one time unit. 
This makes the ages of all tokens in places $\places_\counters$ 
at least equal to $1$.
Thus, in order to avoid resetting the values of the 
counters, we add, for each
counter in $\counters$ a new transition.
In Figure~\ref{constrinc:fig}, we assume that
 $\places_{\counters}=\set{\counter_1,\ldots,\counter_n}$ 
and thus 
we add the transitions $\ltransition_{\imath}^{1},\ltransition_{\imath}^{2},\ldots,\ltransition_{\imath}^{n}$.
These transitions are used to {\it refresh} the ages of the tokens in
the places in $\places_{\counters}$.
Now, if a token in place $\counter_1$ has its age equal to $1$,
 and thus has become too old for firing other transitions ({\it decrements}), it is replaced by a fresh token of age $0$.  
Notice that the refreshment phase either does not take any time at all or 
it deadlocks.
Finally, when the transition $\transition_\imath^3$ is fired, the new control state
will be $\state'$ and there will be a new token of age $0$ in $\counter$.
The resulting marking will therefore correspond to the counter
$\counter$ having an increment by the value $1$.
The refreshing process for the  counters $\counter_1,\ldots,\counter_n$ 
will be stopped
after firing $\transition_\imath^3$, since the token in $r_{\imath}^{2}$ will now be removed.
Notice that some tokens in 
$\counter_1,\counter_2,\ldots,\counter_n$ may be lost 
(i.e., may still have age greater or equal to $1$), since the TPN 
has a lazy semantics and these tokens may not have been refreshed.
Possibly losing tokens is allowed in the simulation of LCM by TPN, since
the semantics of LCM allows spontaneous decreases in counters.
\item
A {\it decrement} $\imath=\tuple{\state,\dec{c},\state'}$ in $\ctransitions$
is simulated by a similar set of transitions  in $\transitions$ which are of the form in
Figure~\ref{constrdec:fig}.
These transitions also move a token from place
$\state$ to place  $\state'$ and remove
a token of age $0$  from place $\counter$. Again,
we let at least one time
 unit pass during these transitions. The description is similar to the case 
for the increment-operation.
\item
For each place $\counter$ in $\places_{\counters}=\set{\counter_1,\ldots,\counter_n}$, there is a
transition which we call $\mathit{loss}_\counter$
(Figure~\ref{loss:fig}).
A transition ${\mathit loss}_{\counter}$ 
removes a token of age $0$ from the counter $\counter\in\places_\counters$ 
and thus simulates the lossiness of  counter $\counter$.
\begin{figure}[htbp]
\begin{center}
\scalebox{0.9}{
\input{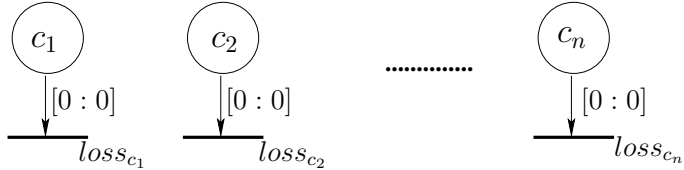}
}
\caption{Simulating losses.}
\label{loss:fig}
\end{center}
\end{figure}
\item
The construction for the {\it reset} instruction 
$\imath=\tuple{\state,\counter:=0,\state'}$
in $\ctransitions$ is shown in Figure~\ref{test:fig}.

\begin{figure}[htbp]
\begin{center}
\scalebox{0.9}{
\input{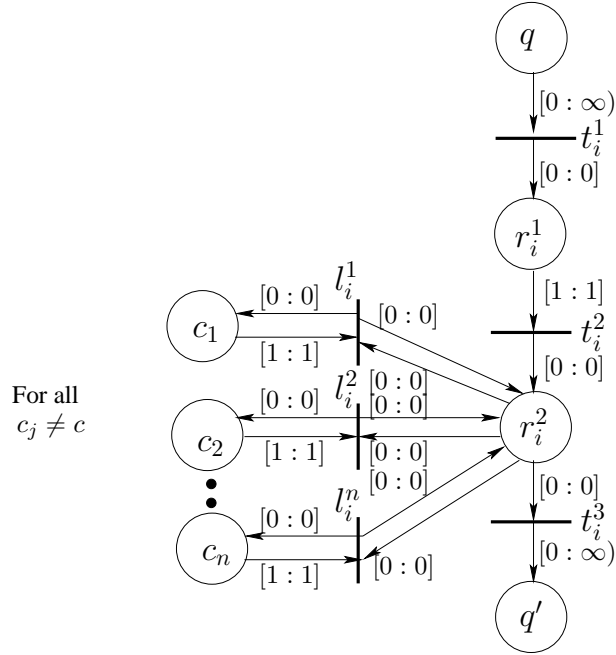}
}
\caption{Simulating the operation of resetting the value of the counter
$\counter$ to $0$. All other counters $c_j$ with $\counter_j \neq c$ can be
refreshed.}
\label{test:fig}
\end{center}
\end{figure}

The idea is that we reset the value of counter $\counter$ to $0$,
by making the ages of all tokens in place $\counter$ at least equal to $1$.
Observe that we simulate resetting the counter in $\lcm$ by 
resetting the counter in $\tpn$.
%
%
All tokens in each of the 
 places in $\places_{\counters}$ which had 
age $0$ have now age equal to $1$.
Thus, in order to avoid resetting the values of the 
counters other than $\counter$, we add, for each
counter in $\counters-\set{\counter}$ a new transition.
In Figure~\ref{test:fig}, we assume that
 $\places_{\counters} - \set{\counter}=\set{\counter_1,\ldots,\counter_n}$ 
and thus 
we add the transitions $\ltransition_{\imath}^{1},\ltransition_{\imath}^{2},\ldots,\ltransition_{\imath}^{n}$.
These transitions are used to {\it refresh} the ages of the tokens in
the places in $\places_{\counters}-\set{\counter}$, i.e., all counters can be
refreshed expect $\counter$.
Now, if a token in place $\counter_i$ has its age equal to $1$,
 and thus has become too old for firing other transitions ({\it decrements}), it is replaced by a fresh token of age $0$.
Finally, when the transition $\transition_\imath^3$ is fired, the new control state
will be $\state'$, and each token in place $\counter$ will have
an age which is at least one.
The resulting marking will therefore correspond to the counter
$\counter$ having the value $0$.
\item
Initialization. To guess the initial value in counter $\counter_1$ of the LCM, 
we add an extra place $\state_{\it init}$ in $\places$ and add two 
transitions in $\transitions$, shown in Figure~\ref{init:fig}. First the transition 
$t_{\imath_1}$
 is enabled if there is a token in $\state_{\it init}$ with age $0$. 
By executing this transition $n$ times (for some $n \ge 0$) without letting any time pass,
 we can produce $n$ tokens in the counter $\counter_1$. 
This simulates an initial value $n$ of $\counter_1$ in LCM. Then, we switch control 
for simulating the usual operations of the LCM by executing the  transition $t_{\imath_2}$
in Figure~\ref{init:fig}, which moves the token from $\state_{\it init}$ 
to $\state_0$.
\begin{figure}[htbp]
\begin{center}
\scalebox{0.9}{
\input{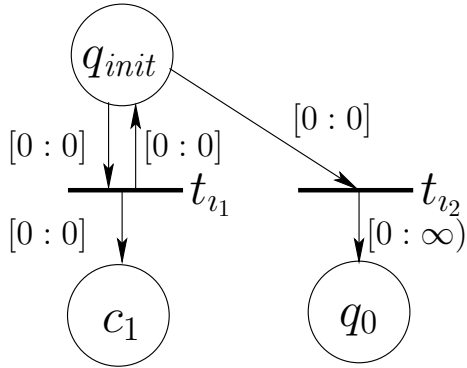}
}
\caption{Initialization.}
\label{init:fig}
\end{center}
\end{figure}
\end{enumerate}

Consider a marking $\marking$ of $\tpn$ and a configuration
$\conf=\tuple{\state,\val}$ of $\lcm$.
We say that $\marking$ is an {\em encoding} of $\conf$ if $\marking$
contains a token in place $\state$ and the number of tokens with
ages equal to $0$
 in place $\counter$ is equal to $\val(\counter)$ for 
each $\counter\in \counters$.
Furthermore, all other places in $\marking$ are empty.

We also use the following notion of intermediate markings. 
A marking 
is called  {\it intermediate }  
if it has
 a token in place $r_\imath^1$ ($r_\imath^2$) where $\imath$ is of the form $(\state,c:=0,\state')$
 and there are no tokens in other intermediate places and in those belonging to $\places_\states$.

We derive $\tpn$ from $\lcm$ as described above.
We define $\marking_0$ to be $\lst{(\state_{\it init},0)}$.

\begin{lem}\label{lem:LCM_TPN}
$\tpn$ has an infinite non-zeno $\marking_0 = \lst{(\state_{\it init},0)}$-computation
if and only if there exists an $n \ge 0$ s.t. the LCM $\lcm$ has an
infinite $\conf_0=(q_0,n,0,0,0)$-computation.  
\end{lem}
\proof\ 

\noindent$\Leftarrow:$\ Let $\conf_0 := (q_0,n,0,0,0)$ and $M_0 := \lst{(\state_{\it init},0)}$.
Given an infinite $\conf_0$-computation $\comp$ 
of $\lcm$,
we show that there is an infinite non-zeno $\marking_0$-computation $\comp'$.

To show this, it is enough to prove the following. 
\begin{enumerate}[(a)]
\item Starting from a marking
$\marking_0$ in TPN, there is a 
zero-time  computation from $\marking_0$ to a marking $M$ which is an encoding
of $\conf_0$. 
In fact, $\marking_0 \stackrel{n}{\longrightarrow}_{t_{\imath_1}} 
\longrightarrow_{t_{\imath_2}} \marking$ (see Figure~\ref{init:fig}).

\item After the initialization step, given two configurations $\conf,\conf'$ of $\lcm$ such that 
$\conf\cmtrans\conf'$ and a marking $\marking$ which is an encoding of $\conf$, there is a sequence in $\tpn$ of the form 
$\marking=\marking_0\wtrans{}\marking_1\wtrans{}\cdots\wtrans{}\marking_k=\marking'$ where $k\geq 1$ and the following holds.
\begin{enumerate}[$\bullet$]
\item  $\marking'$ is an encoding of $\conf'$.
\item $\marking_i$ is an intermediate marking for 
$0 < i < k$.
\end{enumerate}
\end{enumerate}
Since $\conf \cmtrans\conf'$, we know that $\conf'$ is derived from $\conf$,
using one of the four possible types of transitions described for LCMs.
We show the claim only for the first case, namely when
$\conf'$ is derived from $\conf$ by executing an increment
 instruction $\imath$.
The other cases can be explained in a similar manner.
Let $\conf=\tuple{\state,\cvalue}$ and $\conf'=\tuple{\state',\cvalue'}$.
Since $\marking$ is an encoding of $\conf$,
it means that place $\state$ in $\marking$ contains a token.
From the construction described above (Figure~\ref{constrinc:fig})
we know that 
from $\marking$, we can fire $\transition_\imath^1$ and produce 
a marking $\qmarking_1$ such that $\marking\longrightarrow_{\transition_\imath^1}\qmarking_1$. $\qmarking_1$ is obtained from $\marking$ by removing the token from $\state$ and adding a token of age $0$ in $r_\imath^1$. 
This means that
 both $\marking$ and $\qmarking_1$ contains 
exactly equal number of tokens 
of age $0$ at each place in $\places_\counters$.

Next we let time pass by one time unit and obtain a marking $\marking_1$ 
such that $\qmarking_1\longrightarrow_1\marking_1$. This means that
$\marking\wtrans{}\marking_1$. Notice that all the tokens with age $0$ in 
the places of $\places_\counter$ in $\qmarking_1$ have transformed into
tokens of age $1$ in $\marking_1$.
Now,  firing the transition $\transition_\imath^2$  from $\marking_1$
 results in a marking $\qmarking_2$ 
such that $\marking_1\longrightarrow_{\transition_\imath^2}\qmarking_2$.
The transition $\transition_\imath^2$ removes the token of age $1$ from 
$r_\imath^1$ and adds a token of age $0$ in $r_\imath^2$.
Here, for each place in $\places_\counters$,
there are no tokens with age less than $1$. Furthermore, 
the number of tokens of age $1$ in each place $\counter'\in\places_\counters$
is the same in both $\marking_1$ and $\qmarking_2$.
We define $\marking_2=\qmarking_2$. So, $\marking_1\wtrans{}\marking_2$.

To restore the ages of the tokens of age $0$ at each 
place in $\places_\counters$ in the marking $M_0$ (these 
tokens correspond to the values of the counters in $\conf$), 
 we start a {\em refreshment} phase.
Suppose for a counter $\counter_1\in\places_\counters$, 
$\cvalue(\counter_1)=x$. Then we fire the transition $\ltransition_\imath^1$ 
$x$ times from $M_2$ and {\it refresh}
all $x$ tokens of age $1$ in $\counter_1$ to age $0$. 
Similarly we refresh all tokens of age $1$ in the 
other counters in $\places_\counters$. 
Notice that we do not let time pass between 
these discrete transitions.

The markings $M_1,M_2,\ldots$, etc. in the above are all intermediate markings.
Now we fire 
 the transition $\transition_\imath^3$ by moving the token from $r_\imath^2$ 
to $\state'$ and adding a token of age $0$ to place $\counter$, 
yielding a marking $\marking'$.
This means that for each counter $\counter'\in\places_\counters\setminus\set{\counter}$, the number of tokens of age $0$ in $\counter'$ for $\marking'$
is the same as that for $\marking$.
Furthermore, in comparison to marking $\marking$, 
there is exactly one extra token of age $0$ at place $\counter$ in $\marking'$.
This means that the new marking $\marking'$ 
will be an encoding of $\conf'$ and $\marking \stackrel{*}{\longrightarrow} \marking'$.

The simulation of other operations can be explained in a similar manner.

Now, if there exists some number n s.t. the LCM has an infinite computation from $(q_0, n, 0, 0, 0)$
then the TPN has an infinite non-zeno computation from an initial marking that
corresponds to 
$(q_0, n, 0, 0, 0)$. This is ensured by the initialization step and 
the above simulation of operations in LCM. The non-zenoness of the computation 
in TPN is ensured by passage of time during each operation of LCM.
Notice that the initialization step takes zero-time.

\noindent$\Rightarrow:$
Suppose that there is an infinite $\marking_0$-computation $\comp$ of $\tpn$ 
taking infinite time. It follows that $\comp$ must contain the transition
${t_{\imath_2}}$, since the initial ${t_{\imath_1}}$-loop takes zero time.
Consider the maximal subsequence
$\comp'$ of $\comp$,
where each marking in $\comp$ is an encoding of some configuration
of $\lcm$.
The sequence $\comp'$ exists for the following reasons.
\begin{enumerate}[$\bullet$]
\item Since $\comp$ is non-zeno and infinite, 
the computation $\comp$ is infinite even after the zeno initialization step.
\item
Furthermore, each operation (increment, decrement, etc) takes 
a finite non-zero amount of time (this follows from the constructions (see the Figures) 
for increment, decrement and resetting). 
\end{enumerate}

From the initialization step, it is straightforward that in 
zero time we reach a marking which is an encoding of $\conf_0 = (q_0,n,0,0,0)$ for 
some $n \ge 0$, i.e., the encoding of $\conf_0$ is the configuration reached immediately after
firing transition ${t_{\imath_2}}$ at the end of the initial guessing-phase.
In the following, we show
that there is an infinite $\conf_0$-computation.

To prove this, it is enough to show that
given two consecutive encodings $\marking$ and $\marking'$ 
(with only intermediate markings in between) in $\comp'$ 
and a configuration $\conf$
 which is an encoding of $\marking$, there is a configuration $\conf'$
 such that $\conf \cmtransclosure \conf'$.
 Let $\conf=\tuple{\state,\cvalue}$.

Since $\marking\stackrel{*}{\longrightarrow}\marking'$ we know that there are markings
$\marking_0,\ldots,\marking_k$  such that
$\marking=\marking_0\wtrans{}\marking_1\wtrans{}\cdots\wtrans{}\marking_k=\marking'$ 
where $k\geq 1$ and $\marking_1,\ldots,\marking_{k-1}$ are
intermediate markings.

There are two cases.
Either $k=1$ or $k>1$.

If $k=1$, i.e., $\marking\wtrans{}\marking'$, we know that
$\marking'$ can be derived from $\marking$ by firing  a 
discrete transition.
This means that 
there is a marking $\qmarking$ such that $\marking\ttrans\marking'$ where
the discrete transition $t$ corresponds to Figure~\ref{loss:fig}.

If $k>1$, then $\marking'$ is obtained from $\marking$ by
firing transitions corresponding to those in Figure~\ref{constrinc:fig},~\ref{constrdec:fig}, and ~\ref{test:fig}.
For instance, consider that $\imath=\tuple{\state,\inc{\counter},\state'}$
is an instruction in $\lcm$, for some counter $\counter$.
From the construction of Figure~\ref{constrinc:fig}, we know that 
 the ages of some of the tokens in
$\places_{\counters}$ may exceed $1$, since not
all tokens need to be refreshed.
We can derive $\conf'$ from $\conf$ by first performing
loss transitions corresponding to tokens which become too old followed
by  executing the 
instruction $\tuple{\state,\inc{\counter},\state'}$. 
Similarly, we can perform loss transitions followed by a decrement or a
reset instruction of the LCM.\qed

\begin{thm}\label{thm:non_zeno}
The Non-Zenoness-Problem for TPN is undecidable.
\end{thm}
\proof
Directly from Lemma~\ref{lem:LCM_TPN} and Theorem~\ref{lcm:theorem}.\qed

Since Non-Zenoness-Problem is the negation of the Universal Zenoness Problem, 
this implies the following result.

\begin{thm}
The Universal Zenoness Problem for TPN is undecidable.
\end{thm}

\hide{
The point is that one must make sure that the TPN which simulates the LCM does not have *any* zeno-runs at all.
One can achieve this by enforcing a 1-time-unit wait after every single computation step.
During that extra wait, all tokens can be refreshed (reset to age zero). Those tokens which are not refreshed
become too old and count as lost (after all we simulate an LCM).

The TPN that we construct in this way has the following properties:
If there exists some number n s.t. the LCM has an infinite computation from $(q_0, n, 0, 0, 0)$
then the TPN has an infinite non-zeno computation from an initial marking that
corresponds to 
$(q_0, n, 0, 0, 0)$.
Otherwise the TPN does not have *any* infinite computation at all from *any* initial marking.

Now, in a second step, we extend the TPN by adding an extra initial control-state and an initial
zeno-time loop that effectively guesses the initial marking (i.e., we guess the number n).
When this guessing loop ends (if it does) we change to the normal control-state of the TPN above.

Now our so constructed TPN has the following properties.
If there exists some number n s.t. the LCM has an infinite computation from 
$(q_0, n, 0, 0, 0)$ then
the TPN has an infinite non-zeno computation. Just do the initial zero-time loop until you reach n and then
switch to the infinite *non-zeno* simulation of the TPN.
Otherwise (if no n is big enough), the TPN does not have *any* infinite non-zeno computation.  Why?
There are two cases:
1. The initial zero-time guessing loop does not stop. Then the computation is zeno.
2. The initial zero-time guessing loop does stop at some number m and switches to normal non-zeno LCM simulation mode.
Then this non-zeno computation will be finite, since the m is not big enough.
}

\section{Token Liveness}
\label{token_liveness:section}

First, we define the {\it liveness} of a token in a marking.

Let $M$ be a marking in a TPN $N$. A token in $M$ is called
{\em syntactically $k$-dead} if its age is $\ge k$. 
It is trivial to decide whether a token is $k$-dead from a marking.

A token is called {\em semantically live} from a marking
$\marking$, if we can fire a sequence
of transitions starting from $\marking$ which eventually consumes the token.
Formally, given
a token $(p,x)$ and a marking $\marking$, we say that $(p,x)$ can be 
{\it consumed} in $\marking$
if there
is a transition $\transition$ 
satisfying the following properties:
\begin{enumerate}[$\bullet$]
\item
$\transition$ is enabled in $\marking$.
\item
$\inputs(t,p)$ is defined and $x\in\inputs(\transition,\place)$.
\end{enumerate}

\begin{defi}
\label{live:definition}
A token $(p,x)$ in a marking $\marking$ is {\it semantically live} 
if there is a finite $\marking$-computation $\comp=\marking\marking_1\cdots\marking_r$ 
 such that the aged token
$(\place,x+\Delta(\comp))$ can be consumed in $\marking_r$.
By $L(\marking)$ we denote set of of all live tokens in $M$.
\end{defi}

Note that token liveness is defined here for individual tokens, not sets of
tokens. There are nets and markings where two tokens $(p,x)$
and $(q,y)$ are both live, but where it is impossible to consume both of them.

\bigskip
\problemx{Semantic liveness of tokens in TPN}
{A timed Petri net $\tpn$ with marking $\marking$ and a token $(\place,x)\in\marking$.}
{Is $(\place,x)$ live, i.e., $(\place,x) \in L(\marking)$ ?}

\bigskip
We show decidability of the semantic token
liveness problem by reducing it to the {\it coverability problem}
for TPNs (which is decidable due to Lemma~\ref{regions:prestar}).

\bigskip
\problemx{Coverability problem}
{
A TPN $\tpn$,  a finite set of initial markings $\markings_{\it init}$ of $\tpn$,
and an upward closed set $\markings_{\it fin}\uparrow$ of
markings of $\tpn$, where $\markings_{\it fin}$ is finite.}
{
$\markings_{\it init}\trans{*}\markings_{\it fin}\uparrow$?
}
\smallskip

%
\begin{thm}
\label{coverability:theorem}
The coverability problem is decidable for TPN \cite{Parosh:Aletta:bqoTPN}.
\end{thm}

Suppose that we are given a TPN $\tpn=\tpntuple$ with marking $\marking$ and a token 
$(\place,x)\in\marking$.
We shall translate the question of whether $(\place,x) \in L(\marking)$
into (several instances of) the coverability problem.
To do that, we construct a new TPN $\tpn'$ by adding a new place $\place^*$ to
the set $\places$.
The new place is not input or output of any transition.
Either there is no transition in $\tpn$ which has $\place$ as its 
input place. Then it is trivial that $(p,x)\not\in L(\marking)$.
Otherwise,
we consider all instances of the coverability problem defined on
$\tpn'$ such that 
\begin{enumerate}[$\bullet$]
\item
$\markings_{\it init}$ contains a single marking 
$\marking- (\place,x)+ (\place^*,x)$. 
\item $\markings_{\it fin}$ is the set of markings of the form 
$\lst{(\place_1,x_1),\ldots,(\place_n,x_n),(p^*,x')}$  
such that 
there is a transition $t$ and
\begin{enumerate}[-]
\item the set of input places of $t$ is given by $\set{p,p_1,\ldots,p_n}$.
\item   $x'\in\inputs(t,p)$ and 
$x_i\in\inputs(t,p_i)$ for each $i:1\leq i \leq n$.
\end{enumerate}

\end{enumerate}
In the construction above, we replace a token 
$(\place,x)$ in the initial marking 
by a token $(\place^*,x)$; we also replace 
 a token $(\place,x')$ 
in the final marking where  $x'\in\inputs(t,p)$ 
by a token $(\place^*,x')$. 
The fact that  the token in the question 
is not consumed in any predecessor of 
a marking in $\markings_{fin}$, is simulated by moving the token into 
the place $\place^*$ (in both the initial and final markings), since 
$\place^*\not\in\places$ and not an input or output place in $\tpn'$. 
Therefore, the token is live in $\marking$ of $\tpn$ 
iff the answer to the coverability problem is 'yes'.

From Theorem~\ref{coverability:theorem}, we get the following.
\begin{thm}
\label{token:liveness:theorem}
The token liveness problem is decidable.
\end{thm}

\section{Boundedness}
\label{boundedness:section}

Given a system and an initial configuration, 
the boundedness problem is the question whether the size of any reachable
configuration is bounded by a constant.
In the context of a TPN, this is the question whether the number of tokens
in any reachable marking is bounded.

Every marking $\marking$ is a multiset of timed tokens.
The size of a marking $\marking$ is defined as the size of this multiset, 
denoted as $\sizeof\marking$ (see Def.~\ref{def:multiset}). In other words,
$\sizeof\marking$ denotes the number of timed tokens in $\marking$.
Given a set of markings $\markings$, we define
$\maxsize\markings := \max\{\sizeof\marking\,|\, \marking\in\markings\}$
as the maximal size of any marking in $\markings$.

In Section~\ref{defs:section} we defined
$\reachable(\marking_0) := \{\marking'\,|\, \marking_0 \trans* \marking'\}$ as the set of markings 
reachable from $\marking_0$.

The boundedness problem for a TPN with an initial marking $\marking_0$ is then
the question whether $\maxsize{\reachable(\marking_0)}$ is bounded.

\begin{rem}
Note that, unlike for normal untimed Petri nets, the boundedness problem
for TPNs is {\em not} equivalent to the question whether 
$\sizeof{\reachable(\marking_0)}$ is bounded.
By the lazy semantics of our TPNs (see Section~\ref{defs:section})
time can always pass and increase the values of the clocks of the tokens.
Thus (unless the initial marking is empty) one obtains 
infinitely many (even uncountably many) different reachable
markings, even if the number of tokens stays constant.
For example, consider a TPN with just one place $p$ and no discrete
transitions and initial marking $\marking_0 := \{(p,0)\}$.
Then $\reachable(\marking_0) = \{\{(p,x)\}\ |\ x \in \nnreals\}$ is infinite,
but $\maxsize{\reachable(\marking_0)}=1$.
\end{rem}  

In this section we consider two different variants of the 
boundedness problem for TPNs. In {\em syntactic boundedness}
all tokens in a marking count towards its size,
while in {\em semantic boundedness} only semantically live tokens
(see Section~\ref{token_liveness:section}) count.

\bigskip
\problemx{Syntactic Boundedness of TPN}
{A timed Petri net $N$ with initial marking $\marking_0$.}
{Is $\maxsize{\reachable(\marking_0)}$ bounded ?}

\bigskip
We give an algorithm similar to the
Karp-Miller algorithm \cite{KaMi:schemata} 
for solving the syntactic boundedness
problem for TPNs.
The algorithm builds a tree, where each node
of the tree is labeled with a region.
We build the tree successively, starting from the root,
which is labeled with  $\region_{\marking_0}$:
the unique region satisfied by
$\marking_0$ (it is easy to compute this region).
At each step we pick a leaf with label $\region$ and perform one of the following
operations:
\begin{enumerate}[(1)]
\item
If $\post(\region)$ is empty we declare the current
node {\it unsuccessful} and 
close the node.
\item
If there is a previous node on the branch which is labeled with
$\region$ then declare the current node {\it duplicate} and 
close the node.
\item
\label{unboundedness:condition}
If there is a predecessor of the current node labeled with
$\region'<^r\region$ then we declare \\
$\maxsize{\reachable(\marking_0)}$ infinite
(the TPN is unbounded), and terminate the procedure.
\item
Otherwise, declare the current node as an {\it interior} node,
add a set of successors to it, each labeled with an element
in  $\post(\region)$.
This step is possible due to Lemma~\ref{post:lemma}.
\end{enumerate}
If  the condition of step \ref{unboundedness:condition}
is never satisfied during the construction of the tree, then we declare
$\maxsize{\reachable(\marking_0)}$ finite
(the TPN is bounded).

The proof of correctness of the above algorithm is 
similar to that of original Karp-Miller construction \cite{KaMi:schemata}.
The termination of the algorithm is 
guaranteed due to the fact that the ordering $\rleq$ on
the set of regions is a well-quasi-ordering (follows from \cite{Higman:divisibility}).

\begin{thm}
\label{theorem:syntactic_boundedness}
Syntactic boundedness of TPN is decidable.
\end{thm}

A consequence of this result is that we can solve the 
non-termination problem for TPNs, i.e., the problem whether
a given marking $\marking$ has at least one infinite run.
(Remember that, by our definition of TPN computations 
(see Section~\ref{defs:section}), every infinite run must contain 
infinitely many discrete transitions.)

\bigskip
\problemx{Non-Termination of TPN}
{A timed Petri net $N$, and a marking $\marking$ of $N$.}
{Does there exist an infinite $\marking$-computation?}

\bigskip
A marking $\marking$ is called a {\it non-terminating marking} of $N$ iff the answer to 
the above problem is 'yes'.
For a given timed Petri net $N$
we let ${\it NONTERM}$ denote the set of the non-terminating markings of $N$.

\begin{thm}\label{thm:nonterm}
Non-Termination of TPN is decidable.
\end{thm}
\proof
By Theorem~\ref{theorem:syntactic_boundedness} we can decide 
syntactic boundedness. If the system is syntactically 
unbounded then it is certainly non-terminating.
If the system is syntactically bounded, then 
all the markings in $\reachable(\marking_0)$ can be
symbolically represented by the finitely many regions computed by the algorithm
above.
In this case 
we have non-termination iff there exists a cyclic (and thus repeatable) 
path among
these regions which contains at least one discrete transition.
(Cyclic paths containing only timed transitions do not induce 
valid infinite runs, since we require that every infinite run contains 
infinitely many discrete transitions.)

This condition can easily be checked in the algorithm above as follows.
If condition (3) is true on some branch then the system is non-terminating.
If some branch stops with condition (2), then check if at least one step
on the path from the previous node $\region$ to the duplicate node 
$\region$ was a discrete
step. If yes, then there exists a repeatable path from $\region$ to 
$\region$ which contains at least one discrete transition and thus
the system is non-terminating.\qed

Since semantically dead tokens cannot influence the behavior of 
a TPN (see Section~\ref{token_liveness:section}), 
one would like to abstract from them. 

Let $N$ be a TPN with marking $\marking$.  Then 
we define the live part of the TPN marking $\marking$ as 
$\reachable^l(\marking) := 
\{L(\marking')\;|\;\marking \trans{*}\marking'\}$, i.e, $\reachable^l(\marking)$ is the set of 
reachable markings where the semantically dead tokens have been removed.

\bigskip
\problemx{Semantic Boundedness of TPN}
{A timed Petri net $N$ with initial marking $\marking_0$.}
{Is $\maxsize{\reachable^l(\marking_0)}$ bounded ?}

\bigskip

\begin{thm}
\label{theorem:semantic_boundedness}
Semantic boundedness of TPN is undecidable.
\end{thm}
\proof
Using slightly modified constructions of
\cite{Escrig:etal:TPN:nondecidability} 
or \cite{Parosh:Aletta:infinity}, we can easily derive
the undecidability of semantic boundedness even for dense-timed Petri nets
(see \cite{Pritha:thesis}).
The idea is to use the same encoding of lossy counter machines (LCM) into TPN as in
Section~\ref{sec:all_zeno} (or a similar encoding, as shown in
\cite{Pritha:thesis}). 
In this encoding, the semantically live tokens
(with age $< 1$) correspond to the counter values of the LCM while the older
(semantically dead) tokens count as lost. Thus the TPN is semantically bounded iff the LCM is
bounded. Since boundedness of LCM is undecidable \cite{Mayr:LCM:TCS}, the result follows.\qed

\section{Summary and Conclusions}\label{sec:conclusion}

\subsection{Problems and their Relation to each other}

We considered the following sets of markings of a given TPN.
\begin{enumerate}[$\bullet$]
\item
${\it NONTERM}$, the set of markings which have an infinite run.
\item
${\it NONZENO}$, the set of markings which have an infinite non-zeno run.
\item
${\it ZENO}$, the set of markings which have an infinite zeno run.
\item
${\it ALLZENO}$, the set of markings which have arbitrarily fast infinite runs.
\item
${\it ZEROTIME}$, the set of markings which have an infinite run taking no time at all. 
\end{enumerate}
Note that ${\it NONZENO}$ is not the complement of ${\it ZENO}$. A marking of
a TPN can have infinite zeno runs or infinite non-zeno runs or both or
neither. However,
${\it NONTERM} = {\it NONZENO}\, \cup\, {\it ZENO}$.

First we consider the relationships between these sets, both for dense-timed
Petri nets and discrete-timed Petri nets.

For discrete-timed Petri nets, we trivially have ${\it ALLZENO} = {\it ZEROTIME}$,
but for dense-timed nets ${\it ZEROTIME} \subset {\it ALLZENO}$, in general.
For example, in the TPN of Figure~\ref{fig:nasty} we have
that the marking $[(X,1),(Y,1),(A,1),(B,1)] \in {\it ALLZENO}$, but 
the marking $[(X,1),(Y,1),(A,1),(B,1)] \notin {\it ZEROTIME}$.

For discrete-timed nets, every zeno-computation has an infinite suffix
that takes no time at all and thus ${\it Pre}^*({\it ZEROTIME}) = {\it ZENO}$.
However, for dense-timed Petri nets, it was shown in Lemma~\ref{lem:zeno_allzeno} that
there exist instances (e.g., Figure~\ref{fig:nasty}) 
where ${\it Pre}^*({\it ALLZENO}) \subset {\it ZENO}$, i.e., a strict subset.

The inclusion ${\it ALLZENO} \subseteq {\it Pre}^*({\it ALLZENO})$ follows
directly from the definition of ${\it Pre}^*$. 
The following example shows that there exist instances where the inclusion is
strict, i.e., ${\it ALLZENO} \subset {\it Pre}^*({\it ALLZENO})$. (This works
for both dense- and discrete time.)
One constructs a TPN and marking $M_0$ such that at $M_0$ one must first wait 1 time
unit before the first transition can fire. This transition then creates 
a marking $M_1 \in {\it ALLZENO}$. Thus 
$M_0 \in {\it Pre}^*({\it ALLZENO})$, but $M_0 \notin {\it ALLZENO}$.

Furthermore, it is trivial (for both dense- and discrete time) that 
${\it ZENO} \subseteq {\it NONTERM}$ and ${\it NONZENO} \subseteq {\it
  NONTERM}$, and that there exist instances where these inclusions are strict.
In general, the sets ${\it ZENO}$ and ${\it NONZENO}$ are incomparable.
Finally, ${\it ZENO}\, \cup\, {\it NONZENO} = {\it NONTERM}$.
The following theorem summarizes these results.

\begin{thm}
In general for dense-timed Petri nets 
{\small
\[
{\it ZEROTIME} \subseteq {\it ALLZENO} \subseteq {\it Pre}^*({\it ALLZENO})
\subseteq {\it ZENO} \subseteq {\it ZENO} \cup {\it NONZENO} = {\it
  NONTERM}
\]
}
and for each inclusion there is an instance where it is strict.

In general for discrete-timed Petri nets
{\small
\[
{\it ZEROTIME} = {\it ALLZENO} \subseteq {\it Pre}^*({\it ALLZENO}) = {\it ZENO}
\subseteq {\it ZENO} \cup {\it NONZENO} = {\it NONTERM}
\]
}
and for each inclusion there is an instance where it is strict.
\end{thm}

\subsection{Decidability Results}

It has been shown in this paper that the sets ${\it ZEROTIME}$, 
${\it ALLZENO}$, ${\it Pre}^*({\it ALLZENO})$, and ${\it ZENO}$ are
effectively constructible as MRUC 
(multi-region upward closures; see Def.~\ref{def:MRUC})
for dense-timed nets and thus also for
discrete-timed nets.
Furthermore, we have shown in Section~\ref{sec:non_zeno} that ${\it NONZENO}$
is undecidable for dense-timed nets. This undecidability proof carries over directly to
discrete-timed nets, since all delays are of length $\ge 1$.

The situation is slightly more complex for the set ${\it NONTERM}$. 
Theorem~\ref{thm:nonterm} showed the decidability of the non-termination
problem for dense-timed nets.
This decidability result trivially carries over to discrete-timed nets.
Like all the other sets of markings considered here, 
the set ${\it NONTERM}$ is closed under the relation $\equiv$ on regions (see
Def.~\ref{region:definition}) and it is also upward-closed. Thus it is
representable as a MRUC.\/ However, this MRUC is not effectively 
constructible. It has been shown by Escrig et al. \cite{Escrig:etal:TPN}
that ${\it NONTERM}$ is not effectively constructible even for discrete-timed
Petri nets. Their proof is similar to the construction in
Section~\ref{sec:non_zeno} (except for the initial guessing phase).
A timed Petri net can simulate a lossy counter machine (or a reset Petri
net). Thus, if one could effectively construct ${\it NONTERM}$, then one could
decide the universal termination problem for lossy counter machines 
$\exists n.\, {\it LCM}^\omega$ 
(see Section~\ref{sec:non_zeno}) which is known to be undecidable
\cite{Mayr:LCM:TCS}.

The following table summarizes the results on decidability and effective
constructibility of the considered sets of markings of TPN.\/ Note that
all those results coincide for discrete-timed nets and dense-timed nets.
However, the proofs are harder for dense-timed nets.

\bigskip
\begin{center}
\begin{tabular}{|l||c|c|}\hline
Set               & Decidable? & Effectively constructible?\\ \hline\hline
${\it NONTERM}$   & Yes (Thm. \ref{thm:nonterm}) & No
(\cite{Escrig:etal:TPN})\\ \hline
${\it NONZENO}$   & No (Thm. \ref{thm:non_zeno}) & No
(Thm. \ref{thm:non_zeno})\\ \hline
${\it ZENO}$      & Yes (Thm. \ref{thm:main_zeno}) & Yes
(Thm. \ref{thm:main_zeno}) \\ \hline
${\it Pre}^*({\it ALLZENO})$ & Yes (Thm. \ref{thm:main_allzeno} and
Lemma \ref{regions:prestar}) & Yes (Thm. \ref{thm:main_allzeno} and
Lemma \ref{regions:prestar}) \\ \hline
${\it ALLZENO}$   & Yes (Thm. \ref{thm:main_allzeno}) & Yes
(Thm. \ref{thm:main_allzeno}) \\ \hline
${\it ZEROTIME}$  & Yes (Thm. \ref{thm:main_zerotime}) & Yes
(Thm. \ref{thm:main_zerotime})
\\ \hline
\end{tabular}
\end{center}
\bigskip

\subsection{Conclusion and Future Work}

We have solved several open problems about the
verification of {\it dense-timed Petri nets (TPNs)} in which each
token has an age represented by a real number, where the
transitions are constrained by the ages of the tokens and the firing semantics
is lazy.
This class is closely related to the class of parameterized
systems of timed processes where each process is
restricted to have a single clock \cite{Parosh:Bengt:Timed:Networks:journal}.

We have shown decidability of zenoness, existence of arbitrarily fast
computations, token-liveness and syntactic 
boundedness for TPNs, as well as the undecidability of universal zenoness.

To solve the zenoness problem, we defined a new class of untimed
Petri nets (SD-TN) which is more general than standard Petri nets, but
which is a subclass of transfer nets. For these SD-TN,
we gave a method to compute a characterization of the
set of markings from which there are infinite computations.
This is interesting in itself, since for  general transfer
nets such a characterization 
is not computable \cite{DJS:ICALP99,Mayr:LCM:TCS}.

We have considered TPNs with just one real-valued clock per token.
For all the problems studied so far, the decidability results 
coincide for dense-time and discrete-time (although the proofs for 
dense-time are harder).

However, if we consider TPNs with {\it two} clocks per token, 
there is a decidability gap between the dense-time and the 
discrete-time domain. The coverability problem becomes undecidable 
for dense-timed TPNs with only {\em two} clocks per token, while 
it remains decidable for discrete-timed TPNs with any finite number 
of clocks per token \cite{ADM:TN2}. 
The class of TPNs with multiple clocks per token is related to
parameterized systems of timed processes, with multiple 
clocks per process \cite{ADM:TN2}.
It is therefore worth investigating whether this more general class
induces a similar gap for the problems we have considered in this paper.

\section*{Acknowledgement}
The authors wish to thank the anonymous referees for their detailed comments.

\bibliographystyle{alpha}
\bibliography{bibdatabase,base,thesis}
\end{document}